%% file: main.tex
\documentclass[journal,11pt,onecolumn,web]{ieeecolor}

\let\labelindent\relax 

\usepackage{enumitem}
\usepackage{generic}
\usepackage{float}
\usepackage{cite}
\usepackage{amsmath,amssymb,amsfonts}
\usepackage{mathtools}
\usepackage{bbm}
\usepackage{bm}
\usepackage{url}
\usepackage{algorithmic}
\usepackage{graphicx}
\usepackage[normalem]{ulem}
\usepackage{textcomp}

\newtheorem{theorem}{Theorem}

\newtheorem{remark}{Remark}
\newtheorem{lemma}{Lemma}
\newtheorem{assumption}{Assumption}
\newtheorem{definition}{Definition}

\begin{document}

\title{On Optimizing the Conditional Value-at-Risk of a Maximum Cost for Risk-Averse Safety Analysis*}

\author{Margaret P. Chapman$^\dagger$, \IEEEmembership{Member, IEEE}, Michael Fau{\ss}$^\ddagger$, \IEEEmembership{Member, IEEE}, and Kevin M. Smith$^{**}$ 
\thanks{A short version of this work has been accepted conditionally by \emph{IEEE Transactions on Automatic Control} in May 2022. This work was supported in part by the Computational Hydraulics International University Grant Program for complementary use of PCSWMM Professional software. K. M. Smith was supported in part by the U.S.  National Science Foundation under Grant NSF-NRT 2021874. The work of M.~Fau{\ss} was supported by the German Research Foundation (DFG) under grant number 424522268. M. P. Chapman acknowledges support from the University of Toronto and the Natural Sciences and Engineering Research Council of Canada (NSERC) Discovery Grants Program, [RGPIN-2022-04140]. Cette recherche a \'{e}t\'{e} financée par le Conseil de recherches en sciences naturelles et en g\'{e}nie du Canada (CRSNG).}
\thanks{$^\dagger$M. P. Chapman is with the Edward S. Rogers Sr. Department of Electrical and Computer Engineering, University of Toronto, Toronto, Ontario M5S 3G4 Canada (email: mchapman@ece.utoronto.ca).}
\thanks{$^\ddagger$M. Fau{\ss} is with the Department of Electrical and Computer Engineering, Princeton University, Princeton, New Jersey 08544 USA (email: mfauss@princeton.edu).}
\thanks{$^{**}$K. M. Smith is with the Department of Civil and Environmental Engineering, Tufts University, Medford, MA 02155 USA and OptiRTC, Inc., Boston, MA 02116 USA (email: kevin.smith@tufts.edu).}
\thanks{*This work solves the risk-averse safety analysis problem. Our prior works \cite{chapmanACC, chapmantac2021} offer approximations.}
}

\maketitle
\pagestyle{empty}
\thispagestyle{empty}
\begin{abstract}
The popularity of Conditional Value-at-Risk (CVaR), a risk functional from finance, has been growing in the control systems community due to its intuitive interpretation and axiomatic foundation. We consider a nonstandard optimal control problem in which the goal is to minimize the CVaR of a maximum random cost subject to a Borel-space Markov decision process. The objective represents the maximum departure from a desired operating region averaged over a given fraction of the worst cases. This problem provides a safety criterion for a stochastic system that is informed by both the \emph{probability} and \emph{severity} of the potential consequences of the system's behavior. In contrast, existing safety analysis frameworks apply stage-wise risk constraints or assess the probability of constraint violation without quantifying the potential severity of the violation.
To the best of our knowledge, the problem of interest has not been solved. To solve the problem, we propose and study a family of stochastic dynamic programs on an augmented state space. We prove that the optimal CVaR of a maximum random cost enjoys an equivalent representation in terms of the solutions to these dynamic programs under appropriate assumptions. For each dynamic program, we show the existence of an optimal policy that depends on the dynamics of an augmented state under the assumptions. In a numerical example, we illustrate how our safety analysis framework is useful for assessing the severity of combined sewer overflows under precipitation uncertainty.
\end{abstract}

\begin{IEEEkeywords}
Conditional Value-at-Risk, Risk-averse optimal control, Safety analysis, Markov decision processes.
\vspace{-3mm}
\end{IEEEkeywords}

\section{Introduction}\label{secI}
\input{1_introduction}
\section{Control System Model}\label{secII}
\input{2_dynamical_system_model}
\section{Risk-Averse Safety Analysis}\label{secIII}
\input{3_risk_sensitive_safety_analysis_problem}

\section{Characterization of Risk-averse Safe Sets using Stochastic Dynamic Programs}\label{secIV}
\input{4_computation_method}
\section{Numerical Example}\label{secV}
\input{5_numerical_example_extended}

\newpage
\section{Conclusions}\label{secVI}
\input{6_conclusions}
\section{Appendix}\label{secVII}
\input{7_appendix_for_extended}
\bibliographystyle{ieeecolor}

\section*{Acknowledgement}
The authors would like to thank Dr. H. Vincent Poor and Mr. Chuanning Wei for discussions.

\end{document}

%% file: 1_introduction.tex
\begin{figure}[ht]
\centerline{\includegraphics[width=0.75\columnwidth]{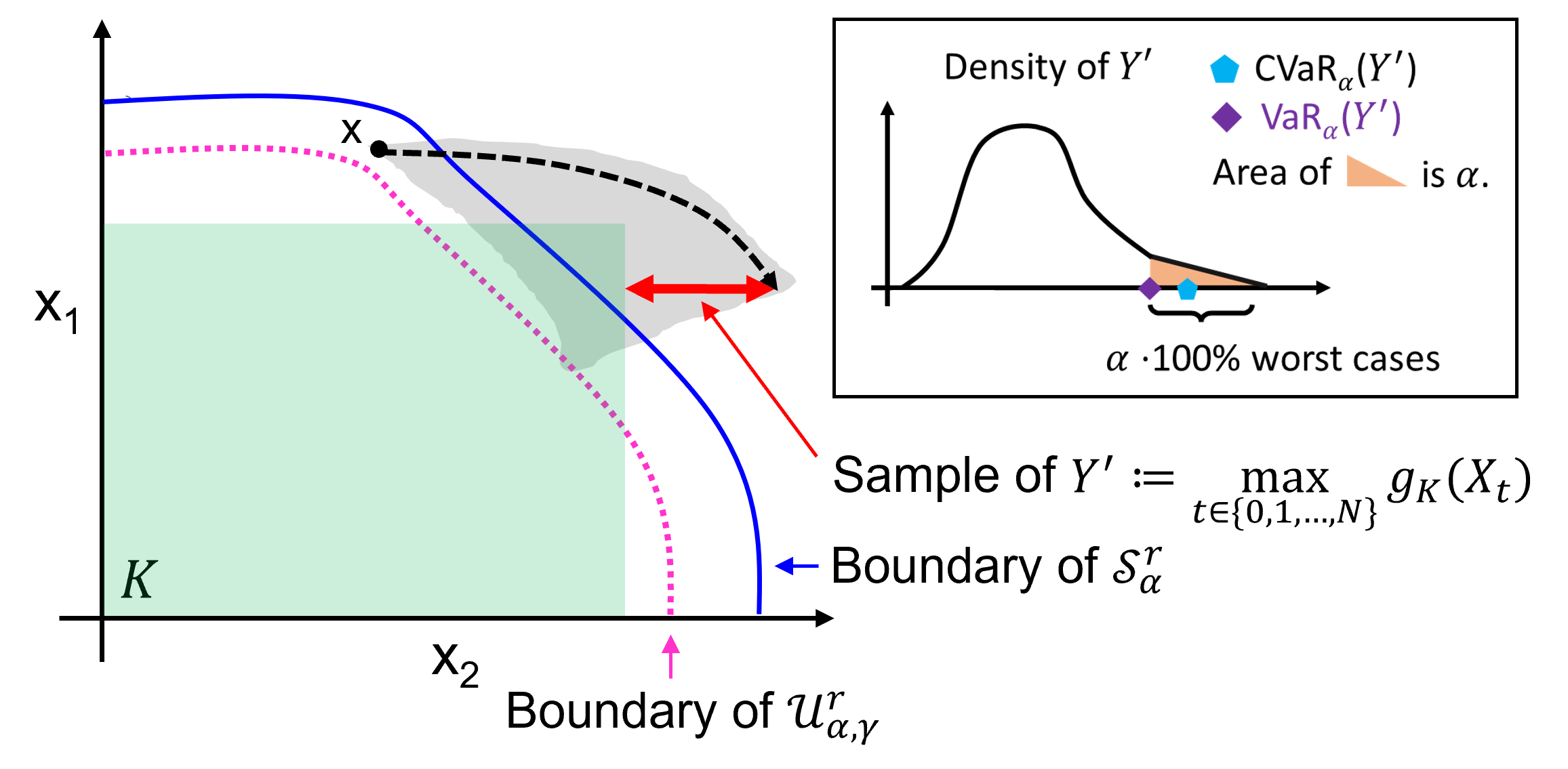}}\vspace{-2mm}
\caption{A risk-averse safe set $\mathcal{S}_\alpha^r$ is the set of initial states from which the Conditional Value-at-Risk (CVaR) at level $\alpha \in (0,1]$ of a trajectory-wise maximum random cost can be reduced to a threshold $r \in \mathbb{R}$. \textcolor{black}{(A random cost is a random variable in which smaller realizations are preferred.)} While we depict a state-dependent maximum random cost $Y'$ in this figure, our theory permits control-dependent random costs as well. Our framework applies to settings in which leaving a desired operating region $K$ may be inevitable, but the extent of a departure should be limited when possible. ($K$ need not be a polytope. However, we require stage and terminal cost functions to be continuous and bounded. In this figure, $g_K(x)$ is a signed distance between a state $x$ and the boundary of $K$.) In this work, we prove that any collection of $\mathcal{S}_\alpha^r$ is given by the solutions to a family of stochastic dynamic programs under a measurable selection assumption. In a numerical example, we compare this characterization to our underapproximation method from \cite{chapmantac2021}. An underapproximation set $\mathcal{U}_{\alpha,\gamma}^r \subseteq \mathcal{S}_\alpha^r$ depends on a soft-maximum parameter $\gamma$ that requires tuning \cite{chapmantac2021}.}\vspace{-5mm}
\label{intpic}
\end{figure}

Control system safety is often assessed through minimax optimal control problems \cite{bertsekas1971minimax, lygeros2011, chen2018hamilton, sylviathesis}, which assume bounded nonstochastic adversarial disturbances that try to inhibit safe or efficient operation. In cases where disturbances are not well-modeled as bounded inputs (e.g., Gaussian noise), then it is standard to define safety in terms of a stochastic optimal control problem, whose optimal value is a probability of satisfactory operation. This framework, called stochastic safety analysis, can accommodate either adversarial \cite{ding2013, yang2018dynamic} or nonadversarial \cite{abate2008probabilistic, summers2010verification} stochastic disturbances. However, a minimax approach may lead to controllers that are too cautious in practice. On the other hand, a purely probabilistic risk assessment indicates the likelihood of a harmful event but has a limited capacity to quantify the amount of harm the event would cause. These different limitations have motivated a growing body of research that lies in the intersection of formal methods and risk analysis for control systems \cite{samuelson2018safety, chapmanACC, Safaoui2020, chapmantac2021, lindemann2021reactive}. 

Here, we study a nonstandard safety analysis problem, which concerns the notion of a \emph{risk-averse safe set} $\mathcal{S}_\alpha^r \coloneqq \{\mathbf{x} \in S : \mathcal{J}_\alpha^*(\mathbf{x}) \leq r\}$. $\mathcal{S}_\alpha^r$ represents the set of initial states from which the maximum distance between the trajectory and a desired operating region averaged over the $\alpha \cdot 100 \%$ worst cases can be reduced to a threshold $r$ (Fig.~\ref{intpic}). \textcolor{black}{The system of interest is a Markov decision process (MDP) with Borel spaces of states, controls, and disturbances, operating on a discrete-time horizon of length $N$, a natural number}. $\mathcal{J}_\alpha^*(\mathbf{x})$ is the optimal value of a stochastic optimal control problem with a Conditional Value-at-Risk maximum cost objective:
\begin{subequations}\label{keyproblem}
\begin{align}
    \mathcal{J}_\alpha^*(\mathbf{x}) & \coloneqq  \inf_{\pi \in \Pi} \text{CVaR}_{\alpha,\mathbf{x}}^\pi(Y), \\
    Y & \coloneqq \max_{t \in \{0,1,\dots,N-1\}}\{ c_t(X_t,U_t), c_N(X_N)\}.  \label{myY}
\end{align}
\end{subequations}
\textcolor{black}{The random variable $Y$ depends on stage and terminal cost functions $c_t$, random states $X_t$, and random controls $U_t$. The quantity $\text{CVaR}_{\alpha,\mathbf{x}}^\pi(Y)$ represents the average value of $Y$ in the $\alpha \cdot 100\%$ worst cases when the initial state is $\mathbf{x}$ and the system uses the control policy $\pi$.} A control policy provides distributions for the realizations of $U_0,U_1,\dots,U_{N-1}$. \textcolor{black}{(We will formalize $\text{CVaR}_{\alpha,\mathbf{x}}^\pi(Y)$ and $\pi$ in Sec. \ref{cvardefsec} and Sec. \ref{wherepolicydef}, respectively.)} The setting is fairly general in theory. It permits nonlinear dynamics, nonconvex bounded cost functions, continuous spaces, and non-Gaussian stochastic disturbances. First, we will explain why \eqref{keyproblem} is an important problem to solve, and then we will explain the novelty of our contribution.\vspace{-3mm}

\subsection{Relevance of the CVaR}\vspace{-1mm}
The CVaR functional, which defines the objective of \eqref{keyproblem}, provides an \emph{intuitive and quantitative interpretation for risk} because it represents the average value of a random variable in a fraction $\alpha$ of the worst cases \cite[Th. 6.2]{shapiro2009lectures}. Other common risk functionals do not have interpretations that are as consistent or clear. Expected utility risk functionals encode risk preferences using utility functions and their parameters \cite{howardmat1972, jacobson1973,whittle1981, bauerlerieder, saldi2020, chapmansmith2021}. It is challenging to provide a precise meaning for the parameter of the classical expected exponential utility functional, which limits its applicability to control systems with specific safety or performance requirements \cite{smithchapman2021}. It may be difficult to interpret a recursive risk functional because it takes the form $\rho_1(C_1 + \rho_2(C_2 + \cdots +  \rho_{N-1}(C_{N-1} + \rho_N(C_N))\cdots))$, where $C_i$ is a random variable and $\rho_i$ is a map between spaces of random variables \cite{ruszczynski2010risk, singh2018, bauerle2020markov}. A weighted sum of the mean and a moment-based dispersion functional, e.g., variance, standard deviation, and upper-semideviation \cite{shapiro2009lectures}, provides an heuristic for the probability and severity of more rare and harmful outcomes. 
%
%
%
%
The CVaR is arguably more intuitive than the broader class of spectral risk functionals, which are ``mixtures'' of the CVaR$_\alpha$ over the values of $\alpha$ \cite[Prop. 2.5]{bauerle2020minimizing}. The Value-at-Risk (VaR) at level $\alpha$, which is the left-side $(1-\alpha)$-quantile, has a clear quantitative interpretation. 
However, the VaR's ability to summarize the severity of harmful outcomes is limited because it is insensitive to the shape of the distribution beyond the ($1-\alpha$)-quantile. From a decision-theoretic perspective, the VaR has the disadvantage of lacking a desirable property called subadditivity \cite{Artzner}. Both of these shortcomings are overcome by the CVaR \cite{rockafellar2002conditional, shapiro2009lectures}.\vspace{-3mm}

\subsection{Relevance of the Maximum Cost} \label{relevanceofmaximumcost}\vspace{-1mm}
We focus on a maximum cost \eqref{myY} generated by an MDP rather than a cumulative cost. While a cumulative cost is typical for MDP problems \cite{bauerlerieder, bauerleott, ruszczynski2010risk, borkar, haskell, bauerle2020markov, bauerle2020minimizing}, a maximum cost is typical for robust safety and reachability analysis problems for nonstochastic systems, e.g., see \cite{chen2018hamilton, sylviathesis}, and the references therein. Maximum costs have natural roles in systems theory, beyond robust safety and reachability analysis. The theory of the long-term behavior of normalized maxima of random variables, i.e., extreme value theory, has applications in finance, the study of human longevity, and hydrology \cite{extremevaluetheory}. 

A maximum cost is appropriate for applications in which the \emph{extent} of a constraint violation over a brief time interval is more critical to assess than its accumulation.\footnote{A constraint violation means that a state or control leaves a desired operating region, and its extent refers to the severity of the violation.} For example, in stormwater management, the \emph{maximum} water level can be a useful surrogate for the maximum flood extent (in more extreme cases) and the maximum discharge rate (in general). These are instantaneous rather than cumulative properties. For gravity-drained stormwater systems, the instantaneous discharge rate through an uncontrolled outlet into open atmosphere is a function of the water level behind the outlet. Therefore, from water levels, we can estimate instantaneous demands on downstream conveyance infrastructure (i.e., infrastructure to transport water rather than to store it). Designing this infrastructure for the worst maximum discharge rate may be cost-prohibitive. However, assessing the average maximum water level in the worst $\alpha \cdot 100\%$ of cases from historical data would allow designers to estimate downstream conveyance capacity demands along a spectrum of worst cases. \vspace{-3mm}


\subsection{Related Literature} 
The problem of computing risk-averse safe sets $\mathcal{S}_\alpha^r$ is distinct from established problems in the stochastic and risk-averse control theory literature and necessitates different techniques. 
Classical discrete-time stochastic control theory, e.g., \cite{bertsekas2004stochastic}, studies the problem of optimizing the expectation of a cumulative cost. In contrast, our focus is optimizing the CVaR of a maximum cost \eqref{keyproblem}. The dynamic programming (DP) proofs from stochastic control theory do not apply to our problem directly. Theoretical challenges arise because the CVaR satisfies only some of the properties that are enjoyed by the expectation. Moreover, while sums and integrals of nonnegative Borel-measurable functions can be interchanged, this is not the case for maxima and integrals in general. Such technical differences between our problem and the scenarios that prevail in the literature make it necessary to
build a pathway from measure-theoretic first principles. Doing so enables us to solve for the sets $\mathcal{S}_\alpha^r$ and the associated optimal control policies under appropriate assumptions.

We take inspiration from a technique called \emph{state-space augmentation}, which has been used to solve risk-averse MDP problems with cumulative costs \cite{bauerleott, bauerlerieder, borkar, haskell, bauerle2020minimizing}.
The problem of minimizing the expectation of a cumulative cost subject to an upper bound on the CVaR of a cumulative cost has been studied in \cite{borkar}. The authors propose offline and online algorithms on augmented state spaces to update a Lagrange multiplier and a lower bound on a cumulative cost \cite{borkar}. Several risk-averse control problems with cumulative costs over an infinite time horizon have been investigated using infinite-dimensional linear programming and state-space augmentation \cite{haskell}. B{\"a}uerle and Ott provide a DP solution to the problem of minimizing the CVaR of a cumulative cost \cite{bauerleott}. While we also use DP, our approach requires different proof techniques to manage a maximum cost \eqref{myY} and to study our proposed algorithm, which we define in terms of dynamics functions $x_{t+1} = f_t(x_t,u_t,w_t)$, stage and terminal cost functions $c_t$, and disturbance distributions $p_t(\mathrm{d}w_t|x_t,u_t)$. 

%

Most literature about risk-averse MDPs concerns exponential utility, taking inspiration from decision theory in economics and extending from 1972 to present-day \cite{howardmat1972,jacobson1973,whittle1981, jaskiew2017, saldi2020, chapmansmith2021}. B{\"a}uerle and Rieder study the problem of optimizing an expected utility for systems on Borel spaces with state-space augmentation, analyzing exponential utility as a special case \cite{bauerlerieder}. Another line of work considers the optimization of recursive risk functionals \cite{ruszczynski2010risk, jaskiew2017, singh2018, bauerle2020markov}; the basic approach is to replace a conditional expectation with a ``conditional risk functional'' to derive a risk-based Bellman equation. 
The problem of minimizing an expected cumulative cost subject to a risk constraint has been studied by, e.g., \cite{borkar, haskell, van2015distributionally, Tsiamis, Safaoui2020, samuelson2018safety}. Linear-quadratic settings have been studied in \cite{van2015distributionally, Tsiamis, Safaoui2020}, and a safety analysis problem with CVaR has been proposed by \cite{samuelson2018safety}. Our problem \eqref{keyproblem} assesses the risk of the entire trajectory, whereas the framework in \cite{samuelson2018safety} is concerned with the risk of each state in the trajectory separately, i.e., $\text{CVaR}_\alpha(\psi(X_t))$ must be small for every $t$. An emerging research direction proposes risk-averse signal temporal logic specifications for linear-quadratic model predictive control \cite{Safaoui2020} and for a setting with continuous-time systems of the form $\dot x = f(x) + g(x) u$ \cite{lindemann2021reactive}. 
We refer the reader to our survey about risk-averse autonomous systems \cite{wangchapman} and the references therein for additional literature. 
%

\emph{Contributions.} 
We show that any collection of risk-averse safe sets is characterized exactly using the solutions to a family of stochastic dynamic programs on an
augmented state space under a measurable selection assumption. We derive this characterization by expressing the minimum CVaR (for a given initial state $\mathbf{x}$ and a given level $\alpha$) as a nested optimization problem with respect to a control policy and a dual parameter $s$. We propose a nonstandard stochastic dynamic program that is parametrized by $s$ to assess a maximum random cost. We show that the algorithm returns an optimal $s$-dependent value function and policy under regularity conditions on the dynamics functions, stage and terminal cost functions, and disturbance distributions. Subsequently, we perform an outer minimization over $s$ to obtain $\mathcal{J}_\alpha^*(\mathbf{x})$ \eqref{keyproblem}. The framework permits nonlinear dynamics, non-Gaussian noise, nonconvex bounded cost functions, and continuous spaces. We solve the risk-averse safety analysis problem, whereas our prior works \cite{chapmanACC,chapmantac2021} provide approximations.
%
For detailed derivations of our theory, we refer the interested reader to the Appendix.

\textcolor{black}{The numerical tractability of the method is limited due to its reliance on DP and an augmented state space. In this work, we provide a nonlinear two-dimensional example motivated by a stormwater management application and offer a comparison to our underapproximation method from \cite{chapmantac2021}. Our on-going and future work involves developing more scalable approaches using extreme value theory and value function approximations.}

\emph{Notation.} We define $\mathbb{R}^* \coloneqq \mathbb{R} \cup \{+\infty,-\infty\}$ and $\mathbb{N}\coloneqq \{1,2,\dots\}$. Given $N \in \mathbb{N}$, we define $\mathbb{T}\coloneqq \{0,1,\dots,N-1\}$ and $\mathbb{T}_N \coloneqq \mathbb{T} \cup \{N\}$. If $\mathcal{M}$ is a metrizable space, then $\mathcal{B}_{\mathcal{M}}$ is the Borel sigma algebra on $\mathcal{M}$. 
If $g : \mathcal{M} \rightarrow \mathbb{R}^*$, then $\min_{x \in \mathcal{M}} g(x)$ means that there is a point $x^* \in \mathcal{M}$ such that $g(x^*) =\inf_{x \in \mathcal{M}}  g(x)$; i.e., $g$ attains its infimum, and $x^*$ is a minimizer. If $g' : \mathcal{M}' \rightarrow \mathcal{M}$, where $\mathcal{M}'$ is a metrizable space, then we define $g \circ g' : \mathcal{M}' \rightarrow \mathbb{R}^*$ by $(g \circ g')(y) \coloneqq g(g'(y))$. If $\mathcal{M}$ is a Borel space, then $\mathcal{P}(\mathcal{M})$ is the space of probability measures on $(\mathcal{M},\mathcal{B}_{\mathcal{M}})$ with the weak topology; if $x \in \mathcal{M}$, then $\delta_x$ is the Dirac measure in $\mathcal{P}(\mathcal{M})$ that is concentrated at $x$. We distinguish between random objects and their realizations (i.e., values) using capital letters and lower-case letters, respectively. The abbreviation l.s.c. means lower semi-continuous. 

%% file: 2_dynamical_system_model.tex
We consider a fully observable MDP operating on a finite discrete-time horizon $\mathbb{T}_N$, where $N \in \mathbb{N}$ is given. The state space $S$, control space $C$, and disturbance space $D$ are nonempty Borel spaces. $X_t$, $U_t$, and $W_t$ are random objects, whose co-domains are $S$, $C$, and $D$, respectively.\footnote{The realizations of $X_t$, $U_t$, and $W_t$ include the possible states, controls, and disturbances at time $t$, respectively.}
The disturbance process $(W_0,W_1,\dots,W_{N-1})$ satisfies the following property: for every $t \in \mathbb{T}$, given $(X_t, U_t)$, $W_t$ is conditionally independent of $W_\tau$ for every $\tau \neq t$. 
The realizations of $X_0$ are concentrated at an arbitrary element $\mathbf{x}$ of $S$.
%
For every $t \in \mathbb{T}$, $p_t(\cdot|\cdot,\cdot)$ is a Borel-measurable stochastic kernel on $D$ given $S \times C$, providing a conditional distribution for the realizations of $W_t$.
For every $t \in \mathbb{T}$, if $(x,u) \in S \times C$ is the realization of $(X_t,U_t)$, then the probability that $X_{t+1}$ is in $\underline{S} \in \mathcal{B}_{S}$ is defined by
    \begin{equation}\label{myQ}
        q_t(\underline{S}|x,u) \coloneqq p_t\bigr( \{w \in D: f_t(x,u,w) \in \underline{S}\} \big| x,u \bigl),
    \end{equation}
where $f_t : S \times C \times D \rightarrow S$ is a Borel-measurable function for the dynamics.
The stage cost function $c_t : S \times C \rightarrow \mathbb{R}$ for every $t \in \mathbb{T}$ and the terminal cost function $c_N : S \rightarrow \mathbb{R}$ are Borel-measurable. 
\begin{assumption}[Measurable selection]\label{Assumption1} We assume:
\begin{enumerate}
\item There exist $a \in \mathbb{R}$ and $b \in \mathbb{R}$ such that $a \leq c_t \leq b$ for every $t \in \mathbb{T}_N$. (We define $\mathcal{Z} \coloneqq [a,b]$.) 
    \item The control space $C$ is compact.
    \item For every $t$, $f_t$ and $c_t$ are continuous functions, and $p_t(\cdot|\cdot,\cdot)$ is a continuous stochastic kernel.
\end{enumerate}
\end{assumption}

We will show that Assumption \ref{Assumption1} guarantees the existence of an optimal policy that depends on the dynamics of a running maximum (Sec. \ref{secIV}). It is standard to impose a measurable selection assumption for stochastic optimal control problems on Borel spaces, e.g., see \cite{bertsekas2004stochastic}. 
As risk-aware MDP problems can pose additional technical challenges, it is common to assume bounded costs, e.g., \cite{bauerleott, bauerlerieder, haskell, jaskiew2017}. 
We assume continuous cost functions $c_t$ because our cost-update operation is a composition of two functions (rather than a sum). \textcolor{black}{Hence, we replace the typical l.s.c. assumption by a property that is preserved under compositions.} \textcolor{black}{In the theoretical sections of this work, we assume that Assumption \ref{Assumption1} holds, even without an explicit statement.} 

%% file: 3_risk_sensitive_safety_analysis_problem.tex
First, we will present an example for the maximum random cost $Y$ \eqref{myY} in terms of a desired operating region $K$. Then, we will provide measure-theoretic definitions for $Y$ and CVaR to formalize our risk-averse safety specification $\mathcal{S}_\alpha^r$.\vspace{-3mm}

\subsection{$Y$ as a Distance between the State Trajectory and $K$} \vspace{-1mm}
Suppose that $K \in \mathcal{B}_S$ is a desired operating region. While we would like the state trajectory to remain inside $K$ always, this may not be possible due to disturbances that may arise. We will explain how one may choose $Y$ \eqref{myY} to represent a distance between the state trajectory and $K$.

Let $g_K : S \rightarrow \mathbb{R}$ be bounded and continuous, where $g_K(x)$ quantifies a signed distance between a state $x$ and the boundary of $K$. For example, if $S \in \mathcal{B}_{\mathbb{R}^2}$ is bounded and $K = [0,k_1] \times [0,k_2] \subset S$ is the set of desired water levels in two storage tanks, then $\max\{x_{1} - k_1, x_{2} - k_2, 0 \}$ or $\max\{x_{1} - k_1, x_{2} - k_2\}$ are suitable choices for $g_K(x)$ with $x = [x_1,x_2]^T \in S$. More generally, if $x$ is outside $K$ and far from its boundary, then $g_K(x)$ has a large positive value. Otherwise, if $x$ is inside $K$, then there are two options: 1) $g_K(x)$ equals zero, or 2) $g_K(x)$ equals a more negative value if $x$ is located more deeply inside $K$.
The former applies when there is no preference for certain trajectories inside $K$. The latter applies when there is a preference for trajectories that are inside $K$ \emph{and} farther from its boundary. 
%

To quantify the extent of the state trajectory's departure relative to $K$, we can choose the terminal and stage cost functions to be $g_K$. That is, we can choose $c_N = g_K$ and $c_t(x,u) = g_K(x)$ for every $t \in \mathbb{T}$ and $(x,u) \in S \times C$.
In this case, if $(x_0,x_1,\dots,x_N) \in S^{N+1}$ is the realization of $(X_0,X_1,\dots,X_N)$, then
    $y = \max \{ g_K(x_t) : t \in \mathbb{T}_N \}$
is the realization of $Y$ \eqref{myY}. In this example, $Y$ represents the extent of the state trajectory's departure from $K$, and we use the notation $Y' = Y$ (Fig. \ref{intpic}).\vspace{-3mm} 
%
%
%

\subsection{A CVaR-based Trajectory-wise Safety Specification}\label{cvardefsec}\vspace{-1mm}
To define risk-averse safe sets formally, we must describe $Y$ \eqref{myY} in measure-theoretic terms. Let $\mathbf{x} \in S$ be an initial state and $\pi \in \Pi$ be a control policy. (We will specify the control policy class $\Pi$ in Sec. \ref{secIV}.) $Y$ is a random variable defined on a probability space $(\Omega, \mathcal{B}_\Omega, P_{\mathbf{x}}^\pi)$. The sample space $\Omega$ contains all possible trajectories; a trajectory is a tuple of states, maximum stage costs, and controls over time. From Assumption 1, every $c_t$ is bounded below by $a \in \mathbb{R}$. Given $(\mathbf{x},a)$, $\pi$, and the system dynamics, there exists a unique probability measure $P_{\mathbf{x}}^\pi \in \mathcal{P}(\Omega)$ (Ionescu-Tulcea Theorem). We write $P_{\mathbf{x}}^\pi$ instead of $P_{\mathbf{x},a}^\pi$ for brevity. 
$E_{\mathbf{x}}^\pi(\cdot)$ denotes the expectation operator with respect to $P_{\mathbf{x}}^\pi$. Since the stage and terminal cost functions are bounded (Assumption 1), $Y$ is bounded everywhere. This is one way to ensure that $E_{\mathbf{x}}^\pi(|Y|)$ is finite, which will allow us to define $\text{CVaR}_{\alpha,\mathbf{x}}^\pi(Y)$. 

As we have mentioned, $\text{CVaR}_{\alpha,\mathbf{x}}^\pi(Y)$ represents the average value of $Y$ in the $\alpha \cdot 100 \%$ worst cases when the initial state is $\mathbf{x}$ and the system uses the control policy $\pi$. The meaning of the $\alpha \cdot 100 \%$ worst cases is made precise using a quantity called the Value-at-Risk of $Y$ at level $\alpha$, which we denote by $\text{VaR}_{\alpha,\mathbf{x}}^\pi(Y)$.
Formally, $\text{CVaR}_{\alpha,\mathbf{x}}^\pi(Y)$ is the expectation of $Y$ conditioned on the event that $Y$ exceeds $\text{VaR}_{\alpha,\mathbf{x}}^\pi(Y)$, provided that $\alpha \in (0,1)$ and the distribution function of $Y$ is continuous at $\text{VaR}_{\alpha,\mathbf{x}}^\pi(Y)$ \cite[Th. 6.2]{shapiro2009lectures}. The Value-at-Risk of $Y$ at level $\alpha \in (0,1)$ is defined by
\begin{equation}
    \text{VaR}_{\alpha,\mathbf{x}}^\pi(Y) \coloneqq \inf\{ y \in \mathbb{R} : P_{\mathbf{x}}^\pi(\{ Y \leq y \}) \geq 1-\alpha \},
\end{equation}
where $y \mapsto P_{\mathbf{x}}^\pi(\{ Y \leq y \})$ is the distribution function of $Y$. Now, for every $\alpha \in (0,1]$, we define $\text{CVaR}_{\alpha,\mathbf{x}}^\pi(Y)$ by
\begin{equation}\label{defcvar}
    \text{CVaR}_{\alpha,\mathbf{x}}^\pi(Y) \coloneqq \inf_{s \in \mathbb{R}} \Big( s + \textstyle \frac{1}{\alpha} E_{\mathbf{x}}^\pi(\max\{Y - s, 0\}) \Big),
\end{equation}
following Shapiro et al. \cite[Eq. (6.22)]{shapiro2009lectures}. We call $s \in \mathbb{R}$ a \emph{dual parameter}. Using the derivation from \cite[p. 258]{shapiro2009lectures}, one can show that if $\alpha \in (0,1)$, then $\text{CVaR}_{\alpha,\mathbf{x}}^\pi(Y)$ equals
\begin{align}\label{empstat}
    \text{VaR}_{\alpha,\mathbf{x}}^\pi(Y) + \textstyle \frac{1}{\alpha} E_{\mathbf{x}}^\pi(\max\{Y - \text{VaR}_{\alpha,\mathbf{x}}^\pi(Y), 0\}).
\end{align}
This relation implies that $\text{CVaR}_{\alpha,\mathbf{x}}^\pi(Y)$ assesses a probability-weighted average of the realizations of $Y$ above $\text{VaR}_{\alpha,\mathbf{x}}^\pi(Y)$. 

CVaR is an attractive choice for defining safety specifications for two reasons. First, the parameter $\alpha$ has a quantitative interpretation as a fraction of the worst cases. Second, CVaR assesses the part of a distribution \emph{above} a particular quantile and therefore is designed to assess more rare and harmful outcomes. We define risk-averse safe sets $\mathcal{S}_\alpha^r$ as the sublevel sets of the optimal CVaR of the maximum random cost $Y$.
\begin{definition}[$\mathcal{S}_\alpha^r$]\label{salphar}
For every $\alpha \in (0,1]$ and $r \in \mathbb{R}$, we define the $(\alpha,r)$-risk-averse safe set by $\mathcal{S}_\alpha^r \coloneqq \{ \mathbf{x} \in S : \mathcal{J}_{\alpha}^*(\mathbf{x}) \leq r \}$ with $\mathcal{J}_{\alpha}^*(\mathbf{x}) \coloneqq \inf_{\pi \in \Pi} \text{CVaR}_{\alpha,\mathbf{x}}^\pi(Y)$ \eqref{keyproblem}.
\end{definition}

In the next section, we will show that risk-averse safe sets can be characterized exactly using stochastic dynamic programs on an augmented state space. 

%% file: 4_computation_method.tex
Unlike the minimum expectation of a cumulative cost, $\mathcal{J}_{\alpha}^*$ cannot be computed using a DP recursion on the state space $S$ alone. Such a recursion holds in special cases due to the structure inherent in certain problems, but it does not hold universally. 
To alleviate the challenge of optimizing the CVaR of a maximum cost, we will construct an augmented state space to record the running maximum. \textcolor{black}{Recall that $\mathcal{Z} \hspace{-.3mm} = \hspace{-.3mm} [a,b]$.} \vspace{-3mm}

\subsection{Construction of an Augmented State Space}\vspace{-1mm}\label{wherepolicydef}
We define the random augmented state by $\mathcal{X}_t \coloneqq (X_t,Z_t)$ for every $t \in \mathbb{T}_N$. $X_t$ is the original $S$-valued random state. $Z_t$ is a $\mathcal{Z}$-valued random object that records the maximum stage cost up to time $t$ (to be further described). 
\textcolor{black}{The realizations of $\mathcal{X}_0 = (X_0,Z_0)$ are concentrated at $(\mathbf{x},a)$, where we recall that $\mathbf{x} \in S$ is arbitrary. $Z_{t+1}$ depends on $X_t$, $U_t$, and $Z_t$ as follows: $Z_{t+1} = \max\{c_t(X_t,U_t),Z_t\}$ for every $t \in \mathbb{T}$. We define $\mathbb{S} \coloneqq S \times \mathcal{Z}$ for brevity.} 
%

$\mathcal{X}_t$ and $U_t$ are functions defined on $\Omega \coloneqq (\mathbb{S} \times C)^N \times \mathbb{S}$. 
Every $\omega \in \Omega$ takes the form
\begin{equation}\label{myomega}
    \omega = (x_0,z_0,u_0,\dots,x_{N-1},z_{N-1},u_{N-1},x_N,z_N)
\end{equation}
with $(x_t,z_t) \in \mathbb{S}$ for every $t \in \mathbb{T}_N$ and $u_t \in C$ for every $t \in \mathbb{T}$. We define $\mathcal{X}_t(\omega) \coloneqq (X_t(\omega),Z_t(\omega)) \coloneqq (x_t,z_t)$ and $U_t(\omega) \coloneqq u_t$ for every $\omega \in \Omega$ whose coordinates are specified by \eqref{myomega}. It follows that $\mathcal{X}_t$ and $U_t$ are Borel-measurable functions. While these definitions are general enough to capture arbitrary dependencies between the coordinates of $\omega$, we restrict ourselves to particular casual dependencies, which we have discussed and will continue to present. Next, we will define the class $\Pi$ of control policies using the augmented state space $\mathbb{S}$.
\begin{definition}[$\Pi$]\label{def:control_policy}
Every control policy $\pi \in \Pi$ takes the form $\pi = (\pi_0,\pi_1,\dots,\pi_{N-1})$, where $\pi_t(\cdot|\cdot,\cdot)$ is a Borel-measurable stochastic kernel on $C$ given $\mathbb{S}$ for every $t \in \mathbb{T}$.
\end{definition}
\begin{remark}[$\Pi$ is history-dependent]
\textcolor{black}{Let $\pi \in \Pi$ be given, and suppose that $(x_t,z_t) \in \mathbb{S}$ is the realization of $\mathcal{X}_t = (X_t,Z_t)$.} 
\textcolor{black}{The distribution $\pi_t(\cdot|x_t,z_t) \in \mathcal{P}(C)$ for the realizations of $U_t$ depends on $(x_t,z_t)$, which depends on the previous states and controls.} 
\end{remark}
\begin{remark}[A deterministic control law $\delta_{\kappa}$]\label{detcontrollaw}
Let $\kappa : \mathbb{S} \rightarrow C$ be Borel-measurable. We use the notation $\delta_{\kappa}$ to denote the following Borel-measurable stochastic kernel on $C$ given $\mathbb{S}$: for every $(x,z) \in \mathbb{S}$, $\delta_{\kappa(x,z)}$ is the Dirac measure in $\mathcal{P}(C)$ that is concentrated at the point $\kappa(x,z) \in C$.
\end{remark}

\textcolor{black}{The next remark presents a convenient notation for an element of $\mathbb{S}$ and a transition law for the realizations of $\mathcal{X}_{t+1}$.}
\begin{remark}[$\chi_t$, $\tilde{q}_t$]
\textcolor{black}{The notation $\chi_t = (x_t,z_t)$ denotes an element of $\mathbb{S}$. For every $t \in \mathbb{T}$ and $(\chi_t,u_t) \in \mathbb{S} \times C$, let $\tilde{q}_t(\cdot|\chi_t,u_t)$ be the product measure of $q_t(\cdot|x_t,u_t)$ \eqref{myQ} and $\delta_{\max\{c_t(x_t,u_t),z_t\}}$. $\tilde{q}_t$ is a continuous stochastic kernel on $\mathbb{S}$ given $\mathbb{S} \times C$ by applying Assumption \ref{Assumption1} (see Appendix).}
\end{remark}

Now, we are ready to formalize the expectation operator $E_{\mathbf{x}}^\pi(\cdot)$. Let $\mathbf{x} \in S$ and $\pi \in \Pi$ be given. If $G : \Omega \rightarrow \mathbb{R}^*$ is Borel-measurable and $E_{\mathbf{x}}^\pi(G) \coloneqq \int_{\Omega} G \; \mathrm{d}P_{\mathbf{x}}^\pi$ exists, then
\begin{align}
 E_{\mathbf{x}}^\pi(G) =   \textstyle \int_{\mathbb{S}} \int_{C} \hspace{-.3mm} \cdots\hspace{-.3mm} \int_{\mathbb{S}} \hspace{-.1mm} G(\chi_0,u_0,\dots,\chi_N)   \; \tilde{q}_{N-1}(\mathrm{d}\chi_{N}|\chi_{N-1},u_{N-1})  \cdots \pi_0(\mathrm{d}u_0|\chi_0) \;  \delta_{\mathbf{x},a}(\mathrm{d}\chi_0), \label{expectation} 
\end{align}
by applying \cite[Prop. 7.28]{bertsekas2004stochastic} and Assumption \ref{Assumption1} (see Appendix). \textcolor{black}{The kernels in \eqref{expectation} describe how an augmented state $\chi_0 = (x_0,z_0)$ may lead to a control $u_0$, how $(\chi_0,u_0)$ may lead to a subsequent augmented state $\chi_1 = (x_1,z_1)$, and so on. The point $(\mathbf{x},a)$ serves as the initial augmented state.}\vspace{-3mm}

%
\subsection{Characterization of Risk-Averse Safe Sets}\vspace{-1mm}
Here, we show that risk-averse safe sets enjoy an equivalent representation in terms of a family of stochastic dynamic programs on the augmented state space under Assumption \ref{Assumption1}.
For convenience, for every $s \in \mathbb{R}$, we define $h^s : \mathbb{R} \rightarrow \mathbb{R}$ by
\begin{align}\label{myh}
h^s(y) \coloneqq \max\{y - s,0\}.
\end{align}
Let $\mathbf{x} \in S$ and $\alpha \in (0,1]$ be given. The optimal value $\mathcal{J}_\alpha^*(\mathbf{x})$ \eqref{keyproblem} can be expressed using the definitions of $\text{CVaR}_{\alpha,\mathbf{x}}^\pi(Y)$ \eqref{defcvar} and $h^s$ \eqref{myh} as follows: 
\begin{align}
     \mathcal{J}_\alpha^*(\mathbf{x}) 
      = \inf_{s \in \mathbb{R}}\Big( s + {\textstyle \frac{1}{\alpha}} \inf_{\pi \in \Pi} E_{\mathbf{x}}^\pi(h^s(Y)) \Big),\label{my10}
\end{align}
where we exchange the order of the infima over $\mathbb{R}$ and $\Pi$. By the definition of $Y$ \eqref{myY} and Assumption 1, we have that $Y(\omega) \in \mathcal{Z}$ for every $\omega \in \Omega$. Consequently, a minimizer in $\mathcal{Z}$ exists for the outer problem of \eqref{my10} by the next lemma. 
\begin{lemma}[Existence of a minimizer]\label{inflemma1}
\textcolor{black}{Let Assumption \ref{Assumption1} hold}, $\mathbf{x} \in S$, $\alpha \in (0,1]$, $G : \Omega \rightarrow \mathbb{R}$ be Borel-measurable, and $G(\omega) \in [a,b]$ for every $\omega \in \Omega$. Define $L_{\mathbf{x}}^\alpha(s) \coloneqq s + {\textstyle\frac{1}{\alpha}} \inf_{\pi \in \Pi} E_{\mathbf{x}}^\pi(h^s(G))$ for every $s \in \mathbb{R}$. Then,
  $  \inf_{s \in \mathbb{R}} L_{\mathbf{x}}^\alpha(s) 
  = \min_{s \in [a,b]} L_{\mathbf{x}}^\alpha(s)$, 
i.e., a minimizer $s_{\mathbf{x},\alpha}^*  \in [a,b]$ exists.
\end{lemma}
\hspace{-4mm}\begin{proof}
Define $\ell \coloneqq \inf_{s \in [a,b]} L_{\mathbf{x}}^\alpha(s)$. Then, for every $s \in [a,b]$, $L_{\mathbf{x}}^\alpha(s) \geq \ell$. Now, if $s \leq a$, then $h^s(G) = G - s$, and hence, $L_{\mathbf{x}}^\alpha(s) \geq L_{\mathbf{x}}^\alpha(a) \geq \ell$. However, if $s \geq b$, then $h^s(G) = 0$, and thus, $L_{\mathbf{x}}^\alpha(s) \geq L_{\mathbf{x}}^\alpha(b) \geq \ell$. Since $L_{\mathbf{x}}^\alpha(s) \geq \ell$ for every $s \in \mathbb{R}$, $\ell = \inf_{s \in \mathbb{R}} L_{\mathbf{x}}^\alpha(s)$ holds.
Since $L_{\mathbf{x}}^\alpha(s)$ is continuous in $s$ and $[a,b]$ is compact, the infimum $\ell$ is attained by a point $s_{\mathbf{x},\alpha}^*  \in [a,b]$ \cite[Th. A6.3]{ash1972}. 
\end{proof}

For every $s \in \mathbb{R}$, we define $V^{s} : S \rightarrow \mathbb{R}^*$ by
\begin{equation}\label{myVs}
    V^{s}(\mathbf{x}) \coloneqq \inf_{\pi \in \Pi} E_{\mathbf{x}}^\pi( h^s(Y) ).
\end{equation}
By Lemma \ref{inflemma1}, there exists a point $s_{\mathbf{x},\alpha}^* \in \mathcal{Z}$ such that
\begin{align}\label{10}
    \mathcal{J}_\alpha^*(\mathbf{x}) & = \min_{s \in \mathcal{Z}} \, \bigl( \, s + {\textstyle\frac{1}{\alpha}}V^{s}(\mathbf{x}) \, \bigr) = s_{\mathbf{x},\alpha}^* + {\textstyle\frac{1}{\alpha}}V^{s_{\mathbf{x},\alpha}^*}(\mathbf{x}).
\end{align}
We will develop a dynamic programming-based solution for $V^{s}$ to characterize $\mathcal{J}_\alpha^*$. Toward this aim, we define extended random variables that represent costs-to-go. For every $s \in \mathbb{R}$ and $t \in \mathbb{T}_N$, we define $Y_t^s : \Omega \rightarrow \mathbb{R}^*$ by
\begin{align}
       Y_t^s & \coloneqq \begin{cases} h^s(\max\{ c_N(X_N), A_t, Z_t \}), & \text{if }t \in \mathbb{T}, \\ h^s(\max\{ c_N(X_N), Z_N \}), & \text{if }t = N,\end{cases}  \label{42}
\end{align}
with $A_t : \Omega \rightarrow \mathbb{R}$,
   $A_t  \coloneqq \max_{i \in \{t,\dots,N-1\}} c_i(X_i,U_i)$,  $t \in \mathbb{T}$.
The next theorem specifies some properties of a conditional expectation $\phi_t^{\pi,s}(x,z) = E^\pi(Y_t^s|\mathcal{X}_t = (x,z))$ of $Y_t^s$ given $\mathcal{X}_t$. \textcolor{black}{The theorem is based on the definition of conditional expectation \cite[Th. 6.3.3]{ash1972} and a basic change-of-measure theorem \cite[Th. 1.6.12]{ash1972}. For brevity, we use the notation $\int_{\Omega} \varphi \circ \mathcal{X}_t\; \mathrm{d}P_{\mathbf{x}}^\pi \coloneqq \int_{\Omega} \varphi(\mathcal{X}_t(\omega))\; \mathrm{d}P_{\mathbf{x}}^\pi(\omega)$, where $\varphi : \mathbb{S} \rightarrow \mathbb{R}^*$ is Borel-measurable.}
\begin{theorem}[Properties of $\phi_t^{\pi,s}$]\label{thm1}
\textcolor{black}{Let Assumption \ref{Assumption1} hold}, and let $\mathbf{x} \in S$, $\pi \in \Pi$, and $s \in \mathbb{R}$ be given. Define the function $J_N^s : \mathbb{S} \rightarrow \mathbb{R}^*$ by
\begin{equation}\label{defJNs}
    J_N^s(x,z) \coloneqq h^s( \max\{ c_N(x), z \}).
\end{equation}
Then, the following statements hold:
\begin{align}
  E_{\mathbf{x}}^\pi( h^s(Y) ) & =  \textstyle \int_{\Omega}  \phi_0^{\pi,s} \circ \mathcal{X}_0\; \mathrm{d}P_{\mathbf{x}}^\pi = \phi_0^{\pi,s}(\mathbf{x},a),
     \label{46}\\
  \textstyle  \int_{\Omega} \phi_N^{\pi,s} \circ \mathcal{X}_N \; \mathrm{d}P_{\mathbf{x}}^\pi & = \textstyle\int_{\Omega} J_N^s \circ \mathcal{X}_N \; \mathrm{d}P_{\mathbf{x}}^\pi,  \label{47} \\
  \textstyle \int_{\Omega} \phi_t^{\pi,s} \circ \mathcal{X}_t\; \mathrm{d}P_{\mathbf{x}}^\pi & = \textstyle\int_{\Omega}  \phi_{t+1}^{\pi,s} \circ \mathcal{X}_{t+1}\; \mathrm{d}P_{\mathbf{x}}^\pi, \quad t \in \mathbb{T}. \label{66}
\end{align}
\end{theorem}\vspace{2mm}
\hspace{-4mm}\begin{proof}
For every $t \in \mathbb{T}_N$, $Y_t^s$ is an extended random variable on $(\Omega,\mathcal{B}_\Omega,P_{\mathbf{x}}^\pi)$, $\mathcal{X}_t : \Omega \rightarrow \mathbb{S}$ is Borel-measurable, and $\int_{\Omega} Y_t^s \; \mathrm{d}P_{\mathbf{x}}^\pi$ exists (recall that $Y_t^s$ is nonnegative). The probability measure induced by $\mathcal{X}_t$ is defined by $P_{\mathbf{x},\mathcal{X}_t}^\pi(\underline{\mathbb{S}}) \coloneqq P_{\mathbf{x}}^\pi(\mathcal{X}_t^{-1}(\underline{\mathbb{S}}))$ for every $\underline{\mathbb{S}} \in \mathcal{B}_{\mathbb{S}}$. 
By the definition of conditional expectation \cite[Th. 6.3.3]{ash1972} and the change-of-measure theorem \cite[Th. 1.6.12]{ash1972}, we have
\begin{equation}\label{20}
  \textstyle \int_{\Omega} Y_t^s \; \mathrm{d}P_{\mathbf{x}}^\pi  = \int_{\Omega} \phi_t^{\pi,s}\circ \mathcal{X}_t \; \mathrm{d}P_{\mathbf{x}}^\pi, \quad  t \in \mathbb{T}_N, 
\end{equation} 
where the integrals exist. Now,
\begin{equation}\label{21}
   \textstyle \int_{\Omega} Y_t^s \; \mathrm{d}P_{\mathbf{x}}^\pi = \int_{\Omega} Y_{t+1}^s \; \mathrm{d}P_{\mathbf{x}}^\pi, \quad  t \in \mathbb{T}, 
\end{equation}
as a consequence of $Z_{t+1} = \max\{c_t(X_t,U_t), Z_t\}$. The relations \eqref{20}--\eqref{21} imply the relation \eqref{66}. The relation \eqref{46} is derived using \eqref{expectation} and \eqref{20} with $t = 0$; note that $E_{\mathbf{x}}^\pi(Y_0^s) = E_{\mathbf{x}}^\pi( h^s(Y) )$ because $a \leq c_t$ for every $t \in \mathbb{T}_N$ and the realizations of $(X_0,Z_0)$ are concentrated at $(\mathbf{x},a)$. The relation \eqref{47} holds by \eqref{20} with $t = N$ and by $Y_N^s = J_N^s \circ \mathcal{X}_N$.
\end{proof}

Subsequently, we will use Theorem \ref{thm1} to derive a DP-based solution for $V^s$ \eqref{myVs}, and we will show the existence of a control policy that is optimal for $V^s$ under Assumption 1. 
%
\begin{theorem}[DP on $\mathbb{S}$]\label{thm2}
Let Assumption 1 hold, and let $s \in \mathbb{R}$ be given. Recall the definition of $J_N^s$ \eqref{defJNs}. For $t = N-1,\dots,1,0$, we define $J_t^s : \mathbb{S} \rightarrow \mathbb{R}^*$ recursively by
\begin{subequations}
\begin{equation}
    J_t^s(x, z) \coloneqq \inf_{u \in C} v_t^s(x,z,u),
\end{equation}
where we define $v_t^s : \mathbb{S} \times C \rightarrow \mathbb{R}^*$ by $v_t^s(x,z,u) \coloneqq$
\begin{align}
   \textstyle \int_{D} J_{t+1}^s\bigl(f_t(x,u,w),  \max\{c_t(x, u),z \}\bigr) \; p_t(\mathrm{d} w|x,u). 
\end{align}
\end{subequations}
Then, for every $t \in \mathbb{T}_N$, $J_t^s$ is l.s.c. and bounded below by zero. Moreover, for every $t \in \mathbb{T}$, there is a Borel-measurable function $\kappa_t^s : \mathbb{S} \rightarrow C$ such that
\begin{equation}\label{mykappaeq}
J_t^s(x, z) = v_t^s(x,z,\kappa_t^s(x,z)), \quad  (x,z) \in \mathbb{S}.
\end{equation}
We define $\pi^s \coloneqq (\delta_{\kappa_0^s},\delta_{\kappa_1^s},\dots,\delta_{\kappa_{N-1}^s})$, which is an element of $\Pi$. Then, for every $\mathbf{x} \in S$, we have
\begin{equation}\label{key}
         J_0^s(\mathbf{x}, a) = V^{s}(\mathbf{x}) = E_{\mathbf{x}}^{\pi^s}(h^s(Y)).
\end{equation}
\end{theorem}\vspace{2mm}
\hspace{-3mm}\begin{proof} $J_t^s$ being l.s.c. and bounded below by zero for every $t \in \mathbb{T}_N$ and the existence of a Borel-measurable function $\kappa_t^s : \mathbb{S} \rightarrow C$ that satisfies \eqref{mykappaeq} for every $t \in \mathbb{T}$ follow from standard induction arguments. These arguments use Assumption 1, properties that are preserved under integration with respect to a continuous stochastic kernel \cite[Prop. 7.30]{bertsekas2004stochastic}, and a measurable selection result \cite[Prop. 7.33]{bertsekas2004stochastic}.

Next, we prove \eqref{key}. We work on the probability spaces $\{(\Omega,\mathcal{B}_{\Omega}, P_{\mathbf{x}}^\pi) : \mathbf{x} \in S, \; \pi \in \Pi\}$. For \eqref{key}, it suffices to show that for every $t \in \mathbb{T}_N$ and $\mathbf{x} \in S$,
\begin{subequations}\label{toshowshow}
\begin{align}
   \forall \pi \in \Pi, \;\;\;\textstyle \int_{\Omega} \phi_t^{\pi,s} \circ \mathcal{X}_t\; \mathrm{d}P_{\mathbf{x}}^\pi & \geq  \textstyle\int_{\Omega}  J_t^s\circ \mathcal{X}_t\; \mathrm{d}P_{\mathbf{x}}^\pi, \label{toshow11} \\
    \textstyle\int_{\Omega}  \phi_t^{\pi^s,s} \circ \mathcal{X}_t\; \mathrm{d}P_{\mathbf{x}}^{\pi^s} & =  \textstyle\int_{\Omega}  J_t^s \circ \mathcal{X}_t \; \mathrm{d}P_{\mathbf{x}}^{\pi^s}. \label{toshow22}
\end{align}
\end{subequations}
Indeed, if $t = 0$, then the above statement implies that for every $\mathbf{x} \in S$ and $\pi \in \Pi$,
\begin{equation}\label{my25}
     E_{\mathbf{x}}^\pi( h^s(Y) ) \geq J_0^s(\mathbf{x},a) = E_{\mathbf{x}}^{\pi^s}( h^s(Y) ),
\end{equation}
using \eqref{46} from Theorem \ref{thm1} and the realizations of $\mathcal{X}_0$ being concentrated at $(\mathbf{x},a)$ \eqref{expectation}. Then, we take the infimum of the expression in \eqref{my25} with respect to $\pi \in \Pi$ to derive \eqref{key}. The function $\phi_t^{\pi,s}$ appears inside an integral in \eqref{toshowshow} because a conditional expectation is not unique everywhere in general \cite[Th. 6.3.3]{ash1972}. We proceed by induction to prove \eqref{toshowshow}. The base cases $(t=N)$ for \eqref{toshowshow} hold by \eqref{47} from Theorem \ref{thm1}. Now, suppose that for some $t \in \mathbb{T}$, we have: for every $\mathbf{x} \in S$,
\begin{equation}\label{induct1}
    \forall \pi \in \Pi, \; \; \;\;\textstyle \int_{\Omega} \phi_{t+1}^{\pi,s} \circ \mathcal{X}_{t+1}\; \mathrm{d}P_{\mathbf{x}}^\pi  \geq  \textstyle\int_{\Omega}  J_{t+1}^s\circ \mathcal{X}_{t+1}\; \mathrm{d}P_{\mathbf{x}}^\pi.
\end{equation}
Let $\mathbf{x} \in S$ and $\pi \in \Pi$ be given. To show the induction step for \eqref{toshow11}, it suffices to show that
\begin{equation}\label{28}
    \textstyle\int_{\Omega}  J_{t+1}^s\circ \mathcal{X}_{t+1}\; \mathrm{d}P_{\mathbf{x}}^\pi \geq \int_{\Omega}  J_{t}^s\circ \mathcal{X}_{t}\; \mathrm{d}P_{\mathbf{x}}^\pi ,
\end{equation}
by applying \eqref{66} from Theorem \ref{thm1} and the induction hypothesis \eqref{induct1}. Noting that $J_{t+1}^s \circ \mathcal{X}_{t+1} : \Omega \rightarrow \mathbb{R}^*$ is Borel-measurable and nonnegative,  
we use \eqref{expectation}, the change-of-measure result \cite[Th. 1.6.12]{ash1972}, and the Fubini Theorem \cite[Th. 2.6.6]{ash1972} to derive
\begin{equation}\label{27}
   \textstyle \int_{\Omega} J_{t+1}^s \circ \mathcal{X}_{t+1} \; \mathrm{d}P_{\mathbf{x}}^\pi = \int_{\Omega} v_t^{s,\pi} \circ \mathcal{X}_{t} \; \mathrm{d}P_{\mathbf{x}}^\pi,
\end{equation}
where $v_t^{s,\pi} : \mathbb{S} \rightarrow \mathbb{R}^*$ is given by
\begin{equation}
  \textstyle v_t^{s,\pi}(x,z) \coloneqq \int_{C} v_t^s(x,z,u)\; \pi_t(\mathrm{d}u|x,z).
\end{equation}
Since $v_t^{s,\pi} \circ \mathcal{X}_{t} : \Omega \rightarrow \mathbb{R}^*$ and $J_t^{s} \circ \mathcal{X}_{t} : \Omega \rightarrow \mathbb{R}^*$ are Borel-measurable and satisfy $v_t^{s,\pi} \circ \mathcal{X}_{t} \geq J_t^{s} \circ \mathcal{X}_{t} \geq 0$ and \eqref{27} holds, the relation \eqref{28} follows. An induction argument for \eqref{toshow22} is similar. A key step is using \eqref{mykappaeq} to find that $v_t^{s,\pi^s} = J_t^s$.
\end{proof}

In particular, 
by letting any $s \in \mathbb{R}$ be the dual parameter's value and any $\mathbf{x} \in S$ be the initial state, we have shown that \eqref{key} holds under Assumption 1.
Therefore, under Assumption 1, we conclude that for every $s \in \mathbb{R}$ and $\mathbf{x} \in S$, $J_0^s(\mathbf{x}, a) = V^{s}(\mathbf{x})$.
%
This conclusion permits a useful characterization for risk-averse safe sets (Def. \ref{salphar}) in terms of the family $\{J_0^s : s \in \mathcal{Z}\}$ under Assumption 1:
\begin{equation}\label{13}
    \mathcal{S}_\alpha^r = \Big\{ \mathbf{x} \in S : \min_{s \in \mathcal{Z}} \, \bigl( \, s + {\textstyle\frac{1}{\alpha}}J_0^s(\mathbf{x},a) \, \bigr) \leq r\Big\}.
\end{equation}
To derive \eqref{13}, we use \eqref{10} as well. Since $\{J_0^s : s \in \mathcal{Z}\}$ does not depend on $\alpha$ or $r$, the family $\{J_0^s : s \in \mathcal{Z}\}$ characterizes any collection of risk-averse safe sets $\{ \mathcal{S}_\alpha^r : \alpha \in \Lambda, r \in R\}$, where $\Lambda$ is a subset of $(0,1]$ and $R$ is a subset of $\mathbb{R}$.

\textcolor{black}{The results in this section provide a nonunique optimal policy on the augmented state space $\pi^{s_{\mathbf{x},\alpha}^*} \in \Pi$ under Assumption 1.} 
\textcolor{black}{Policies on augmented state spaces} have also been developed by, e.g., \cite{bauerleott, bauerlerieder, haskell, bauerle2020minimizing}. Nonunique optimal policies are typical in stochastic optimal nonlinear control. 
\begin{remark}[Policy deployment]\label{policydeployment}
Let $\alpha \in (0,1]$ and $\mathbf{x} \in S$ be given. Let $\pi^{s_{\mathbf{x},\alpha}^*} \in \Pi$ satisfy \eqref{key}, where $s_{\mathbf{x},\alpha}^* \in \mathcal{Z}$ satisfies \eqref{10}.  
Let $\kappa_{t}^{s_{\mathbf{x},\alpha}^*}$ be the control law for time $t \in \mathbb{T}$ associated with $\pi^{s_{\mathbf{x},\alpha}^*}$.
Let $(x_0,z_0) = (\mathbf{x},a)$ and $t = 0$. For $t = 0,1,\dots,N-1$, repeat the following four steps: 1) choose $u_t = \kappa_{t}^{s_{\mathbf{x},\alpha}^*}(x_t,z_t)$; 2) nature provides a realization $w_t$ of $W_t$ according to the distribution $p_t(\cdot|x_t,u_t)$; 3) the realization $(x_{t+1},z_{t+1})$ of $(X_{t+1},Z_{t+1})$ is $(f_t(x_t,u_t,w_t), \max\{c_t(x_t,u_t),z_t\})$; 4) $t$ updates by 1.
\end{remark}

%% file: 5_numerical_example_extended.tex
Risk-averse safety analysis, as presented here, suffers from the curse of dimensionality inherent to DP and requires an augmented state space. Despite these computational challenges, risk-averse safety analysis may be a useful tool for designing control systems. At the design stage, large-scale off-line simulations may be commonplace, and designers may be required to assess multiple alternatives in light of uncertainty.

\subsection{Description of the Application} 
\begin{figure*}[ht]
\centerline{\includegraphics[width=0.9\textwidth]{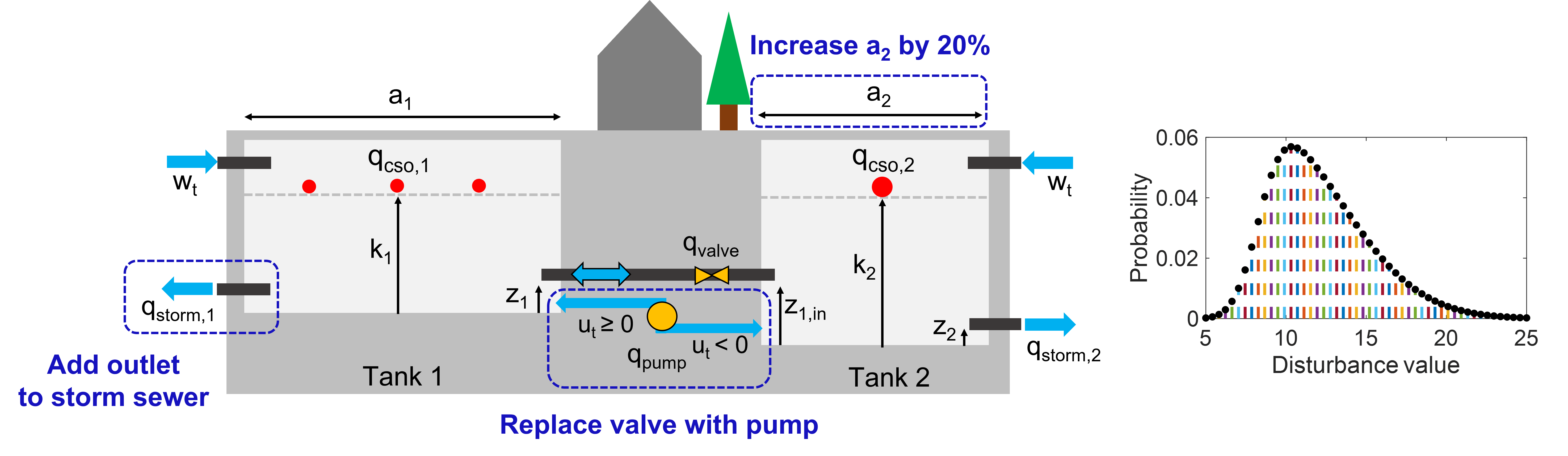}}
\caption{A schematic of the stormwater system and a probability mass function for the surface runoff disturbance (cubic feet per second, cfs). The baseline design and three alternative designs are shown.}
\label{watersysdrawing}
\end{figure*}

We consider the problem of modifying the design of an urban \emph{stormwater system} (i.e., a network of pipes, storage tanks, natural streams, etc., near an urban area). 
Apart from being actively controlled, the stormwater system that we consider is otherwise typical.\footnote{Actively controlled stormwater systems are becoming more common but are relatively novel technologies, e.g., see \cite{smartwater}.} 
The system consists of two tanks connected by a valve, and water flows by gravity between the tanks based on the relative difference in water
levels and the position of the valve (Fig.~\ref{watersysdrawing}). Water enters the system through a random process of surface runoff. Water exits the system through a \emph{storm sewer} drain that is connected to tank 2 or through outlets that lead to a \emph{combined sewer}. 
The storm sewer directs stormwater to a nearby water body; this is the desired outcome and occurs without penalty. Unfortunately, the storm sewer's capacity is limited, and when water levels become too high, excess flows are directed to a combined sewer. In drier periods, a combined sewer carries a mixture of untreated wastewater and stormwater to a wastewater treatment plant. However, when storm events cause the flow in a combined sewer to exceed its design capacity, a flow regulator (downstream from our system) will divert some of the untreated mixture of stormwater and sewage into a nearby water body. This event is known as a \emph{combined sewer overflow}. Combined sewers are present in older cities, such as Toronto and San Francisco, and overflows from these sewers can harm local ecosystems. We aim to use risk-averse safety analysis to examine how design modifications to the system above may reduce the risk of combined sewer overflows by managing the maximum water levels in the system.
\subsection{System Model for the Baseline Design} The state $X_t = [X_{t1}, X_{t2}]^T$ is a vector of the random water levels in tank 1 and tank 2 at time $t$. The co-domain of $X_t$ is $S = [0, \bar{k}_1] \times [0, \bar{k}_2]$ ft\textsuperscript{2}, where $\bar{k}_i = k_i + 2$ ft and $k_i$ is the maximum water level that tank $i$ can hold without releasing water into the combined sewer. The control input is the valve position at time $t$, and the co-domain of $U_t$ is $C = [0,1]$ (closed to open, unitless). The tuple $(W_0,W_1,\dots,W_{N-1})$ represents surface runoff that arises due to precipitation uncertainty. The disturbances are independent and identically distributed, and their distribution does not depend on the current state or control. The realizations of $W_t$ have units of cubic feet per second (cfs). The disturbance space is a subset of the non-negative orthant in $\mathbb{R}$ containing finitely many elements, $D = \{w^{(j)} : j = 1,2,\dots,N_W\}$ with $N_W \in \mathbb{N}$. In prior work \cite{sustech}, we simulated a design storm in PCSWMM software (Computational Hydraulics International), which is an extension of the US Environmental Protection Agency's Stormwater Management Model \cite{swmm}. A \emph{design storm} is a synthetic precipitation time series based on historical data that a local government uses to specify regulations for new or retrofitted stormwater systems. The empirical distribution from our simulations had positive skew, and the mean was approximately 12.2 cfs. We used these characteristics to inform the choice of the disturbance distribution shown in Fig.~\ref{watersysdrawing}: mean (12.2 cfs), variance (9.9 cfs$^2$), and skew (0.74).
%

Let $t \in \mathbb{T}$ be given. If $x_t \in S$, $u_t \in C$, and $w_t \in D$ are the values of $X_t$, $U_t$, and $W_t$, respectively, then the value of $X_{t+1}$ is given by
\begin{subequations}\label{dynwater}
\begin{equation}\begin{aligned}
    x_{t+1} = f(x_t,u_t,w_t) = x_t + \triangle \cdot \bar{f}(x_t,u_t,w_t),
\end{aligned}\end{equation}
such that if $x_{t+1,i} \geq \bar{k}_i$, then we redefine $x_{t+1,i} = \bar{k}_i$. The symbol $\triangle$ is the duration between time $t$ and time $t+1$, which is constant for all $t \in \mathbb{T}$. In this model, $f = f_t$ for all $t \in \mathbb{T}$.
%
The function $\bar{f}$ is chosen according to simplified Newtonian physics:
\begin{equation}\label{physics}\begin{aligned}
\bar{f}(x,u,w) & \coloneqq \left[ \bar{f}_1(x,u,w), \bar{f}_2(x,u,w) \right]^{T}\\
\bar{f}_1(x,u,w) & \coloneqq  \frac{w - q_{\text{cso,1}}(x) - q_\text{valve}(x,u)}{\mathbf{a}_1}\\
\bar{f}_2(x,u,w) & \coloneqq  \frac{w - q_{\text{cso,2}}(x) + q_\text{valve}(x,u)  - q_\text{storm,2}(x)}{\mathbf{a}_2} \\
q_\text{valve}(x,u) & \coloneqq u \cdot \bar{\pi} \mathbf{r}_{\text{v}}^2 \cdot \text{sign}{(h(x))} \cdot \sqrt{2 \mathbf{g} |h(x)|}  \\
h(x) & \coloneqq \max\{x_1 - \mathbf{z}_1,0\} - \max\{x_2 - \mathbf{z}_{1,\text{in}},0\}.\\
\end{aligned}\end{equation}
Table \ref{tanksysinfo} lists model parameters. The outlets to the combined sewer and to the storm sewer are equipped with outflow regulation devices that produce a linear outflow rate. For example, $q_\text{storm,2}(x)$ with $x \in S$ is given by
\begin{equation}\label{stormq}\begin{aligned}
   q_\text{storm,2}(x) & \coloneqq q_\text{max,2} - \frac{q_\text{max,2}}{\bar{k}_2 - \mathbf{z}_2} \min\{\bar{k}_2-x_2, \bar{k}_2 - \mathbf{z}_2\}\\
   q_\text{max,2} & \coloneqq \mathbf{c}_\text{d} \bar{\pi} \mathbf{r}_{\text{s}}^2  \sqrt{2 \mathbf{g} (\bar{k}_2 - \mathbf{z}_2)} ,
\end{aligned}\end{equation}
\end{subequations}
where $q_\text{max,2}$ is tank 2's maximum outflow rate to the storm sewer, $\mathbf{c}_\text{d}$ is a discharge coefficient, $\bar{\pi} \approx 3.14$, $\mathbf{g}$ is gravitational acceleration, $\mathbf{r}_{\text{s}}$ is the storm sewer outlet radius, and $\mathbf{z}_2$ is the storm sewer outlet elevation. The outflow rates to the combined sewer, $q_{\text{cso,1}}$ and $q_{\text{cso,2}}$, are defined similarly to \eqref{stormq}. The constraint set $K = [0, k_1] \times [0, k_2]$ specifies the invert elevations of the combined sewer outlets (i.e., the maximum water levels that the tanks can hold without releasing water into the combined sewer). The function $g_K$ quantifies the maximum water elevation above the combined sewer invert elevations, 
\begin{equation}\label{mygksw}
    g_K(x) = \max\{x_{1} - k_1, x_{2} - k_2, 0\} \;\;\;\;\; \forall x \in S.
\end{equation}

\begin{table}[ht]
\centering
\caption{Stormwater System Parameters (Baseline)}
\label{table}
\setlength{\tabcolsep}{3pt}
\begin{tabular}{|p{50pt}|p{175pt}|p{70pt}|}
\hline
\textbf{Symbol} &  \textbf{Description} & \textbf{Value} \vspace{.5mm}\\
\hline
$\mathbf{a}_1$ & Surface area of tank 1 & 30000 ft\textsuperscript{2} \vspace{.6mm}\\
$\mathbf{a}_2$ & Surface area of tank 2 & 10000 ft\textsuperscript{2} \vspace{.6mm}\\
$\mathbf{c}_\text{d}$ & Discharge coefficient & 0.61 (no units) \vspace{.6mm}\\
$\mathbf{g}$ & Acceleration due to gravity & 32.2 $\frac{\text{ft}}{\text{s}^2}$\vspace{.6mm}\\
$a$ & Minimum of $c_t = g_K$ \eqref{mygksw} & 0 ft\vspace{1mm}\\
$b$ & Maximum of $c_t = g_K$ \eqref{mygksw} & 2 ft\vspace{.6mm}\\
$k_1$ & Combined sewer outlet elevation, tank 1 & 3 ft \vspace{.6mm}\\
$k_2$ & Combined sewer outlet elevation, tank 2 & 4 ft \vspace{.6mm}\\
$\bar{k}_1$ & Maximum value of $x_1$ & 5 ft \vspace{.6mm}\\
$\bar{k}_2$ & Maximum value of $x_2$ & 6 ft \vspace{.6mm}\\
$N$ & Length of discrete time horizon & 20 ($=$ 1 h) \vspace{.6mm}\\
$\bar{\pi}$ & Circle circumference-to-diameter ratio & $\approx$ 3.14 \vspace{.6mm}\\
$\mathbf{r}_\text{s}$ & Storm sewer outlet radius & $1/3$ ft \vspace{.6mm}\\
$\mathbf{r}_{\text{v}}$ & Valve radius & $1/3$ ft\vspace{.6mm}\\
$ \triangle $ & Duration of $[t,t+1)$ & 3 min \vspace{.6mm}\\
$\mathbf{z}_1$ & Pipe elevation with respect to base of tank 1 & 1 ft\vspace{.6mm}\\
$\mathbf{z}_{1,\text{in}}$ & Pipe elevation with respect to base of tank 2 & 2 ft\vspace{.6mm}\\
$\mathbf{z}_2$ & Storm sewer outlet elevation & 1 ft\vspace{.6mm}\\
N/A & Number of combined sewer outlets, tank 1 & 3 \vspace{.6mm}\\
N/A & Number of combined sewer outlets, tank 2 & 1 \vspace{.6mm}\\
N/A & Combined sewer outlet radius, tank 1 & 1/4 ft \vspace{.6mm}\\
N/A & Combined sewer outlet radius, tank 2 & 3/8 ft \vspace{.5mm}\\
\hline
\multicolumn{3}{p{295pt}}{ft $=$ feet, s $=$ 
seconds, min $=$ minutes, h $=$ hours.}
\end{tabular}
\label{tanksysinfo}
\end{table}
%

\subsection{Verification of Assumption \ref{Assumption1}} It holds that $g_K(x) \in \mathcal{Z} = [a,b] = [0,2]$ for all $x \in S$, where $g_K$ is defined by \eqref{mygksw}. We choose $c_t = g_K$ for all $t \in \mathbb{T}_N$. Thus, $c_t$ is $\mathcal{Z}$-valued and continuous. The control space $C = [0,1]$ is compact. In our example, the stochastic kernel for the disturbance process does not depend on $(x,u)$, which implies that it is constant and therefore continuous in $(x,u)$. 
The dynamics function $f$ \eqref{dynwater} is continuous because it is a composition of continuous functions. Recall that the function $\lambda(x) \coloneqq \text{sign}(x) \, \sqrt{\lvert x \rvert}$ is continuous since $\lim_{x \uparrow 0} \lambda(x) = \lim_{x \downarrow 0} \lambda(x) = 0$.
\subsection{Designs} We investigate the effect of different designs on the system's safety, as quantified in terms of risk-averse safe sets. The designs are listed below:
\begin{enumerate}[label=\alph*)] 
     \item Baseline;
    \item Replace the valve with a \textcolor{black}{controllable bidirectional} pump, whose maximum pumping rate is $\bar{q}_{\text{pump}}$;
    \item Retrofit tank 1 with an outlet that drains to a storm sewer without penalty; or
    \item Increase the surface area of tank 2 by 20\%.
\end{enumerate}
We modify the baseline system model to obtain a model representing design b, c, or d. For design d, we set $\mathbf{a}_2 = 12000$ ft\textsuperscript{2}. For design c, the equation for the flow through tank 1 changes to the following:
\begin{equation}
    \bar{f}_1(x,u,w)  \coloneqq  \frac{w - q_{\text{cso,1}}(x) - q_\text{valve}(x,u) - q_\text{storm,1}(x)}{\mathbf{a}_1},\\
\end{equation}
where $q_\text{storm,1}(x)$ takes the same form as $q_\text{storm,2}(x)$ in \eqref{stormq}. 

For design b, the control space becomes $C = [-1, 1]$, and the term $q_{\text{valve}}(x,u)$ is replaced by $q_\text{pump}(x,u)$, which models the flow rate generated by a pump. Prior to presenting the form of $q_\text{pump}(x,u)$, we introduce its dependencies, $\mathcal{I}_i(x,u)$ and $\ell(x_i,u)$. $\mathcal{I}_i(x,u)$ is a Boolean variable that determines whether the water level is too low to permit pumping, and the function $\ell(x_i,u)$ represents a start-up phase. $\mathcal{I}_1(x,u)$ is true if and only if the pump attempts to push water from tank 1 to tank 2 ($u < 0$), but the water level in tank 1 is too low. $\mathcal{I}_2(x,u)$ has an analogous interpretation. Formally, we define $\mathcal{I}_1(x,u)$ and $\mathcal{I}_2(x,u)$ as follows:
\begin{subequations}\label{pumpeq}
\begin{equation}\begin{aligned}
    \mathcal{I}_1(x,u) & \coloneqq x_1 < \mathbf{z}_\text{p} - \epsilon \text{ and } u < 0\\
    \mathcal{I}_2(x,u) & \coloneqq x_2 < \mathbf{z}_\text{p} - \epsilon \text{ and }u \geq 0,
\end{aligned}\end{equation}
where $\mathbf{z}_\text{p}$ is a threshold elevation and $\epsilon$ is a small positive number. We define the function $\ell(x_i,u)$ as follows:
\begin{equation}
    \ell(x_i,u) \coloneqq \frac{\bar{q}_{\text{pump}} \cdot u}{2 \epsilon} (x_i + \epsilon - \mathbf{z}_\text{p} ).
\end{equation}
We define $q_\text{pump}(x,u)$ as follows:
\begin{equation}\label{qpump1}
    q_\text{pump}(x,u) \coloneqq \begin{cases}0 & \text{if }\mathcal{I}_1(x,u) \text{ or } \mathcal{I}_2(x,u) \\ -\ell(x_1,u) & \text{if } x_1 \in [\mathbf{z}_\text{p} - \epsilon,\mathbf{z}_\text{p} + \epsilon] \text{ and } u < 0\\
    -\ell(x_2,u) & \text{if } x_2 \in [\mathbf{z}_\text{p} - \epsilon,\mathbf{z}_\text{p} + \epsilon] \text{ and } u \geq 0\\
    -u \cdot \bar{q}_{\text{pump}} & \text{otherwise}.\end{cases}
\end{equation}
\end{subequations}
One can show that $q_\text{pump}$ is a composition of continuous functions by replacing the case statements in \eqref{qpump1} with minimum and maximum operators:
    \begin{align*}
        q_\text{pump}(x,u) &= \frac{-\bar{q}_{\text{pump}}}{2 \epsilon} \Bigl( \min\{ 0, u\} \nu(y_1) + \max\{ 0, u\} \nu(y_2) \Bigr) \\
        \nu(y) &\coloneqq \max\{ 0 , \min\{y, 2\epsilon \}\} \\
        y_i &\coloneqq x_i + \epsilon - \mathbf{z}_\text{p}.
    \end{align*}
Table \ref{pumpdesign} lists model parameters for the pump design (b).
\begin{table}[ht]
\centering
\caption{Stormwater System Parameters (Pump design)}
\setlength{\tabcolsep}{3pt}
\begin{tabular}{|p{50pt}|p{125pt}|p{60pt}|}
\hline
\textbf{Symbol} &  \textbf{Description} & \textbf{Value} \vspace{.5mm}\\
\hline
$\bar{q}_{\text{pump}}$ & Maximum pumping rate & 10 cfs \vspace{.5mm}\\
$\epsilon$ & Slack variable & $\frac{1}{12}$ ft \vspace{.6mm}\\
$\mathbf{z}_\text{p}$ & Threshold pumping elevation & 1 ft \vspace{.6mm}\\
\hline
\multicolumn{3}{p{190pt}}{ft $=$ feet, cfs $=$ 
cubic feet per second.}
\end{tabular}
\label{pumpdesign}
\end{table}
\subsection{Current Method vs. Under-approximation Method} Computations of $\mathcal{S}_\alpha^r$ for the four designs are shown in Fig.~\ref{exactvsunderapprox}. For comparison, we provide computations of the under-approximation set $\mathcal{U}_{\alpha,\gamma}^r$ ($\gamma = 20$) using the method from our prior work \cite{chapmantac2021}. The under-approximation method uses a $\gamma$-dependent soft-maximum and an $\alpha$-dependent upper bound for the CVaR to derive a $(\gamma,\alpha)$-dependent upper bound for the optimal value ${\mathcal{J}}_\alpha^*(\mathbf{x})$ \eqref{keyproblem} with $c_t = g_K$. For a fixed $\gamma$, solving one MDP problem is required to compute $\mathcal{U}_{\alpha,\gamma}^r$ for all $\alpha$ and $r$ of interest. The DP iterates for this MDP problem are defined on the original state space $S$, and the objective is an expected cumulative $\gamma$-dependent cost. We have explored values of $\gamma$ between 10 and 120 in increments of roughly 10. We have chosen $\gamma = 20$ because this value provides relatively large estimates of $\mathcal{U}_{\alpha,\gamma}^r$ for more risk-averse values of $\alpha$. The selection of an appropriate $\gamma$ depends on one's preferences, and additional guidance is provided in \cite{chapmantac2021}.

While no parameter tuning is required for the current method, which provides $\mathcal{S}_\alpha^r$ exactly in principle, greater computational resources are required. First, due to time inconsistency and a non-additive cost function, the dynamic program that determines the minimum CVaR is defined on an augmented state space, which is the Cartesian product of the original state space $S$ and the interval $\mathcal{Z}$. Moreover, due to the definition for the CVaR \eqref{defcvar}, a second outer optimization with respect to the dual parameter is required. Consequently, the inner dynamic program is implemented repeatedly for different values of the dual parameter. While this increases the computational complexity significantly, problems with different dual parameters can be solved in parallel to reduce computation time. 

One can run the under-approximation method on a standard laptop (2--4 CPU cores) in approximately 10 minutes for a fixed $\gamma$ and a fixed design. However, this approach is not suitable for the current method. In particular, we used a high-performance computing cluster. The complete job (four designs) required about 54 hours and 30 CPU cores.\footnote{In the main paper, we state that about 13.5 hours is required for each design because (54 hours)/(4 designs) = 13.5 hours per design. We used the Tufts Linux Research Cluster (Medford, MA) running MATLAB (The Mathworks, Inc.), and our code is available from https://github.com/risk-sensitive-reachability/RSSAVSA-2021.}
These run-time and CPU values should be considered a rough comparison of the resources that naive implementations of the two methods require; we have made no attempt to optimize computational efficiency beyond parallelizing the operations in a given DP recursion.
Table \ref{tradeoffsbetweenmethods} summarizes the main trade-offs between the current method and under-approximation method.
\begin{table}[ht]
\centering
\caption{Trade-offs between Methods}
\setlength{\tabcolsep}{3pt}
\begin{tabular}{|p{252pt}|p{252pt}|}
\hline
\textbf{Current Method} &  \textbf{Under-Approximation Method \cite{chapmantac2021} }\\
\hline
Provides $\mathcal{S}_\alpha^r$ exactly in principle & Provides an under-approximation for $\mathcal{S}_\alpha^r$ in principle \vspace{.5mm}\\
\hline
Does not require parameter tuning & The soft-maximum parameter $\gamma$ requires tuning. \vspace{.5mm}\\
\hline
Requires significantly more computational resources & Requires significantly less computational resources \vspace{.5mm}\\
\hline
Useful for in-depth analysis of a small number of promising designs & Useful as a screening tool to identify more promising designs from a collection of candidate designs \vspace{.5mm}\\
\hline
\end{tabular}
\label{tradeoffsbetweenmethods}
\end{table}

\subsection{Discussion of the Numerical Results} 
\begin{figure*}[ht]
\centerline{\includegraphics[width=\textwidth]{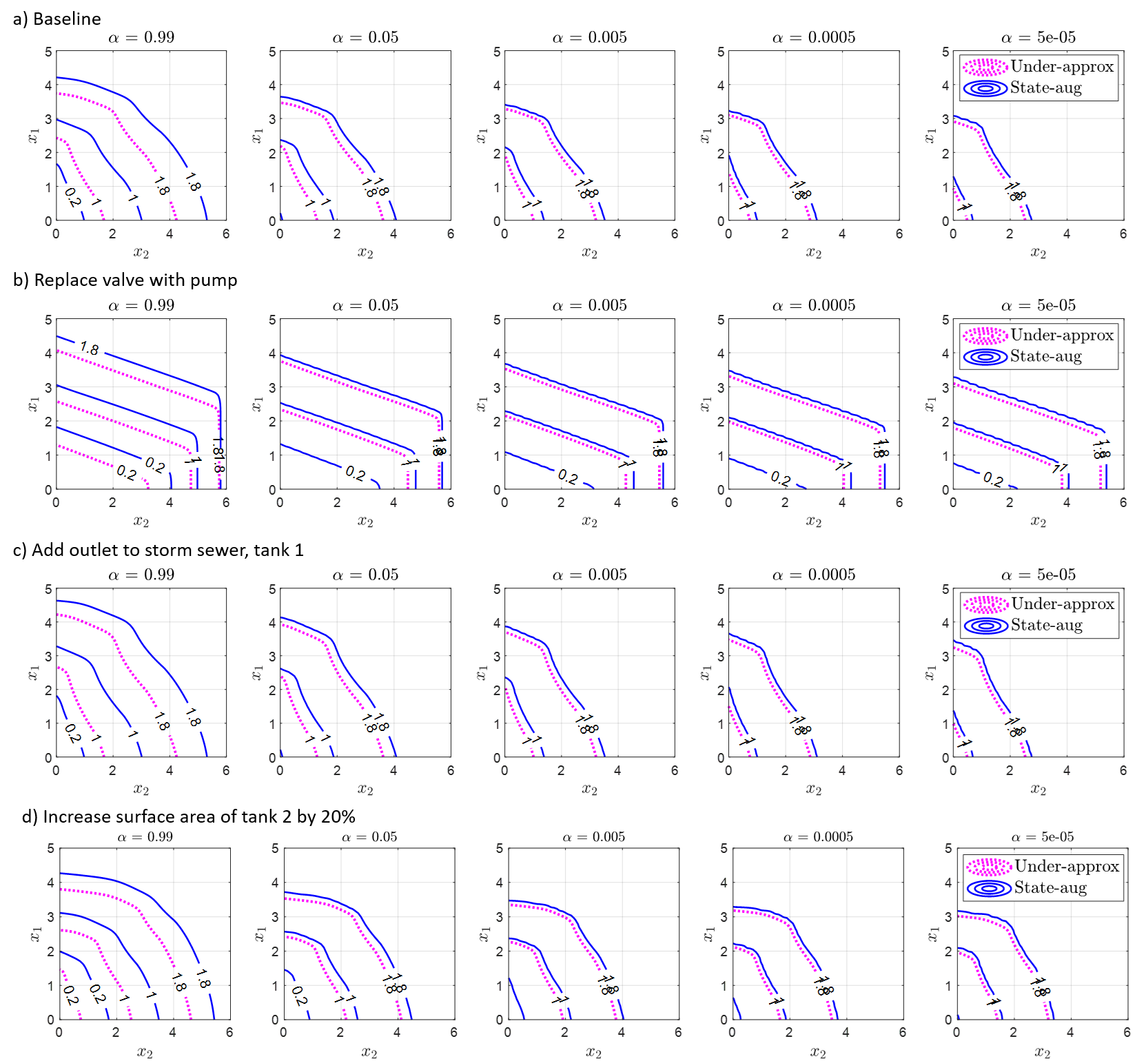}}
\caption{Contours of computations of risk-averse safe sets for $\alpha \in \{0.99, 0.05, 0.005, 0.0005, 5 \cdot 10^{-5}\}$ and $r \in \{0.2, 1, 1.8\}$. Each row pertains to a particular design. Solid blue lines show the numerical results for $\mathcal{S}_\alpha^r$. Pink dotted lines show the numerical results for $\mathcal{U}_{\alpha,\gamma}^r$ ($\gamma = 20$) using the under-approximation method from \cite{chapmantac2021}. 
}
\label{exactvsunderapprox}
\end{figure*}

\begin{figure*}[ht]
\centerline{\includegraphics[width=\textwidth]{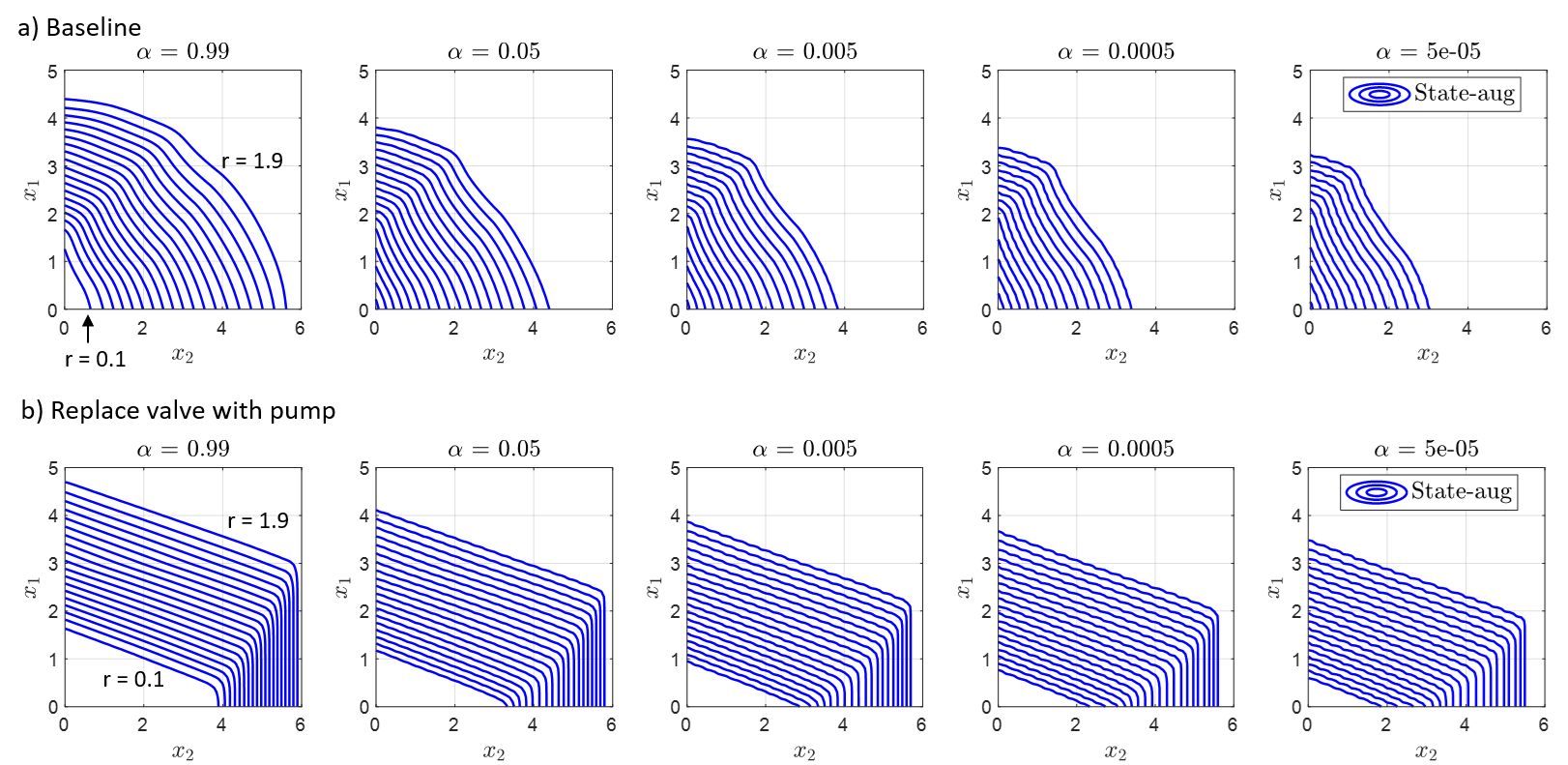}}
\caption{Contours of computations of risk-averse safe sets for $\alpha \in \{0.99, 0.05, 0.005, 0.0005, 5 \cdot 10^{-5}\}$ and $r \in \{0.1, 0.2, \dots, 1.9\}$ when using the current method for two designs: a) baseline and b) replace valve with pump. This figure shows the contour shapes for these two designs in more detail by presenting more values of $r$ compared to Fig.~\ref{exactvsunderapprox}.}
\label{onlystateaug}
\end{figure*}

\begin{figure*}[ht]
\centerline{\includegraphics[width=\textwidth]{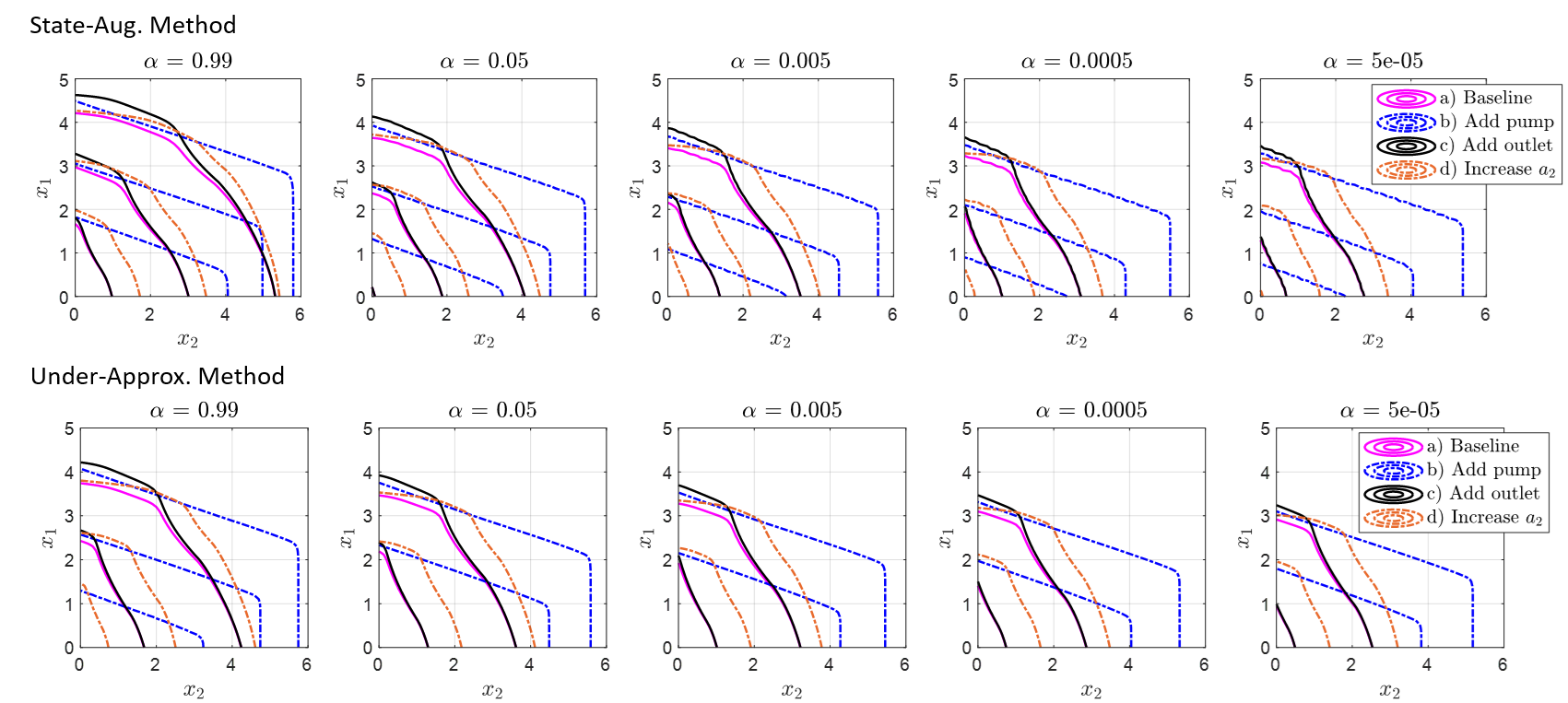}}
\caption{This figure presents numerical results from Fig.~\ref{exactvsunderapprox} from a different perspective to depict how a different design may produce a differently shaped or sized contour, $r \in \{0.2, 1, 1.8\}$. We placed the results for all designs in a single sub-plot pertaining to a particular $\alpha$. We used the current method to obtain the results in the top row, and the under-approximation method ($\gamma = 20$) to obtain the results in the bottom row. Note that the contours for design a (baseline, magenta solid lines) and design c (add outlet, black solid lines) overlap in regions where $x_2$ has larger values.}
\label{alldesigns}
\end{figure*}

We use the notation $\hat{\mathcal{S}}_\alpha^r$ $(\hat{\mathcal{U}}_{\alpha,\gamma}^r)$ to indicate a computation of $\mathcal{S}_\alpha^r$ $(\mathcal{U}_{\alpha,\gamma}^r)$. This notation emphasizes the distinction between an exact mathematical quantity and a computation of this quantity returned by a computer program. The under-approximation method preserves interesting and potentially useful qualitative features that are provided by the current method. For example, the $\hat{\mathcal{S}}_\alpha^r$-contours for the pump design (b) are more rectangular in comparison to those for the baseline design (a) (Fig.~\ref{onlystateaug}). These features are apparent from the $\hat{\mathcal{U}}_{\alpha,\gamma}^r$-contours as well (Fig.~\ref{exactvsunderapprox}, first two rows, pink dotted lines). The $\hat{\mathcal{S}}_\alpha^r$-contours for the outlet design (c) are stretched along the $x_1$-axis in comparison to the baseline design (a) (Fig.~\ref{alldesigns}, top, black vs. pink). This effect is also seen by observing the associated contours of $\hat{\mathcal{U}}_{\alpha,\gamma}^r$ (Fig.~\ref{alldesigns}, bottom, black vs. pink). Increasing the surface area of tank 2 (design d) stretches the contours of $\hat{\mathcal{S}}_\alpha^r$ and $\hat{\mathcal{U}}_{\alpha,\gamma}^r$ along the $x_2$-axis in comparison to the baseline design (Fig.~\ref{alldesigns}, top and bottom, orange dotted vs. pink). As the risk-aversion level $\alpha$ becomes smaller (more pessimistic), the contours of $\hat{\mathcal{S}}_\alpha^r$ and $\hat{\mathcal{U}}_{\alpha,\gamma}^r$ contract, as we expect, while the qualitative features are preserved (Fig.~\ref{exactvsunderapprox}).

While the under-approximation method recovers qualitative features and requires reduced computational resources, it tends to over-estimate the effect of making a design change (Table \ref{comparingdesignstable}). Consequently, we see the under-approximation method as a preliminary screening tool to identify more promising designs from a collection of candidate designs. On the other hand, we see the current method as a tool for in-depth analysis of a small number of promising designs that have been selected \emph{a priori} by preliminary screening.

\textcolor{black}{The risk-aversion level $\alpha$ allows one to specify a degree of pessimism in terms of a fraction of worst cases, which has benefits for designing systems in practice. Stormwater systems are often required to satisfy precise regulatory criteria. For example, an outflow rate must be no more than a given threshold when simulating the outcome from a (non-stochastic) design storm via hydrology and hydraulic modeling software, e.g., \cite{swmm}. Our framework could be used in parallel with standard practices to quantify the effect of stochastic surface runoff on low-dimensional models of proposed designs, as the degree of risk aversion $\alpha$ varies. The value of $\alpha$ provides a systematic and interpretable way to assess a design with respect to varying degrees of pessimism about the future. While the typical minimax approach to control systems leads to robust designs by adopting a worst-case perspective, designing for the worst case may not be financially feasible, especially given the limited budgets afforded to ``ordinary'' rather than ``safety-critical'' infrastructure. Therefore, the flexibility afforded by $\alpha$ may be useful for assessing trade-offs between system performance and financial considerations in practice.} 


%
%
%
%

\begin{table}[ht]
\centering
\caption{Quantitative comparison of designs ($r = 1$)}
\setlength{\tabcolsep}{3pt}
\begin{tabular}{|p{125pt}|p{50pt}|p{50pt}|p{50pt}|}
\hline
\textbf{$\alpha =$ 0.99} &  \textbf{b vs. a} &  \textbf{c vs. a} & \textbf{d vs. a} \vspace{.5mm}\\
\hline
Current Method & 0.93 & 0.079 & 0.34\vspace{.5mm}\\
Under-approximation Method & 2.6 & 0.069 & 0.93\vspace{.5mm}\\
\hline
\textbf{$\alpha =$ 0.05} &  \textbf{b vs. a} &  \textbf{c vs. a} & \textbf{d vs. a} \vspace{.5mm}\\
\hline
Current Method & 2.1 & 0.068 & 0.71\vspace{.5mm}\\
Under-approximation Method & 3.6 & 0.059 & 1.3\vspace{.5mm}\\
\hline
\textbf{$\alpha =$ 0.005} &  \textbf{b vs. a} &  \textbf{c vs. a} & \textbf{d vs. a} \vspace{.5mm}\\
\hline
Current Method & 3.1 & 0.059 & 1.1\vspace{.5mm}\\
Under-approximation Method & 5.1 & 0.072 & 1.9\vspace{.5mm}\\
\hline
\textbf{$\alpha =$ 0.0005} &  \textbf{b vs. a} &  \textbf{c vs. a} & \textbf{d vs. a} \vspace{.5mm}\\
\hline
Current Method & 4.9 & 0.055 & 1.8\vspace{.5mm}\\
Under-approximation Method & 7.6 & 0.03 & 2.8\vspace{.5mm}\\
\hline
\textbf{$\alpha =$ 0.00005} &  \textbf{b vs. a} &  \textbf{c vs. a} & \textbf{d vs. a} \vspace{.5mm}\\
\hline
Current Method & 9.0 & 0.054 & 3.3\vspace{.5mm}\\
Under-approximation Method & 14 & 0.031 & 5.3\vspace{.5mm}\\
\hline
\multicolumn{4}{p{290pt}}{We list the increase in the size of a risk-averse safe set for design b, c, or d compared to the baseline (design a). This is a quantitative depiction of the some of the results in Fig.~\ref{alldesigns}. Let $\text{N}_{\text{y},\alpha}$ denote the number of states in the computation of $\mathcal{S}_\alpha^r$ for design y $\in \{\text{a},\text{b},\text{c},\text{d}\}$ and $r = 1$. Let $\hat{\text{N}}_{\text{y},\alpha}$ denote the number of states in the computation of $\mathcal{U}_{\alpha,\gamma}^r$ for design y, $r = 1$, and $\gamma = 20$. Each quantity in a row labeled Current Method takes the form $(\text{N}_{\text{y},\alpha}-\text{N}_{\text{a},\alpha})/\text{N}_{\text{a},\alpha}$. Each quantity in a row labeled Under-approximation Method takes the form $(\hat{\text{N}}_{\text{y},\alpha}-\hat{\text{N}}_{\text{a},\alpha})/\hat{\text{N}}_{\text{a},\alpha}$.}
\end{tabular}
\label{comparingdesignstable}
\end{table}

%% file: 6_conclusions.tex
By overcoming theoretical challenges attributed to optimizing the CVaR of a trajectory-wise maximum cost, we have shown that risk-averse safe sets enjoy an equivalent representation in terms of the solutions to a family of stochastic dynamic programs. \textcolor{black}{We are investigating extensions to higher-dimensional systems in the finite-time case using extreme value theory \cite{extremevaluetheory} and in the infinite-time case using value function approximations.} \textcolor{black}{In the future, we hope to study new problems that combine performance and risk-averse safety criteria, such as optimizing a utility functional subject to a constraint on the CVaR of a maximum cost.}

%% file: 7_appendix_for_extended.tex
Here, we provide step-by-step technical details underlying the theoretical results of the main paper. Throughout the Appendix, we assume that Assumption 1 holds, even if this is not explicitly stated. This assumption is useful for ensuring that integrals are well-defined, optimal policies exist, etc. In particular, while a milder assumption may be sufficient, we construct $P_{\mathbf{x}}^\pi$ using Assumption 1 (Sec. \ref{appendixremark}).
\subsection{An Extended Proof for Lemma 1}
\emph{Lemma 1 (Existence of a minimizer in $[a,b]$):} Let Assumption 1 hold. Let $\mathbf{x} \in S$ and $\alpha \in (0,1]$ be given. Suppose that $G : \Omega \rightarrow \mathbb{R}$ is measurable relative to $\mathcal{B}_{\Omega}$ and $\mathcal{B}_{\mathbb{R}}$, and suppose that $G(\omega) \in [a,b]$ for every $\omega \in \Omega$. Define $L_{\mathbf{x}}^\alpha(s) \coloneqq s + {\textstyle\frac{1}{\alpha}} \underset{\pi \in \Pi}{\inf} E_{\mathbf{x}}^\pi(h^s(G))$ with $h^s(y) \coloneqq \max\{y-s,0\}$ \eqref{myh}. Then,
\begin{equation}\label{eq:r_bound}\begin{aligned}
    \inf_{s \in \mathbb{R}} L_{\mathbf{x}}^\alpha(s) 
  = \min_{s \in [a,b]} L_{\mathbf{x}}^\alpha(s), 
\end{aligned}\end{equation}
where min means that a minimizer $s_{\mathbf{x},\alpha}^*  \in [a,b]$ exists.

\hspace{-8mm}\begin{proof}
Note that $L_{\mathbf{x}}^\alpha \colon \mathbb{R} \to \mathbb{R}$ is given by
\begin{align}
 L_{\mathbf{x}}^\alpha(s) & = s + {\textstyle\frac{1}{\alpha}} v_{s}^*(\mathbf{x}) \\
    v_{s}^*(\mathbf{x}) & \coloneqq \inf_{\pi \in \Pi} E_{\mathbf{x}}^\pi(\max\{G - s, 0 \}).
\end{align}
$L_{\mathbf{x}}^\alpha$ is $\mathbb{R}$-valued in particular because 
\begin{equation}
  \forall s \in \mathbb{R} \; \;\forall \pi \in \Pi, \;\;\;  \max\{b - s, 0 \} \geq E_{\mathbf{x}}^\pi(\max\{G - s, 0 \}) \geq 0. 
\end{equation}
Define 
\begin{equation}
    \ell \coloneqq \inf_{s \in [a,b]} L_{\mathbf{x}}^\alpha(s).
\end{equation}
$\ell$ is finite because
\begin{equation}
   L_{\mathbf{x}}^\alpha(a) \geq \inf_{s \in [a,b]} L_{\mathbf{x}}^\alpha(s) \geq a.
\end{equation}
To derive the second inequality, note that 
\begin{equation}
  \forall s \in [a,b],\;\;\;  L_{\mathbf{x}}^\alpha(s) = s + {\textstyle\frac{1}{\alpha}}\inf_{\pi \in \Pi} E_{\mathbf{x}}^\pi(\max\{G - s, 0 \}) \geq s + {\textstyle\frac{1}{\alpha}}0 = s \geq a.
\end{equation}

We will show that for any $s \in \mathbb{R}$, $L_{\mathbf{x}}^\alpha(s) \geq \ell$, from which we will conclude that
\begin{equation}\label{9}
    \inf_{s \in \mathbb{R}} L_{\mathbf{x}}^\alpha(s) \geq \ell \coloneqq \inf_{s \in [a,b]} L_{\mathbf{x}}^\alpha(s) \geq \inf_{s \in \mathbb{R}} L_{\mathbf{x}}^\alpha(s).
\end{equation}
First, for any $s \in [a,b]$, $L_{\mathbf{x}}^\alpha(s) \geq \ell$ holds by the definition of $\ell$. Second, consider $s \leq a$, equivalently, $-s \geq -a$. In this case, for every $\omega \in \Omega$,
\begin{equation}
    G(\omega) - s \geq G(\omega) - a \geq 0  \implies \max\{G(\omega)-s,0\} = G(\omega)-s,
\end{equation}
and therefore,
\begin{equation}
    L_{\mathbf{x}}^\alpha(s)\Bigr\rvert_{s \leq a} = s + {\textstyle\frac{1}{\alpha}}\inf_{\pi \in \Pi} E_{\mathbf{x}}^\pi(\max\{G - s, 0 \})\Bigr\rvert_{s \leq a} = s + {\textstyle\frac{1}{\alpha}}\inf_{\pi \in \Pi} E_{\mathbf{x}}^\pi(G-s).
\end{equation}
We continue the algebra to find that
\begin{align}\label{12}
    L_{\mathbf{x}}^\alpha(s)\Bigr\rvert_{s \leq a} = s + {\textstyle\frac{1}{\alpha}}\inf_{\pi \in \Pi} E_{\mathbf{x}}^\pi(G) - {\textstyle\frac{1}{\alpha}}s & = s(1-{\textstyle\frac{1}{\alpha}}) + {\textstyle\frac{1}{\alpha}}\inf_{\pi \in \Pi} E_{\mathbf{x}}^\pi(G).  
\end{align}  
If $s = a$ in \eqref{12}, then we have
\begin{equation}\label{13}
    L_{\mathbf{x}}^\alpha(a) = a(1-{\textstyle\frac{1}{\alpha}}) + {\textstyle\frac{1}{\alpha}}\inf_{\pi \in \Pi} E_{\mathbf{x}}^\pi(G).
\end{equation}
Now, since $-s \geq  -a$ and ${\textstyle\frac{1}{\alpha}}-1 \geq 0$, it holds that
\begin{equation}\label{14}
    s(1-{\textstyle\frac{1}{\alpha}}) = -s ({\textstyle\frac{1}{\alpha}}-1) \geq -a ({\textstyle\frac{1}{\alpha}}-1) = a(1-{\textstyle\frac{1}{\alpha}}).
\end{equation}
Therefore,
\begin{equation}
    L_{\mathbf{x}}^\alpha(s)\Bigr\rvert_{s \leq a} \overset{\eqref{12}}{=} s(1-{\textstyle\frac{1}{\alpha}}) + {\textstyle\frac{1}{\alpha}}\inf_{\pi \in \Pi} E_{\mathbf{x}}^\pi(G) \overset{\eqref{14}}{\geq}  a(1-{\textstyle\frac{1}{\alpha}}) + {\textstyle\frac{1}{\alpha}}\inf_{\pi \in \Pi} E_{\mathbf{x}}^\pi(G) \overset{\eqref{13}}{=} L_{\mathbf{x}}^\alpha(a) \geq \ell. 
\end{equation}
The third and last case is to consider $s \geq b$, equivalently, $-s \leq -b$, from which we deduce that for every $\omega \in \Omega$,
\begin{equation}
    G(\omega) - s \leq G(\omega) - b \leq 0 \implies \max\{G(\omega)-s,0\} = 0.
\end{equation}
It follows that
\begin{equation}
     L_{\mathbf{x}}^\alpha(s)\Bigr\rvert_{s \geq b} = s + {\textstyle\frac{1}{\alpha}}\inf_{\pi \in \Pi} E_{\mathbf{x}}^\pi(\max\{G - s, 0 \})\Bigr\rvert_{s \geq b} = s + {\textstyle\frac{1}{\alpha}} 0 = s \geq b = L_{\mathbf{x}}^\alpha(b) \geq \ell.
\end{equation}
We have shown that for every $s \in \mathbb{R}$, $L_{\mathbf{x}}^\alpha(s) \geq \ell$ holds, and therefore, we conclude that \eqref{9} holds, equivalently,
\begin{equation}
     \inf_{s \in \mathbb{R}} L_{\mathbf{x}}^\alpha(s) = \inf_{s \in [a,b]} L_{\mathbf{x}}^\alpha(s).
\end{equation}

To show that $\underset{s \in [a,b]}{\inf} L_{\mathbf{x}}^\alpha(s) = \underset{s \in [a,b]}{\min} L_{\mathbf{x}}^\alpha(s)$, we prove that $L_{\mathbf{x}}^\alpha$ is continuous in $s$.\footnote{The infimum of a lower semi-continuous function on a compact topological space is attained \cite[Th. A6.3, p. 389]{ash1972}.}
%
Since $L_{\mathbf{x}}^\alpha(s) = s + {\textstyle\frac{1}{\alpha}} v_{s}^*({\mathbf{x}})$, it suffices to show that 
\begin{equation}\label{defvs}
 v_{s}^*({\mathbf{x}}) = \inf_{\pi \in \Pi} E_{\mathbf{x}}^\pi(\max\{G - s, 0 \})
\end{equation}
is continuous in $s$. We will show that for any $s \in \mathbb{R}$ and $s' \in \mathbb{R}$,
\begin{equation}\label{20}
    | v_s^*(\mathbf{x}) - v_{s'}^*(\mathbf{x}) | \leq |s - s'|,
\end{equation}
that is, $v_{s}^*(\mathbf{x})$ is Lipschitz continuous in $s$ with Lipschitz constant equal to one. First, note that for any $\mathbf{c} \in \mathbb{R}$ and $\mathbf{d} \in \mathbb{R}$, we have
\begin{equation}\label{21}
    \max\{ \mathbf{c} + \mathbf{d}, 0\} \leq \max\{\mathbf{c}, 0\} + \max\{\mathbf{d}, 0\}
\end{equation}
because
\begin{equation}
    \mathbf{c} + \mathbf{d} \leq \max\{\mathbf{c},0\} + \max\{\mathbf{d}, 0\},
\end{equation}
$y \mapsto \max\{y,0\}$ is non-decreasing, and thus,
\begin{equation}
     \max\{ \textcolor{blue}{\mathbf{c} + \mathbf{d}}, 0\} \leq \max\{\textcolor{blue}{\max\{\mathbf{c}, 0\} + \max\{\mathbf{d}, 0\}},0 \}= \textcolor{blue}{\max\{\mathbf{c}, 0\} + \max\{\mathbf{d}, 0\}},
\end{equation}
where the equality holds because $\max\{\mathbf{c}, 0\} + \max\{\mathbf{d}, 0\}\geq 0$. In addition, $\max\{\mathbf{c}, 0\} \leq |\mathbf{c}|$ holds because
\begin{equation}
    \max\{\mathbf{c}, 0\} = \begin{cases} \mathbf{c} & \text{if } \mathbf{c} \geq 0 \\ 0 & \text{if } \mathbf{c} < 0 \end{cases}.
\end{equation}
By using \eqref{21} and $\max\{\mathbf{c}, 0\} \leq |\mathbf{c}|$, we have that for any $\omega \in \Omega$,
\begin{align}
    \max\{G(\omega) - s, 0\} &  = \max\{(G(\omega) - s') + (s' - s), 0\}\label{25}\\
    & \leq \max\{G(\omega) - s', 0\} + \max\{s' - s, 0\}\\
    & \leq \max\{G(\omega) - s', 0\} + |s' - s|. \label{2727}
\end{align}
From \eqref{2727}, we conclude that for any $\pi \in \Pi$,
\begin{equation}
    E_{\mathbf{x}}^\pi(\max\{G - s, 0\}) \leq E_{\mathbf{x}}^\pi(\max\{G - s', 0\}) + |s' - s|.
\end{equation}
By taking the infimum over $\Pi$, we have
\begin{equation}\label{29}
   \inf_{\pi \in \Pi} E_{\mathbf{x}}^\pi(\max\{G - s, 0\}) \leq \inf_{\pi \in \Pi} E_{\mathbf{x}}^\pi(\max\{G - s', 0\}) + |s' - s|,
\end{equation}
and by using the definition \eqref{defvs}, it holds that
\begin{equation}\label{30}
    v_{s}^*(\mathbf{x}) \leq v_{s'}^*(\mathbf{x}) + |s' - s|.
\end{equation}
By exchanging the roles of $s$ and $s'$ in \eqref{25}--\eqref{30}, we find that
\begin{equation}\label{31}
    v_{s'}^*(\mathbf{x}) \leq v_{s}^*(\mathbf{x}) + |s - s'|.
\end{equation}
From \eqref{30} and \eqref{31}, we have
\begin{equation}
-|s - s'|   \leq  v_{s}^*(\mathbf{x}) - v_{s'}^*(\mathbf{x}) \leq |s - s'|,
\end{equation}
which proves the desired statement \eqref{20}.
\end{proof}

%
\subsection{About the Dirac measure}\label{aboutDiracsec}
This subsection derives a fact about the Dirac measure. While the content is elementary, we could not find a full explanation in any classical measure theory or real analysis textbook. So, we provide an explanation here.

Let us study a Dirac measure on a Borel space. Suppose that $\mathcal{Y}$ is a Borel space and $y \in \mathcal{Y}$ is given. Let $\delta_y$ be the Dirac measure in $\mathcal{P}(\mathcal{Y})$ concentrated at $y$. $\delta_{y}$ is also called the \emph{unit mass} concentrated at $y$ \cite[p. 17]{rudin1987realcomplex}. If $\phi : \mathcal{Y} \rightarrow \mathbb{R}^*$ is Borel-measurable, then 
\begin{equation}\label{diraceq}
    \int_\mathcal{Y} \phi \; \mathrm{d}\delta_{y} \coloneqq \int_\mathcal{Y} \phi(y_0) \; \delta_{y}(\mathrm{d}y_0) = \phi(y).
\end{equation}
Why does the equation \eqref{diraceq} hold? Recall that $\delta_{y} : \mathcal{B}_{\mathcal{Y}} \rightarrow \{0,1\}$ is defined by \cite[Examples 1.20 (b), p. 17]{rudin1987realcomplex}, 
\cite[p. 130 top]{bertsekas2004stochastic},
\begin{equation}\label{defDirac}
    \delta_{y}(B) \coloneqq \begin{cases} 1, & \text{if }y \in B, \\  0, & \text{if }y \in \mathcal{Y}\setminus B, \end{cases}
\end{equation}
and therefore,
\begin{equation}\label{my69}
    \delta_{y}(B) = I_{B}(y)
\end{equation}
for every $B \in \mathcal{B}_{\mathcal{Y}}$. First, suppose that $\phi \coloneqq I_{B}$, where $B \in \mathcal{B}_{\mathcal{Y}}$. In this case, we have
\begin{equation}
    \int_\mathcal{Y} \phi \; \mathrm{d}\delta_{y} = \int_\mathcal{Y} I_{B} \; \mathrm{d}\delta_{y} \overset{(*)}{=} \delta_y(B) \overset{\eqref{my69}}{=} I_{B}(y) = \phi(y),
\end{equation}
where $(*)$ holds by the definition of the integral \cite[1.5.3, p. 36]{ash1972}. Second, suppose that $\phi \coloneqq \sum_{i=1}^n b_i I_{B_i}$, where $n \in \mathbb{N}$, $\{B_1, B_2,\dots, B_n\}$ is a disjoint collection of sets in $\mathcal{B}_{\mathcal{Y}}$, and $b_i \in (0,+\infty)$ for every $i \in \{1,2,\dots,n\}$. In this second case, $\phi$ is called a nonnegative finite-valued simple function, and we have
\begin{equation}\label{7171}
    \int_\mathcal{Y} \phi \; \mathrm{d}\delta_{y} =  \int_\mathcal{Y} \left(\sum_{i=1}^n b_i I_{B_i}\right) \mathrm{d}\delta_{y} \overset{(*)}{=} \sum_{i=1}^n b_i \delta_{y}(B_i) \overset{\eqref{my69}}{=} \sum_{i=1}^n b_i I_{B_i}(y) = \phi(y).
\end{equation}
Third, suppose that $\phi$ is a nonnegative Borel-measurable function. Then, $\phi$ is the (pointwise) limit of a nondecreasing sequence of nonnegative, finite-valued, simple functions $\phi_i$ \cite[Th. 1.5.5 (a), p. 38]{ash1972}. That is, $\phi_i$ is nonnegative, finite-valued, and simple for every $i \in \mathbb{N}$, $\phi_i \leq \phi_{i+1}$ for every $i \in \mathbb{N}$, and 
\begin{equation}\label{71a}
    \phi(y_0) = \lim_{i \rightarrow +\infty} \phi_i(y_0), \quad \quad y_0 \in \mathcal{Y}.
\end{equation} 
Then, by the Monotone Convergence Theorem \cite[1.6.2, p. 44]{ash1972}, we have
\begin{equation}\label{71b}
    \lim_{i \rightarrow +\infty} \int_{\mathcal{Y}} \phi_i \; \mathrm{d}\delta_y = \int_{\mathcal{Y}} \phi \; \mathrm{d}\delta_y.
\end{equation}
Moreover, since $\phi_i$ is nonnegative, finite-valued, and simple, we have
\begin{equation}\label{71c}
    \int_{\mathcal{Y}} \phi_i \; \mathrm{d}\delta_y \overset{\eqref{7171}}{=} \phi_i(y), \quad \quad i \in \mathbb{N}.
\end{equation}
All together, we have
\begin{equation}\label{75alltogether}
    \int_{\mathcal{Y}} \phi \; \mathrm{d}\delta_y \overset{\eqref{71b}}{=} \lim_{i \rightarrow +\infty} \int_{\mathcal{Y}} \phi_i \; \mathrm{d}\delta_y \overset{\eqref{71c}}{=}\lim_{i \rightarrow +\infty}\phi_i(y) \overset{\eqref{71a}}{=} \phi(y).
\end{equation}
Finally, suppose that $\phi$ is an arbitrary Borel-measurable function. Then, we can write $\phi$ in terms of its positive and negative parts as follows \cite[p. 37]{ash1972}:
\begin{equation}\label{split}
    \phi(y_0) = \underbrace{\max\{\phi(y_0),0\}}_{\phi^+(y_0)} - \underbrace{\max\{-\phi(y_0),0\}}_{\phi^-(y_0)}, \quad \quad y_0 \in \mathcal{Y}.
\end{equation}
The right side of \eqref{split} can \emph{never} have the form $+\infty - \infty$ because
\begin{itemize}
    \item $\phi(y_0) \in \mathbb{R}$ $\implies$ $\phi^+(y_0) \in \mathbb{R}$ and $\phi^-(y_0) \in \mathbb{R}$;
    \item $\phi(y_0) = +\infty$ $\implies$ $\phi^+(y_0) = +\infty$ and $\phi^-(y_0) = 0$;
    \item $\phi(y_0) = -\infty$ $\implies$ $\phi^+(y_0) = 0$ and $\phi^-(y_0) = +\infty$.
\end{itemize}
Since $\phi^+$ and $\phi^-$ are Borel-measurable and nonnegative, we have
\begin{equation}\label{splitint}\begin{aligned}
    \int_{\mathcal{Y}} \phi^+ \; \mathrm{d}\delta_y & \overset{\eqref{75alltogether}}{=} \phi^+(y), \\ \int_{\mathcal{Y}} \phi^- \; \mathrm{d}\delta_y & \overset{\eqref{75alltogether}}{=} \phi^-(y).
\end{aligned}\end{equation}
Then,
\begin{equation}\label{veryclose}
    \phi(y) \overset{\eqref{split}}{=} \phi^+(y) - \phi^-(y) \overset{\eqref{splitint}}{=}\int_{\mathcal{Y}} \phi^+ \; \mathrm{d}\delta_y - \int_{\mathcal{Y}} \phi^- \; \mathrm{d}\delta_y
\end{equation}
does \emph{not} have the form $+\infty-\infty$, as explained above. Lastly, we apply the definition of the integral \cite[p. 37]{ash1972} to conclude that
\begin{equation}
    \phi(y) \overset{\eqref{veryclose}}{=} \int_{\mathcal{Y}} \phi^+ \; \mathrm{d}\delta_y - \int_{\mathcal{Y}} \phi^- \; \mathrm{d}\delta_y = \int_{\mathcal{Y}} \phi \; \mathrm{d}\delta_y,
\end{equation}
which shows the desired statement \eqref{diraceq}. In the next subsection, we derive $P_{\mathbf{x}}^\pi$ and the associated expectation.
\\\\
\subsection{A Derivation for $P_{\mathbf{x}}^\pi$ and the Associated Expectation}\label{appendixremark}
It is well-established that the system model of interest permits the construction of a unique probability measure on a space containing all possible trajectories.\footnote{For example, see \cite[Prop. 7.28, pp. 140--141]{bertsekas2004stochastic}, which is a special case of the Ionescu-Tulcea Theorem.} This measure is used to evaluate expectations of random variables, which can represent costs that may be incurred as the system evolves over time. Recall Assumption 1:
\begin{enumerate}
\item There exist $a \in \mathbb{R}$ and $b \in \mathbb{R}$ such that $a \leq c_t \leq b$ for every $t \in \mathbb{T}_N$. (We define $\mathcal{Z} \coloneqq [a,b]$.)
    \item The control space $C$ is compact.
    \item For every $t$, $f_t$ and $c_t$ are continuous functions, and $p_t(\cdot|\cdot,\cdot)$ is a continuous stochastic kernel.
\end{enumerate}
Note that we are working with Borel spaces:
\begin{itemize}
   \item $S$, $C$, and $D$ are Borel spaces by the assumed system model. $\mathcal{Z} = [a,b]$ is a closed subset of $\mathbb{R}$, which implies that $\mathcal{Z} \in \mathcal{B}_{\mathbb{R}}$. Since $\mathbb{R}$ is a Borel space and $\mathcal{Z} \in \mathcal{B}_{\mathbb{R}}$, $\mathcal{Z}$ is also a Borel space \cite[Prop. 7.12, p. 119]{bertsekas2004stochastic}. $\mathbb{S} \coloneqq S \times \mathcal{Z}$ with the product topology is a Borel space because it is a finite Cartesian product of Borel spaces \cite[Prop. 7.13, p. 119]{bertsekas2004stochastic}. Similarly, $S \times C$ with the product topology is a Borel space. Borel spaces are separable and metrizable \cite[p. 118]{bertsekas2004stochastic}.
\end{itemize}
Often, we use the notation $\mathbb{S} = S \times \mathcal{Z}$, and $\chi_t = (x_t,z_t)$ or $\chi = (x,z)$ denotes an arbitrary element of $\mathbb{S}$. If $\mathcal{M}$ is a metrizable space, we equip the set of probability measures on $(\mathcal{M},\mathcal{B}_{\mathcal{M}})$ with the weak topology, and we denote this topological space by $\mathcal{P}(\mathcal{M})$ \cite[p. 122, p. 127]{bertsekas2004stochastic}. Next, we will define a stochastic kernel on $\mathbb{S}$ given $\mathbb{S} \times C$ that provides the conditional distribution for the realizations of the augmented state. 

\subsubsection{Construction and Analysis of $\tilde{q}_t$}
Recall the definition 
\begin{equation}\label{myQ}
        q_t(\underline{S}|x,u) \coloneqq p_t( \{w \in D: f_t(x,u,w) \in \underline{S}\} | x,u ), \quad \quad \underline{S} \in \mathcal{B}_S, \quad  (x,u) \in S \times C,
\end{equation}
and the notation
\begin{equation}
\overline{q}_t(\underline{\mathcal{Z}}|x,z,u) \coloneqq \delta_{\max\{c_t(x,u),z\}}(\underline{\mathcal{Z}}), \quad \quad \underline{\mathcal{Z}} \in \mathcal{B}_{\mathcal{Z}}, \quad  (x,z,u) \in \mathbb{S} \times C.
\end{equation}
For every $(x,z,u) \in \mathbb{S} \times C$, let $\tilde{q}_t(\cdot|x,z,u)$ be the product measure of $q_t(\cdot|x,u)$ and $\overline{q}_t(\cdot|x,z,u)$. The product measure $\tilde{q}_t(\cdot|x,z,u)$ is the unique measure on $(\mathbb{S},\mathcal{B}_{\mathbb{S}})$ such that
\begin{equation}
    \tilde{q}_t(\underline{S} \times \underline{\mathcal{Z}}|x,z,u) = q_t(\underline{S}|x,u) \cdot \overline{q}_t(\underline{\mathcal{Z}}|x,z,u), \quad \quad \underline{S} \in \mathcal{B}_{S}, \quad \underline{\mathcal{Z}} \in \mathcal{B}_{\mathcal{Z}},
\end{equation}
by \cite[Cor. 2.6.3, p. 100]{ash1972} and \cite[Prop. 7.13, p. 119]{bertsekas2004stochastic}.
When we apply \cite[Cor. 2.6.3]{ash1972}, we are working with the probability spaces $(S,\mathcal{B}_S,q_{t}(\cdot|x,u))$ and $(\mathcal{Z},\mathcal{B}_\mathcal{Z},\overline{q}_{t}(\cdot|x,z,u))$, where the measures are sigma finite because they are finite (as they are probability measures). The domain of the product measure $\tilde{q}_t(\cdot|x,z,u)$ is the product sigma algebra of $\mathcal{B}_S$ and $\mathcal{B}_\mathcal{Z}$ \cite[Cor. 2.6.3]{ash1972}, which is the smallest sigma algebra that \emph{contains} all sets of the form $\underline{S} \times \underline{\mathcal{Z}}$ with $\underline{S} \in \mathcal{B}_{S}$ and $\underline{\mathcal{Z}} \in \mathcal{B}_{\mathcal{Z}}$ \cite[p. 97]{ash1972}. Since $S$ and $\mathcal{Z}$ are Borel spaces, the product sigma algebra of $\mathcal{B}_S$ and $\mathcal{B}_\mathcal{Z}$ is equivalent to $\mathcal{B}_{S \times \mathcal{Z}}$ \cite[Prop. 7.13, p. 119]{bertsekas2004stochastic}.

\begin{lemma}[Analysis of $\tilde{q}_t$]\label{analysistildeq}
Under Assumption 1, $\tilde{q}_t$ is a continuous stochastic kernel on $\mathbb{S}$ given $\mathbb{S} \times C$.
\end{lemma}
\hspace{-4mm}\begin{proof}
For every $(x,z,u) \in \mathbb{S} \times C$, we have that $\tilde{q}_t(\cdot|x,z,u) \in \mathcal{P}(\mathbb{S})$ because $\tilde{q}_t(\cdot|x,z,u)$ is a probability measure on  $(\mathbb{S},\mathcal{B}_{\mathbb{S}})$ and we have equipped the set of probability measures on $(\mathbb{S},\mathcal{B}_{\mathbb{S}})$ with the weak topology. $\tilde{q}_t$ is a stochastic kernel on $\mathbb{S}$ given $\mathbb{S} \times C$ because it provides a family of elements of $\mathcal{P}(\mathbb{S})$, where each element of $\mathcal{P}(\mathbb{S})$ depends on an element of $\mathbb{S} \times C$ \cite[Def. 7.12, p. 134]{bertsekas2004stochastic}. Now, consider the function $\gamma_t : \mathbb{S} \times C \rightarrow \mathcal{P}( \mathbb{S})$ defined by
\begin{equation}\label{mymy84}
    \gamma_t(x,z,u) \coloneqq \tilde{q}_t(\cdot|x,z,u), \quad \quad \quad (x,z,u) \in \mathbb{S} \times C.
\end{equation}
To show that $\tilde{q}_t$ is a continuous stochastic kernel, we need to show that $\gamma_t$ is continuous \cite[Def. 7.12]{bertsekas2004stochastic}. Recall the following properties:
\begin{itemize}
    \item The map $\gamma_{1,t} : S \times C \rightarrow \mathcal{P}(S)$ defined by
    \begin{equation}
        \gamma_{1,t}(x,u) \coloneqq q_t(\cdot|x,u)
    \end{equation}
   is continuous because $f_t$ is continuous, $S$, $C$, and $D$ are Borel spaces, and $p_t$ is a continuous stochastic kernel on $D$ given $S \times C$ \cite[top of p. 209]{bertsekas2004stochastic}.
    \item The map $\gamma_{2,t} : S \times \mathcal{Z} \times C \rightarrow \mathcal{P}(\mathcal{Z})$ defined by 
    \begin{equation}
        \gamma_{2,t}(x,z,u) \coloneqq \overline{q}_t(\cdot|x,z,u) \coloneqq \delta_{\max\{c_t(x,u),z\}}
    \end{equation}
   is continuous. The reason is three-fold: the map $\nu : \mathcal{Z} \rightarrow \mathcal{P}(\mathcal{Z})$ defined by $\nu(z) \coloneqq \delta_z$ is continuous \cite[Cor. 7.21.1, p. 130]{bertsekas2004stochastic},\footnote{In our work, $\delta_z$ denotes the Dirac measure on $(\mathcal{Z},\mathcal{B}_{\mathcal{Z}})$ concentrated at $z$. The reference \cite{bertsekas2004stochastic} uses the notation $p_z$ instead. When reading \cite[Cor. 7.21.1, p. 130]{bertsekas2004stochastic}, note that a homeomorphism is continuous.} the map $(x,z,u) \mapsto \max\{c_t(x,u),z\}$ is continuous due to $\max$ and $c_t$ being continuous, and a composition of two continuous maps on topological spaces is continuous \cite[p. 119]{folland2013real}. 
   \item The map $\gamma_{3,t} : S \times \mathcal{Z} \times C \rightarrow \mathcal{P}(S) \times \mathcal{P}(\mathcal{Z})$ defined by
   \begin{equation}\label{my8686}
       \gamma_{3,t}(x,z,u) \coloneqq (\gamma_{1,t}(x,u), \gamma_{2,t}(x,z,u)) = (q_t(\cdot|x,u), \overline{q}_t(\cdot|x,z,u))
   \end{equation}
   is continuous because $\gamma_{1,t}$ and $\gamma_{2,t}$ are continuous (we equip $\mathcal{P}(S) \times \mathcal{P}(\mathcal{Z})$ with the product topology \cite[Th. A3.2, p. 377]{ash1972}, \cite[p. 120]{bertsekas2004stochastic}).
   \item We paraphrase the following result \cite[Lemma 7.12, p. 144]{bertsekas2004stochastic}: \textcolor{blue}{If $\mathbb{X}$ and $\mathbb{Y}$ are separable metrizable spaces, then the map $\sigma : \mathcal{P}(\mathbb{X}) \times \mathcal{P}(\mathbb{Y}) \rightarrow \mathcal{P}(\mathbb{X} \times \mathbb{Y})$ defined by
   \begin{equation}\label{mymy86}
       \sigma(p,q) \coloneqq p q,
   \end{equation}
   where $pq$ is the product of the measures $p$ and $q$, is  continuous.}
\end{itemize}
Since $S$ and $\mathcal{Z}$ are Borel spaces, they are separable and metrizable \cite[p. 118]{bertsekas2004stochastic}. For every $(x,z,u) \in S \times \mathcal{Z} \times C$, the product of $q_t(\cdot|x,u)$ and $\overline{q}_t(\cdot|x,z,u)$ is $\tilde{q}_t(\cdot|x,z,u)$, and thus,
\begin{equation}\label{mu8787}
    \sigma(q_t(\cdot|x,u), \overline{q}_t(\cdot|x,z,u)) \overset{\eqref{mymy86}}{=}  \tilde{q}_t(\cdot|x,z,u).
\end{equation}
All together, we have
   \begin{equation}
     \sigma(\gamma_{3,t}(x,z,u)) \overset{\eqref{my8686}}{=} \sigma(q_t(\cdot|x,u), \overline{q}_t(\cdot|x,z,u)) \overset{\eqref{mu8787}}{=}  \tilde{q}_t(\cdot|x,z,u) \overset{\eqref{mymy84}}{=} \gamma_t(x,z,u).
   \end{equation}
   Since $\gamma_t = \sigma \circ \gamma_{3,t}$ is a composition of continuous maps, $\gamma_t$ is continuous. We conclude that $\tilde{q}_t$ is a  continuous stochastic kernel under Assumption 1.
\end{proof}

Next, we will use Lemma \ref{analysistildeq} and \cite[Prop. 7.28]{bertsekas2004stochastic} to derive a useful probability measure.

\subsubsection{Construction of $P_{\mathbf{x}}^\pi$ and Definition of $E_{\mathbf{x}}^\pi(\cdot)$}
Let Assumption 1 hold, and let $\mathbf{x} \in S$ and $\pi \in \Pi$ be given. Please note the following items:
\begin{itemize}
    \item $\delta_{\mathbf{x},a}$ is the Dirac measure on $(\mathbb{S}, \mathcal{B}_{\mathbb{S}})$ concentrated at the point $(\mathbf{x},a)$, i.e., for every $\underline{\mathbb{S}} \in \mathcal{B}_{\mathbb{S}}$,
    \begin{equation}\label{diracxsdef}
        \delta_{\mathbf{x},a}(\underline{\mathbb{S}}) \coloneqq \begin{cases} 1, & \text{if }(\mathbf{x},a) \in \underline{\mathbb{S}}, \\ 0, & \text{if }(\mathbf{x},a) \in \mathbb{S} \setminus \underline{\mathbb{S}}. \end{cases}
    \end{equation} 
    \item For every $t \in \mathbb{T} = \{0,1,\dots,N-1\}$ $\tilde{q}_t$ is a continuous stochastic kernel on $\mathbb{S}$ given $\mathbb{S} \times C$ under Assumption 1 (Lemma \ref{analysistildeq}). (We only need $\tilde{q}_t$ to be Borel-measurable to apply \cite[Prop. 7.28]{bertsekas2004stochastic}.) 
    \item For every $t \in \mathbb{T}$, $\pi_t$ is a Borel-measurable stochastic kernel on $C$ given $\mathbb{S}$ by the definition of $\Pi$.
\end{itemize}

%

Next, we use \cite[Prop. 7.28]{bertsekas2004stochastic} to construct a unique probability measure $P_{\mathbf{x}}^\pi \in \mathcal{P}(\Omega)$, where we define $\Omega \coloneqq (\mathbb{S} \times C)^N \times \mathbb{S}$. We translate the notation from \cite[Prop. 7.28]{bertsekas2004stochastic} to our setting in Tables \ref{table1} and \ref{table2}. The text in \textcolor{blue}{blue} denotes the symbols from \cite[Prop. 7.28]{bertsekas2004stochastic}, and the text in black denotes our symbols.
\begin{table}[ht]
\caption{Notation for Borel spaces and samples}
\centering
\setlength{\tabcolsep}{5pt}
\begin{tabular}{|p{50pt}|p{25pt}|p{25pt}|p{25pt}|p{25pt}|p{12pt}|p{25pt}|p{25pt}|p{25pt}|}
\hline
\textcolor{blue}{Borel space} &  \textcolor{blue}{$X_1$} & \textcolor{blue}{$X_2$} & \textcolor{blue}{$X_3$} & \textcolor{blue}{$X_4$} & $\cdots$ & \textcolor{blue}{$X_{n-2}$} & \textcolor{blue}{$X_{n-1}$} & \textcolor{blue}{$X_n$} \vspace{.5mm}\\
\hline
Borel space &  $\mathbb{S}$ & $C$ & $\mathbb{S}$ & $C$ & $\cdots$ & $\mathbb{S}$ & $C$ & $\mathbb{S}$ \vspace{.5mm}\\
\hline
\textcolor{blue}{Sample} &  \textcolor{blue}{$x_1$} & \textcolor{blue}{$x_2$} & \textcolor{blue}{$x_3$} & \textcolor{blue}{$x_4$} & $\cdots$ & \textcolor{blue}{$x_{n-2}$} & \textcolor{blue}{$x_{n-1}$} & \textcolor{blue}{$x_n$} \vspace{.5mm}\\
\hline
Sample &  $\chi_0$ & $u_0$ & $\chi_1$ & $u_1$ & $\cdots$ & $\chi_{N-1}$ & $u_{N-1}$ & $\chi_N$ \vspace{.5mm}\\
\hline
\end{tabular}
\label{table1}
\end{table}
\begin{table}[ht]
\caption{Notation for Borel-measurable stochastic kernels}
\centering
\setlength{\tabcolsep}{5pt}
\begin{tabular}{|p{100pt}|p{250pt}|p{110pt}|}
\hline
 \textcolor{blue}{Stochastic kernel or probability measure} & Samples permitted in conditional statement, \textcolor{blue}{$y_i = (x_1,\dots,x_i)$} &  Stochastic kernel or probability measure\vspace{.5mm}\\
\hline
\textcolor{blue}{$p(\mathrm{d}x_1)$} & N/A & $\delta_{\mathbf{x},a}(\mathrm{d}\chi_0)$ \vspace{.5mm}\\
\hline
\textcolor{blue}{$q_1(\mathrm{d}x_2|y_1)$} & $\textcolor{blue}{y_1 = x_1} = \chi_0$ & $\pi_0(\mathrm{d}u_0|\chi_0)$ \vspace{.5mm}\\
\hline
\textcolor{blue}{$q_2(\mathrm{d}x_3|y_2)$} & $\textcolor{blue}{y_2 = (x_1,x_2)} =  (\chi_0,u_0)$  & $\tilde{q}_0(\mathrm{d}\chi_1|\chi_0,u_0)$ \vspace{.5mm}\\
\hline
 \textcolor{blue}{$q_3(\mathrm{d}x_4|y_3)$} & $\textcolor{blue}{y_3 = (x_1,x_2,x_3)} =  (\chi_0,u_0,\chi_1)$ & $\pi_1(\mathrm{d}u_1|\chi_1)$ \vspace{.5mm}\\
 \hline
 \textcolor{blue}{$q_4(\mathrm{d}x_5|y_4)$} & $\textcolor{blue}{y_4 = (x_1,x_2,x_3,x_4)} =  (\chi_0,u_0,\chi_1,u_1)$ & $\tilde{q}_1(\mathrm{d}\chi_2|\chi_1,u_1)$ \vspace{.5mm}\\
\hline
  \textcolor{blue}{$\cdots$} & $\cdots$ &$\cdots$ \vspace{.5mm}\\
\hline
\textcolor{blue}{$q_{n-2}(\mathrm{d}x_{n-1}|y_{n-2})$}  & $\textcolor{blue}{y_{n-2} = (x_1,x_2,\dots,x_{n-2})} =  (\chi_0,u_0,\dots,\chi_{N-1})$ & $\pi_{N-1}(\mathrm{d}u_{N-1}|\chi_{N-1})$    \vspace{.5mm}\\
\hline
  \textcolor{blue}{$q_{n-1}(\mathrm{d}x_n|y_{n-1})$}  & $\textcolor{blue}{y_{n-1} = (x_1,x_2,\dots,x_{n-2},x_{n-1})} = (\chi_0,u_0,\dots,\chi_{N-1},u_{N-1})$ &
  $\tilde{q}_{N-1}(\mathrm{d}\chi_{N}|\chi_{N-1},u_{N-1})$\vspace{.5mm}\\
\hline
\end{tabular}
\label{table2}
\end{table}
\noindent

The result \cite[Prop. 7.28]{bertsekas2004stochastic}, which is a special case of the Ionescu-Tulcea Theorem, states that for $n = 2, 3, \dots$, there is a unique probability measure \textcolor{blue}{$r_n \in \mathcal{P}(X_1 \times \cdots \times X_n)$} such that
\begin{equation}\begin{aligned}
   & \textcolor{blue}{ r_n(\underline{X}_1 \times \underline{X}_2 \times \cdots \times  \underline{X}_{n-1} \times \underline{X}_{n}) }\\ & = \textcolor{blue}{\textstyle \int_{\underline{X}_1} \int_{\underline{X}_2} \cdots \int_{\underline{X}_{n-1}}  q_{n-1}(\underline{X}_{n}|x_1,x_2, \dots, x_{n-1}) \; q_{n-2}(\mathrm{d}x_{n-1}|x_1,x_2, \dots, x_{n-2}) \cdots q_1(\mathrm{d}x_2|x_1) \; p(\mathrm{d}x_1)}
\end{aligned}\end{equation}
for every \textcolor{blue}{$\underline{X}_1 \in \mathcal{B}_{X_1}$}, $\dots$, \textcolor{blue}{$\underline{X}_n \in \mathcal{B}_{X_n}$}. Using $n = 2N + 1$ and the notation from Tables \ref{table1} and \ref{table2},\footnote{The tuple $(\chi_0,u_0,\dots, \chi_{N-1},u_{N-1},\chi_N)$ has $2N + 1$ entries.} there is a unique probability measure $P_{\mathbf{x}}^\pi \in \mathcal{P}(\Omega)$ such that 
\begin{equation}\label{keyP}\begin{aligned}
    & P_{\mathbf{x}}^\pi(\underline{\mathbb{S}}_0 \times \underline{C}_0 \times \underline{\mathbb{S}}_1 \times \underline{C}_1 \times \underline{\mathbb{S}}_2 \times \cdots \times \underline{C}_{N-1} \times  \underline{\mathbb{S}}_N) \\ & = \textstyle \int_{\underline{\mathbb{S}}_0} \int_{\underline{C}_0} \int_{\underline{\mathbb{S}}_1} \int_{\underline{C}_1}\int_{\underline{\mathbb{S}}_2} \cdots \int_{\underline{C}_{N-1}} \tilde{q}_{N-1}(\underline{\mathbb{S}}_N|\chi_{N-1},u_{N-1}) \; \pi_{N-1}(\mathrm{d}u_{N-1}|\chi_{N-1}) 
     \cdots \\ & \hphantom{==========================} \tilde{q}_1(\mathrm{d}\chi_2|\chi_1,u_1) \; \pi_1(\mathrm{d}u_1|\chi_1) \; \tilde{q}_0(\mathrm{d}\chi_1|\chi_0,u_0) \; \pi_0(\mathrm{d}u_0|\chi_0) \; \delta_{\mathbf{x},a}(\mathrm{d}\chi_0)
\end{aligned}\end{equation}
for every $\underline{\mathbb{S}}_0 \in \mathcal{B}_{\mathbb{S}},\dots,\underline{\mathbb{S}}_N \in \mathcal{B}_{\mathbb{S}}$ and for every $\underline{C}_0 \in \mathcal{B}_C,\dots,\underline{C}_{N-1} \in \mathcal{B}_C$. While $P_{\mathbf{x}}^\pi$ depends on $a$, we do not include $a$ when writing the symbol $P_{\mathbf{x}}^\pi$ for brevity.

Let $G: \Omega \rightarrow \mathbb{R}^*$ be measurable relative to $\mathcal{B}_{\Omega}$ and $\mathcal{B}_{\mathbb{R}^*}$, i.e., $G: (\Omega,\mathcal{B}_{\Omega}) \rightarrow (\mathbb{R}^*,\mathcal{B}_{\mathbb{R}^*})$. Our presentation of the definition of the expectation of $G$ with respect to $P_{\mathbf{x}}^\pi$ follows \cite{ash1972} and \cite{bertsekas2004stochastic}. The positive and negative parts of $G$ are defined by
\begin{align}
    G^+(\omega) & \coloneqq \max\{G(\omega),0\},\\
    G^-(\omega) & \coloneqq \max\{-G(\omega),0\},
\end{align}
respectively, where $G^+ : \Omega \rightarrow \mathbb{R}^*$ and $G^-: \Omega \rightarrow \mathbb{R}^*$ are measurable relative to $\mathcal{B}_{\Omega}$ and $\mathcal{B}_{\mathbb{R}^*}$ \cite[p. 37]{ash1972}, \cite[p. 103]{bertsekas2004stochastic}. The expectation of $G$ with respect to $P_{\mathbf{x}}^\pi$ is defined by
\begin{subequations}\label{expectationdef}
\begin{equation}\label{defintegral}
    E_{\mathbf{x}}^\pi(G) \coloneqq \int_{\Omega} G \; \mathrm{d}P_{\mathbf{x}}^\pi \coloneqq \int_{\Omega} G^+ \; \mathrm{d}P_{\mathbf{x}}^\pi - \int_{\Omega} G^-  \; \mathrm{d}P_{\mathbf{x}}^\pi,
\end{equation}
\emph{if} the right side of \eqref{defintegral} does \emph{not} take the form $+\infty - \infty$; if the right side of \eqref{defintegral} takes the form $+\infty-\infty$, then we say that $E_{\mathbf{x}}^\pi(G)$ does not exist \cite[p. 37]{ash1972}. While $E_{\mathbf{x}}^\pi(\cdot)$ depends on $a$ through $P_{\mathbf{x}}^\pi$, we do not include $a$ when writing the symbol $E_{\mathbf{x}}^\pi(\cdot)$ for brevity.
If $\int_{\Omega} G^+ \; \mathrm{d}P_{\mathbf{x}}^\pi < +\infty$ or  $\int_{\Omega} G^- \; \mathrm{d}P_{\mathbf{x}}^\pi < +\infty$, that is, if $E_{\mathbf{x}}^\pi(G)$ exists,\footnote{If $G : \Omega \rightarrow \mathbb{R}^*$ is Borel-measurable and nonnegative, then $G^-(\omega) \coloneqq \max\{-G(\omega),0\} = 0$ for every $\omega \in \Omega$, which implies that $\int_\Omega G^{-}(\omega) \; \mathrm{d}P_{\mathbf{x}}^\pi(\omega) = 0 < +\infty$.} then we have
\begin{equation}\begin{aligned}
   \int_{\Omega} G \; \mathrm{d}P_{\mathbf{x}}^\pi 
   =  \int_{\mathbb{S}} \int_{C}  \cdots \int_{\mathbb{S}} G(\chi_0,u_0,\dots,\chi_N)  \; \tilde{q}_{N-1}(\mathrm{d}\chi_{N}|\chi_{N-1},u_{N-1})  \cdots \pi_0(\mathrm{d}u_0|\chi_0) \;  \delta_{\mathbf{x},a}(\mathrm{d}\chi_0),
\end{aligned}\end{equation}
\end{subequations}
by \cite[Prop. 7.28, see Eq. (47)]{bertsekas2004stochastic}, where we only write some of the stochastic kernels from \eqref{keyP} for brevity. Many of the subsequent sections will use the definition of the expectation with respect to $P_{\mathbf{x}}^\pi$. 
\subsubsection{Further Discussion about Integration}\label{phiBorelmeas}
Suppose that $g : \Omega \rightarrow \mathbb{R}^*$ is Borel-measurable (i.e., measurable relative to $\mathcal{B}_{\Omega}$ and $\mathcal{B}_{\mathbb{R}^*}$) and bounded below. Let $\pi \in \Pi$ be given. Consider the function $\psi^\pi : \mathbb{S} \rightarrow \mathbb{R}^*$ defined by $\psi^\pi(\chi_0) \coloneqq$
\begin{equation}\label{86}
   \int_{C} \int_{\mathbb{S}} \cdots \int_{C} \underbrace{\int_{\mathbb{S}} g(\chi_0,\dots,u_{N-1},\chi_N) \;
  \tilde{q}_{N-1}(\mathrm{d}\chi_{N}|\chi_{N-1},u_{N-1})}_{g_n(\chi_0,u_0,\dots,\chi_{N-1},u_{N-1})} \; \pi_{N-1}(\mathrm{d}u_{N-1}|\chi_{N-1})  \cdots \tilde{q}_0(\mathrm{d}\chi_1|\chi_0,u_0) \; \pi_0(\mathrm{d}u_0|\chi_0),
\end{equation} 
where we only write some of the integrals for brevity. Why is $\psi^\pi$ \eqref{86} Borel-measurable and bounded below? To address this question, the following fact is useful.

\begin{remark}[Extending stochastic kernels]\label{remarkext}
Suppose that $\mathbb{X}$, $\mathbb{Y}$, and $\mathbb{Z}$ are Borel spaces, and let $q(\mathrm{d}y|x)$ be a Borel-measurable stochastic kernel on $\mathbb{Y}$ given $\mathbb{X}$. Then, the stochastic kernel $q'$ on $\mathbb{Y}$ given $\mathbb{X} \times \mathbb{Z}$ defined by
\begin{equation}\label{my87}
    q'(\mathrm{d}y|x,z) \coloneqq q(\mathrm{d}y|x), \quad \quad (x,z) \in \mathbb{X} \times \mathbb{Z},
\end{equation}
is Borel-measurable. To prove this fact, we need to show that the function $\sigma' : \mathbb{X} \times \mathbb{Z} \rightarrow \mathcal{P}(\mathbb{Y})$ defined by
\begin{equation}\label{my88}
    \sigma'(x,z) \coloneqq q'(\mathrm{d}y|x,z)
\end{equation}
is Borel-measurable, i.e., for every $B \in \mathcal{B}_{\mathcal{P}(\mathbb{Y})}$, it holds that
\begin{equation}
    \{(x,z) \in \mathbb{X} \times \mathbb{Z} :  \sigma'(x,z) \in B \} \in \mathcal{B}_{\mathbb{X} \times \mathbb{Z}}.
\end{equation}
Let $B \in \mathcal{B}_{\mathcal{P}(\mathbb{Y})}$ be given. Since $q$ is a Borel-measurable stochastic kernel on $\mathbb{Y}$ given $\mathbb{X}$, the function $\sigma : \mathbb{X} \rightarrow \mathcal{P}(\mathbb{Y})$ defined by
\begin{equation}\label{my90}
    \sigma(x) \coloneqq q(\mathrm{d}y|x)
\end{equation}
is Borel-measurable, implying that
\begin{equation}
    \{x \in \mathbb{X} :  \sigma(x) \in B \} \in \mathcal{B}_{\mathbb{X}}.
\end{equation}
Since $\mathcal{B}_{\mathbb{X} \times \mathbb{Z}}$ contains all sets of the form $\underline{\mathbb{X}}\times\underline{\mathbb{Z}}$ with $\underline{\mathbb{X}} \in \mathcal{B}_{\mathbb{X}}$ and $\underline{\mathbb{Z}} \in \mathcal{B}_{\mathbb{Z}}$ (for instance, see the proof of \cite[Prop. 7.13, p. 119]{bertsekas2004stochastic}), we have
\begin{equation}\label{my92}
    \{x \in \mathbb{X} :  \sigma(x) \in B \} \times \mathbb{Z} \in \mathcal{B}_{\mathbb{X} \times \mathbb{Z}}.
\end{equation}
By \eqref{my87}, \eqref{my88}, and \eqref{my90}, we have
\begin{equation}\label{my93}
    \sigma'(x,z) = \sigma(x), \quad \quad (x,z) \in \mathbb{X} \times \mathbb{Z}.
\end{equation}
Therefore,
\begin{equation}\label{my94}
     \{(x,z) \in \mathbb{X} \times \mathbb{Z} :  \sigma'(x,z) \in B \} \overset{\eqref{my93}}{=} \{(x,z) \in \mathbb{X} \times \mathbb{Z} :  \sigma(x) \in B \}  = \{x \in \mathbb{X} :  \sigma(x) \in B \} \times \mathbb{Z},
\end{equation}
which is a member of $\mathcal{B}_{\mathbb{X} \times \mathbb{Z}}$ by \eqref{my92}. 
\end{remark}

The desired properties of $\psi^\pi$ \eqref{86} (Borel-measurable, bounded below) follow from Remark \ref{remarkext} and by successive applications of \cite[Prop. 7.29, p. 144]{bertsekas2004stochastic}, which we paraphrase: \textcolor{blue}{Let $\mathbb{X}$ and $\mathbb{Y}$ be Borel spaces, $q(\mathrm{d}y|x)$ be a Borel-measurable stochastic kernel on $\mathbb{Y}$ given $\mathbb{X}$, and $f: \mathbb{X} \times \mathbb{Y} \rightarrow \mathbb{R}^*$ be Borel-measurable and bounded below. Then, the function $\gamma : \mathbb{X} \rightarrow \mathbb{R}^*$ defined by $\gamma(x) \coloneqq \int_{\mathbb{Y}} f(x,y) \; q(\mathrm{d}y|x)$ is Borel-measurable and bounded below.}\footnote{The bounded-below property is not included in the statement of \cite[Prop. 7.29, p. 144]{bertsekas2004stochastic}.} To show that $\psi^\pi$ \eqref{86} is Borel-measurable and bounded below, one applies this proposition to the inner-most integral in \eqref{86} and then proceeds to the outer-most integral. We outline the first two steps below: 
\begin{enumerate}
    \item Consider $\mathbb{X} = (\mathbb{S} \times C)^N$ and $\mathbb{Y} = \mathbb{S}$. Define a stochastic kernel $\tilde{q}_{N-1}'$ on $\mathbb{Y}$ given $\mathbb{X}$ by
\begin{equation}\label{my95}
    \tilde{q}_{N-1}'(\mathrm{d}\chi_N|\chi_0, u_0,\dots, \chi_{N-1},u_{N-1}) \coloneqq \tilde{q}_{N-1}(\mathrm{d}\chi_N|\chi_{N-1},u_{N-1})
\end{equation}
for every $(\chi_0, u_0,\dots, \chi_{N-1},u_{N-1}) \in \mathbb{X}$.
$\tilde{q}_{N-1}'$ is Borel-measurable due to the Borel-measurability of $\tilde{q}_{N-1}$ (Remark \ref{remarkext}). The function $g : \mathbb{X} \times \mathbb{Y} \rightarrow \mathbb{R}^*$ in \eqref{86} is Borel-measurable and bounded below by assumption. Thus, the function $g_n : \mathbb{X} \rightarrow \mathbb{R}^*$ defined by
\begin{equation}\label{mygn}
\begin{aligned}
     g_n(\textcolor{magenta}{\chi_0, u_0,\dots, \chi_{N-1},u_{N-1}})  & \coloneqq \int_{\mathbb{S}} g(\textcolor{magenta}{\chi_0, u_0,\dots, \chi_{N-1},u_{N-1}},\chi_N) \;
  \tilde{q}_{N-1}(\mathrm{d}\chi_N|\chi_{N-1},u_{N-1}) \\
  & = \int_{\mathbb{S}} g(\textcolor{magenta}{\chi_0, u_0,\dots, \chi_{N-1},u_{N-1}},\chi_N) \;
  \tilde{q}_{N-1}'(\mathrm{d}\chi_N|\textcolor{magenta}{\chi_0, u_0,\dots, \chi_{N-1},u_{N-1}})
\end{aligned}
\end{equation}
is Borel-measurable and bounded below by applying \cite[Prop. 7.29, p. 144]{bertsekas2004stochastic}. We also use the fact that a finite Cartesian product of Borel spaces with the product topology is a Borel space \cite[Prop. 7.13, p. 119]{bertsekas2004stochastic}.
\item By substituting $g_n$ \eqref{mygn} into the definition of $\psi^\pi$ \eqref{86}, we have
\begin{equation}
   \psi^\pi(\chi_0) = \int_{C} \int_{\mathbb{S}} \cdots \underbrace{\int_{C} g_n(\chi_0,u_0,\dots,\chi_{N-1},u_{N-1}) \; \pi_{N-1}(\mathrm{d}u_{N-1}|\chi_{N-1})}_{g_{n-1}(\chi_0, u_0,\dots, \chi_{N-1})}  \cdots \tilde{q}_0(\mathrm{d}\chi_1|\chi_0,u_0) \; \pi_0(\mathrm{d}u_0|\chi_0).
\end{equation}
By an analogous argument, $g_{n-1}$ is Borel-measurable and bounded below.
\end{enumerate}

\subsection{More Measure-theoretic Fundamentals}\label{dprecursion}
First, we recall some preliminaries. 
Every $\omega \in \Omega \coloneqq (\mathbb{S} \times C)^N \times \mathbb{S}$ takes the form
\begin{equation}\label{myomega}
    \omega = (x_0,z_0,u_0,\dots,x_{N-1},z_{N-1},u_{N-1},x_N,z_N),
\end{equation}
and we recall the notation $\chi_t = (x_t,z_t) \in S \times \mathcal{Z} = \mathbb{S}$. 
We define $X_t$, $Z_t$, and $U_t$ to be projections from $\Omega$ to $S$, $\mathcal{Z}$, and $C$, respectively, such that for every $\omega \in \Omega$ of the form in \eqref{myomega},
\begin{subequations}\label{ro}
\begin{align}
    X_t(\omega) &\coloneqq x_t, \;\;\;\;\;\;\;\; t \in \mathbb{T}_N, \\
    Z_t(\omega) &\coloneqq z_t, \;\;\;\;\;\;\;\;  t \in \mathbb{T}_N, \\
    U_t(\omega) &\coloneqq u_t, \;\;\;\;\;\;\;\; t \in \mathbb{T}.
\end{align}
\end{subequations}
%
$Z_{t+1}$ depends on $X_t$, $Z_t$, and $U_t$ as follows:
\begin{equation}\label{myZt}
    Z_{t+1}(\textcolor{magenta}{\omega}) = \max\{c_t(X_t(\textcolor{magenta}{\omega}),U_t(\textcolor{magenta}{\omega})), Z_t(\textcolor{magenta}{\omega})\}, \quad \quad \textcolor{magenta}{\omega} \in \Omega, \quad t \in \mathbb{T}.
\end{equation}
The realizations of $(X_0,Z_0)$ are concentrated at $(\mathbf{x}, a)$, where $\mathbf{x}$ can be any element of $S$. 
%
%
For every $s \in \mathbb{R}$ and $\textcolor{magenta}{\omega} \in \Omega$, we define
\begin{align}
Y(\textcolor{magenta}{\omega}) & \coloneqq \max\Big\{ c_N(X_N(\textcolor{magenta}{\omega})), \max_{i \in \mathbb{T}} c_i(X_i(\textcolor{magenta}{\omega}),U_i(\textcolor{magenta}{\omega})) \Big\}, \label{myYbar}\\
   Y_t^s(\textcolor{magenta}{\omega}) & \coloneqq h^s\left(\max\Big\{ c_N(X_N(\textcolor{magenta}{\omega})), \max_{i \in \{t,\dots,N-1\}} c_i(X_i(\textcolor{magenta}{\omega}),U_i(\textcolor{magenta}{\omega})), Z_t(\textcolor{magenta}{\omega}) \Big\}\right), \;\;\; \;\;\; t \in \mathbb{T},\label{42}\\
   Y_N^s(\textcolor{magenta}{\omega}) & \coloneqq h^s\left(\max\{ c_N(X_N(\textcolor{magenta}{\omega})), Z_N(\textcolor{magenta}{\omega}) \}\right), \label{YN}
\end{align}
where $h^s : \mathbb{R} \rightarrow \mathbb{R}$ is defined by
\begin{equation}\label{hs}
    h^s(\textcolor{blue}{y}) \coloneqq \max\{\textcolor{blue}{y} - s,0\},
\end{equation}
which is nonnegative and continuous.
\begin{remark}[Equivalent expressions for $Y$ and $Y_t^s$]
For every $s \in \mathbb{R}$ and $\omega \in \Omega$ of the form in \eqref{myomega}, we have
\begin{align}
Y(\omega) & = \max\Big\{ c_N(x_N), \max_{i \in \mathbb{T}} c_i(x_i,u_i) \Big\},\label{44a}\\
Y_t^s(\omega) & = \max\Big\{ \textcolor{blue}{\max\Big\{ c_N(x_N), \max_{i \in \{t,\dots,N-1\}} c_i(x_i,u_i), z_t \Big\}} - s, 0 \Big\}, \quad \quad  t \in \mathbb{T},\\
Y_N^s(\omega) & = \max\{ \textcolor{blue}{\max\{ c_N(x_N), z_N \}} - s, 0 \},
\end{align}
by applying the definitions of $X_t$, $Z_t$, and $U_t$ \eqref{ro} and the definition of $h^s$ \eqref{hs}.
\end{remark}
\subsubsection{Analysis of $(X_t,Z_t)$}\label{randomobj}
Here, we explain why $(X_t,Z_t)$ is a random object. First, we introduce some terminology. If $(\Omega_1, \mathcal{F}_1)$ and $(\Omega_2, \mathcal{F}_2)$ are measurable spaces and $\bar{g} : \Omega_1 \rightarrow \Omega_2$ is measurable relative to $\mathcal{F}_1$ and $\mathcal{F}_2$, then $\bar{g}$ is called a \emph{random object} \cite[p. 214]{ash1972}. If $(\Omega_2,\mathcal{F}_2) = (\mathbb{R},\mathcal{B}_{\mathbb{R}})$, then $\bar{g}$ is called a \emph{random variable}. If $(\Omega_2,\mathcal{F}_2) = (\mathbb{R}^*,\mathcal{B}_{\mathbb{R}^*})$, then $\bar{g}$ is called an \emph{extended random variable}.  

Now, let $t \in \mathbb{T}_N$ be given, and recall that $X_t : \Omega \rightarrow S$ and $Z_t : \Omega \rightarrow \mathcal{Z}$ are defined by
\begin{align}
    X_t(\omega) \coloneqq x_t,\\
    Z_t(\omega) \coloneqq z_t,
\end{align}
for every $\omega \in \Omega$ of the form in \eqref{myomega}. $X_t$ is measurable relative to $\mathcal{B}_{\Omega}$ and $\mathcal{B}_{S}$; i.e., $X_t$ is Borel-measurable. This is because for every $\underline{S} \in \mathcal{B}_{S}$, we have
\begin{align}
    X_t^{-1}(\underline{S}) & \coloneqq \{ \omega \in \Omega : X_t(\omega) \in \underline{S} \} \\
    & = \{ (x_0,z_0,u_0,\dots,x_t,z_t,u_t,\dots,x_N,z_N) \in \Omega : x_t \in \underline{S} \}\\
    & = \mathbb{S} \times C \times \cdots \times \underline{S} \times \mathcal{Z} \times C \times \cdots \times \mathbb{S}.\label{63}
\end{align}
The set $X_t^{-1}(\underline{S})$ is an element of $\mathcal{B}_{\Omega}$ because $X_t^{-1}(\underline{S})$ is a measurable rectangle by \eqref{63}; e.g., see the proof of \cite[Prop. 7.13, p. 119]{bertsekas2004stochastic}. Analogous steps show that $Z_t$ is measurable relative to $\mathcal{B}_{\Omega}$ and $\mathcal{B}_{\mathcal{Z}}$; i.e., $Z_t$ is Borel-measurable.

To show that $(X_t,Z_t)$ is Borel-measurable, we can use \cite[Prop. 7.14, p. 120]{bertsekas2004stochastic}, which we paraphrase: \textcolor{blue}{Let $\bar{X}$, $\bar{Y}_1$, and $\bar{Y}_2$ be Borel spaces, and for $i = 1,2$, let $\bar{f}_i : \bar{X} \rightarrow \bar{Y}_i$ be a function. If $\bar{f}_1$ and $\bar{f}_2$ are Borel-measurable, then the function $\bar{F}_2 : \bar{X} \rightarrow \bar{Y}_1 \times \bar{Y}_2$ defined by
\begin{equation}
    \bar{F}_2(x) \coloneqq (\bar{f}_1(x),\bar{f}_2(x))
\end{equation}
is Borel-measurable.}

In our problem, $\Omega$, $S$, and $\mathcal{Z}$ are Borel spaces, and $X_t : \Omega \rightarrow S$ and $Z_t : \Omega \rightarrow \mathcal{Z}$ are functions. For brevity, we define $\mathcal{X}_t \coloneqq (X_t,Z_t)$. Since $X_t$ and $Z_t$ are Borel-measurable, the function $\mathcal{X}_t : \Omega \rightarrow \mathbb{S}$ defined by
\begin{equation}\label{augstatedef}
    \mathcal{X}_t(\omega) \coloneqq (X_t,Z_t)(\omega) \coloneqq (X_t(\omega),Z_t(\omega))
\end{equation}
is Borel-measurable, i.e., measurable relative to $\mathcal{B}_{\Omega}$ and $\mathcal{B}_{\mathbb{S}}$, by \cite[Prop. 7.14]{bertsekas2004stochastic}. The function $(X_t,Z_t) : \Omega \rightarrow \mathbb{S}$ being Borel-measurable means that $(X_t,Z_t)$ is a random object.

\subsubsection{Studying $Y_0^s$ and $\max\{Y-s,0\}$}
The next lemma provides a relationship between $Y_0^s$ and $\max\{Y-s,0\}$.
\begin{lemma}\label{lemma3pt2}
Let Assumption 1 hold, and let $\mathbf{x} \in S$, $\pi \in \Pi$, and $s \in \mathbb{R}$ be given. Then, we have
\begin{equation}\label{desiredlemma3pt2}
    E_{\mathbf{x}}^\pi(Y_0^s) = E_{\mathbf{x}}^\pi(\max\{Y-s,0\}).
\end{equation}
\end{lemma}\vspace{3mm}
\hspace{-4mm}\begin{proof}
$Y_0^s$ is a random variable on $(\Omega,\mathcal{B}_{\Omega},P_{\mathbf{x}}^\pi)$ because $Y_0^s : \Omega \rightarrow \mathbb{R}$ is measurable relative to $\mathcal{B}_{\Omega}$ and $\mathcal{B}_{\mathbb{R}}$, and $P_{\mathbf{x}}^\pi$ is a probability measure on $(\Omega,\mathcal{B}_{\Omega})$. 
For convenience, we restate $Y_0^s$ \eqref{42} using $\mathbb{T} = \{0,1,\dots,N-1\}$
\begin{equation}\label{mymy116}
    \textcolor{magenta}{Y_0^s} = \textcolor{magenta}{h^s\left(\max\left\{ c_N(X_N), \max_{i \in \mathbb{T}} c_i(X_i,U_i), Z_0 \right\}\right)},
\end{equation}
and $\max\{Y-s,0\}$ using the definition of $h^s$ \eqref{hs} and the definition of $Y$ \eqref{myYbar}:
\begin{equation}\label{mymy117}
    \max\{Y-s,0\} \overset{\eqref{hs}}{=} h^s(Y) \overset{\eqref{myYbar}}{=} h^s\left(\max\left\{ c_N(X_N), \max_{i \in \mathbb{T}} c_i(X_i,U_i) \right\}\right).
\end{equation}
By comparing \eqref{mymy116} and \eqref{mymy117}, to show that $E_{\mathbf{x}}^\pi(Y_0^s) = E_{\mathbf{x}}^\pi(\max\{Y-s,0\})$, it suffices to show that 
\begin{equation}\label{my118118}
    \max\left\{ c_N(X_N), \max_{i \in \mathbb{T}} c_i(X_i,U_i), Z_0 \right\} = Y \quad \quad \text{a.e. w.r.t. $P_{\mathbf{x}}^\pi$}.
\end{equation}
Indeed, if \eqref{my118118} holds, then
\begin{equation}
    \textcolor{magenta}{h^s\left(\max\left\{ c_N(X_N), \max_{i \in \mathbb{T}} c_i(X_i,U_i), Z_0 \right\}\right)} = h^s(Y) \quad \quad \text{a.e. w.r.t. $P_{\mathbf{x}}^\pi$},
\end{equation}
equivalently, using \eqref{mymy116},
\begin{equation}
    \textcolor{magenta}{Y_0^s} =  h^s(Y) \quad \quad \text{a.e. w.r.t. $P_{\mathbf{x}}^\pi$},
\end{equation}
equivalently, using \eqref{mymy117},
\begin{equation}\label{mymy1212}
    Y_0^s =  \max\{Y-s,0\} \quad \quad \text{a.e. w.r.t. $P_{\mathbf{x}}^\pi$}.
\end{equation}
The statement \eqref{mymy1212} implies that
\begin{equation}\label{my122}
    \int_{\Omega} Y_0^s \; \mathrm{d}P_{\mathbf{x}}^\pi = \int_{\Omega} \max\{Y-s,0\} \; \mathrm{d}P_{\mathbf{x}}^\pi,
\end{equation}
which is equivalent to the desired statement \eqref{desiredlemma3pt2}.
The integrals in \eqref{my122} exist because $Y_0^s : \Omega \rightarrow \mathbb{R}$ and $\max\{Y-s,0\} : \Omega \rightarrow \mathbb{R}$ are Borel-measurable and nonnegative. Thus, it suffices to show \eqref{my118118} to complete the proof. 

We need to explain two items before proceeding. The first item concerns the probability measure induced by $\mathcal{X}_0$. This measure is defined by
\begin{equation}\label{mymy777}
    P_{\mathbf{x},\mathcal{X}_0}^\pi(\underline{\mathbb{S}}) \coloneqq P_\mathbf{x}^\pi(\mathcal{X}_0^{-1}(\underline{\mathbb{S}})), \quad \quad  \underline{\mathbb{S}} \in \mathcal{B}_{\mathbb{S}}.
\end{equation}
It holds that $P_{\mathbf{x},\mathcal{X}_0}^\pi = \delta_{\mathbf{x},a}$. Indeed, for every $\underline{\mathbb{S}} \in \mathcal{B}_{\mathbb{S}}$, we have
\begin{equation}\label{P00}\begin{aligned}
    P_{\mathbf{x},\mathcal{X}_0}^\pi(\underline{\mathbb{S}}) & \overset{\eqref{mymy777}}{=}  P_\mathbf{x}^\pi(\{ \omega \in \Omega : \mathcal{X}_0(\omega) \in \underline{\mathbb{S}} \})\\
    & \overset{\hphantom{\eqref{keyP}}}{=}  P_\mathbf{x}^\pi(\underbrace{\underline{\mathbb{S}} \times C}_{\text{stage 0}} \times \cdots \times \underbrace{\mathbb{S} \times C}_{\text{stage $N-1$}} \times \underbrace{\mathbb{S}}_{\text{stage $N$}})\\
    & \overset{\eqref{keyP}}{=} \textstyle \int_{\underline{\mathbb{S}}} \int_{C} \cdots \int_{C}   \tilde{q}_{N-1}(\mathbb{S}|\chi_{N-1},u_{N-1}) \; \pi_{N-1}(\mathrm{d}u_{N-1}|\chi_{N-1})  \cdots \pi_0(\mathrm{d}u_0|\chi_0) \; \delta_{\mathbf{x},a}(\mathrm{d}\chi_0)\\
    & \overset{\hphantom{\eqref{keyP}}}{=} \delta_{\mathbf{x},a}(\underline{\mathbb{S}})
\end{aligned}\end{equation}
because the innermost integrals evaluate to one. Secondly, since the stage and terminal cost functions are bounded below by $a \in \mathbb{R}$, we have
\begin{equation}\label{50}
    \max\left\{c_N(x_N), \max_{i \in \mathbb{T}} c_i(x_i,u_i), \textcolor{magenta}{a} \right\} =  \max\left\{c_N(x_N), \max_{i \in \mathbb{T}} c_i(x_i,u_i) \right\}
\end{equation}
for every $x_0 \in S, \dots, x_N\in S$ and for every $u_0 \in C, \dots, u_{N-1} \in C$. That is, the $a$ is redundant for evaluating the maximum. To see this explicitly, denote $m_0 \coloneqq \max_{i \in \mathbb{T}} c_i(x_i,u_i)$ for brevity. If 
\begin{equation}\label{mymy133}
    \textcolor{blue}{\max \left\{c_N(x_N), m_0 \right\}} \geq \textcolor{magenta}{a},
\end{equation}
then
\begin{equation}
    \max\left\{ \textcolor{blue}{\max \left\{c_N(x_N), m_0 \right\}},\textcolor{magenta}{a}\right\} = \textcolor{blue}{\max \left\{c_N(x_N), m_0\right\}},
\end{equation}
which is equivalent to \eqref{50} because
\begin{equation}
    \max\left\{ \textcolor{blue}{\max \left\{c_N(x_N), m_0 \right\}},\textcolor{magenta}{a}\right\} = \max\left\{ c_N(x_N), m_0, \textcolor{magenta}{a} \right\}.
\end{equation}
The inequality \eqref{mymy133} holds because $c_N(x_N) \geq \textcolor{magenta}{a}$ and $m_0 \geq \textcolor{magenta}{a}$.

Now, suppose that $\omega = (x_0,z_0,u_0,\dots,x_{N},z_{N}) \in \Omega = S \times \mathcal{Z} \times C \times \cdots \times S \times \mathcal{Z}$ satisfies $(x_0,z_0) \in S \times \{a\}$. Then,
\begin{align}
   \textcolor{red}{ \max\left\{ c_N(X_N(\omega)), \max_{i \in \mathbb{T}} c_i(X_i(\omega),U_i(\omega)), Z_0(\omega) \right\}} & \overset{\eqref{ro}}{=} \max\left\{ c_N(x_N), \max_{i \in \mathbb{T}} c_i(x_i,u_i), z_0 \right\} \\ & \overset{z_0 \in \{a\}}{=} \max\left\{ c_N(x_N), \max_{i \in \mathbb{T}} c_i(x_i,u_i), a \right\} \\
    & \overset{\eqref{50}}{=} \max\left\{ c_N(x_N), \max_{i \in \mathbb{T}} c_i(x_i,u_i)\right\}\\
    & \overset{\eqref{ro}}{=} \max\Big\{ c_N(X_N(\omega)), \max_{i \in \mathbb{T}} c_i(X_i(\omega),U_i(\omega)) \Big\} \\
    & \overset{\eqref{myYbar}}{=} Y(\omega).
\end{align}
Therefore,
\begin{equation}\label{myB1}
    B_1 \coloneqq \{\omega = (x_0,z_0,u_0,\dots,x_{N},z_{N}) \in \Omega : (x_0,z_0) \in S \times \{a\}\}
\end{equation}
is a subset of
\begin{equation}
    B_2 \coloneqq \left\{\omega \in \Omega : \textcolor{red}{\max\left\{ c_N(X_N(\omega)), \max_{i \in \mathbb{T}} c_i(X_i(\omega),U_i(\omega)), Z_0(\omega) \right\}} = Y(\omega) \right\},
\end{equation}
and hence,
\begin{equation}
    P_{\mathbf{x}}^\pi(B_1) \leq P_{\mathbf{x}}^\pi(B_2).
\end{equation}
The statement that we desire is \eqref{my118118}, which is the same as $P_{\mathbf{x}}^\pi(B_2) = 1$.
It suffices to show that $P_{\mathbf{x}}^\pi(B_1) = 1$, since then we would have
\begin{equation}
  1 = P_{\mathbf{x}}^\pi(B_1) \leq P_{\mathbf{x}}^\pi(B_2) \leq 1.
\end{equation}
Finally, we have
\begin{align}
    P_{\mathbf{x}}^\pi(B_1) & \overset{\eqref{myB1}}{=} P_{\mathbf{x}}^\pi(\{\omega = (x_0,z_0,u_0,\dots,x_{N},z_{N}) \in \Omega : (x_0,z_0) \in S \times \{a\}\})\\
    & \overset{\eqref{ro},\eqref{augstatedef}}{=} P_{\mathbf{x}}^\pi(\{\omega \in \Omega : \mathcal{X}_0(\omega) \in S \times \{a\}\})\\
    & \overset{\eqref{mymy777}}{=}
    P_{\mathbf{x},\mathcal{X}_0}^\pi(S \times \{a\}) \\
    & \overset{\eqref{P00}}{=} \delta_{\mathbf{x},a}(S \times \{a\})\\
    & \overset{\hphantom{\eqref{P00}}}{=} 1,
\end{align}
where the last line holds because $(\mathbf{x},a) \in S \times \{a\}$ \eqref{diracxsdef}.
\end{proof}
%

\subsubsection{Analysis of $Y_t^s$}\label{analysisYbarts}
We recall from \eqref{42}--\eqref{YN} that for every $s \in \mathbb{R}$ and $\omega \in \Omega$, 
\begin{align*}
   Y_t^s(\textcolor{magenta}{\omega}) & = h^s\left(\max\Big\{ c_N(X_N(\textcolor{magenta}{\omega})), \max_{i \in \{t,\dots,N-1\}} c_i(X_i(\textcolor{magenta}{\omega}),U_i(\textcolor{magenta}{\omega})), Z_t(\textcolor{magenta}{\omega}) \Big\}\right), \;\;\; \;\;\; t \in \mathbb{T},\\
   Y_N^s(\textcolor{magenta}{\omega}) & = h^s\left(\max\{ c_N(X_N(\textcolor{magenta}{\omega})), Z_N(\textcolor{magenta}{\omega}) \}\right),
\end{align*}
where $h^s : \mathbb{R} \rightarrow \mathbb{R}$ is a nonnegative continuous function defined by $h^s(y) = \max\{y - s,0\}$. Each $c_j$ is real-valued and Borel-measurable, and $X_j : \Omega \rightarrow S$, $Z_j : \Omega \rightarrow \mathcal{Z}$, and $U_j : \Omega \rightarrow C$ are Borel-measurable functions \eqref{ro}. As Borel-measurability is preserved under compositions and maximums, $Y_t^s : \Omega \rightarrow \mathbb{R}$ for every $t \in \mathbb{T}_N$ is Borel-measurable, i.e., measurable relative to $\mathcal{B}_{\Omega}$ and $\mathcal{B}_{\mathbb{R}}$. The following lemma verifies additional useful properties.
\begin{lemma}\label{lemma3pt3}
Let Assumption 1 hold. For every $s \in \mathbb{R}$, $t \in \mathbb{T}_N$, $\mathbf{x} \in S$, and $\pi \in \Pi$, $Y_t^s$ is an extended random variable on $(\Omega,\mathcal{B}_{\Omega},P_{\mathbf{x}}^\pi)$. In addition, for every $t \in \mathbb{T}$, it holds that $Y_t^s = Y_{t+1}^s$.
\end{lemma}
\hspace{-4mm}\begin{proof}
Let $s \in \mathbb{R}$ and $t \in \mathbb{T}_N$ be given. For every $\mathbf{x} \in S$ and $\pi \in \Pi$, $(\Omega,\mathcal{B}_{\Omega},P_{\mathbf{x}}^\pi)$ is a probability space because $\mathcal{B}_{\Omega}$ is a sigma algebra of subsets of $\Omega$ and $P_{\mathbf{x}}^\pi$ is a probability measure on $(\Omega,\mathcal{B}_{\Omega})$. To verify that $Y_t^s$ is an extended random variable on $(\Omega,\mathcal{B}_{\Omega},P_{\mathbf{x}}^\pi)$, we must show that $Y_t^s : \Omega \rightarrow \mathbb{R}^*$ is measurable relative to $\mathcal{B}_{\Omega}$ and $\mathcal{B}_{\mathbb{R}^*}$.

For every $\omega \in \Omega$, it holds that $Y_t^s(\omega) \in \mathbb{R} \subseteq \mathbb{R}^*$, so we can view $Y_t^s$ as a function from $\Omega$ to $\mathbb{R}^*$. Next, we explain why $Y_t^s$ is measurable relative to $\mathcal{B}_{\Omega}$ and $\mathcal{B}_{\mathbb{R}^*}$.
\begin{enumerate}
    \item Note that $\mathcal{B}_{\mathbb{R}^*}$ is generated by, for example, the family $\mathcal{E} \coloneqq \{ (r, +\infty] : r \in \mathbb{R} \}$ \cite[p. 45]{folland2013real}. 
    \item Since $\mathcal{B}_{\mathbb{R}^*}$ is generated by $\mathcal{E}$, $Y_t^s : \Omega \rightarrow \mathbb{R}^*$ is measurable relative to $\mathcal{B}_{\Omega}$ and $\mathcal{B}_{\mathbb{R}^*}$ if and only if
    \begin{equation}\label{mymy69}
       \forall \underline{\mathcal{E}} \in \mathcal{E}, \;\;\; \;\; \{\omega \in \Omega : Y_t^s(\omega) \in \underline{\mathcal{E}}\} \in \mathcal{B}_{\Omega}
    \end{equation}
    by \cite[Prop. 2.1, p. 43]{folland2013real}.
    \item Let $\underline{\mathcal{E}} \in \mathcal{E}$ be given. So, $\underline{\mathcal{E}} = (r, +\infty]$ for some $r \in \mathbb{R}$. Then,
    \begin{equation}\label{my70}
        \{\omega \in \Omega : Y_t^s(\omega) \in \underline{\mathcal{E}}\} = \{\omega \in \Omega : Y_t^s(\omega) \in (r, +\infty]\} = \{\omega \in \Omega : Y_t^s(\omega) \in (r, +\infty)\},
    \end{equation}
    where the last step holds because $Y_t^s(\omega) \in \mathbb{R}$ for every $\omega \in \Omega$.
    \item Since $(r, +\infty)$ is an open set in $\mathbb{R}$, it is an element of $\mathcal{B}_{\mathbb{R}}$. Since $(r, +\infty) \in \mathcal{B}_{\mathbb{R}}$ and $Y_t^s$ is measurable relative to $\mathcal{B}_{\Omega}$ and $\mathcal{B}_{\mathbb{R}}$, we have that
    \begin{equation}\label{71}
        \{\omega \in \Omega : Y_t^s(\omega) \in \underline{\mathcal{E}}\} \overset{\eqref{my70}}{=} \{\omega \in \Omega : Y_t^s(\omega) \in (r, +\infty)\} \in \mathcal{B}_{\Omega},
    \end{equation}
    proving \eqref{mymy69}. We conclude that $Y_t^s$ is measurable relative to $\mathcal{B}_{\Omega}$ and $\mathcal{B}_{\mathbb{R}^*}$.
\end{enumerate}

The last part of the proof is to show that $Y_t^s = Y_{t+1}^s$. We will use the following: for any $r_i \in \mathbb{R}$ with $i \in \{1,2,3,4\}$,
\begin{align}
    \max\{r_1, \max\{r_2, r_3\}, r_4\} & =  \max\{r_1, r_2, r_3, r_4\} \label{72} \\ & = \max\{r_1, r_2, \max\{r_3, r_4\}\}\label{73}
\end{align}
and
\begin{align}\label{my74}
    \max\{r_1, r_2, r_3\} = \max\{r_1, \max\{r_2, r_3\}\}.
\end{align}
To show that $Y_t^s = Y_{t+1}^s$ for every $t \in \mathbb{T} = \{0,1,\dots,N-1\}$, first let $t \in \{0,1,\dots,N-2\}$ be given. For every $\omega \in \Omega$, we have
\begin{align}
     Y_{\textcolor{blue}{t}}^s(\textcolor{magenta}{\omega}) & \overset{\eqref{42}}{=} h^s\left(\max\Big\{ c_N(X_N(\textcolor{magenta}{\omega})), \; \;\max_{i \in \{\textcolor{blue}{t},\dots,N-1\}} c_i(X_i(\textcolor{magenta}{\omega}),U_i(\textcolor{magenta}{\omega})), \; \; Z_{\textcolor{blue}{t}}(\textcolor{magenta}{\omega}) \Big\}\right) \\
     & \overset{\eqref{72}}{=} h^s\left(\max\Big\{ c_N(X_N(\textcolor{magenta}{\omega})), \; \; \max_{i \in \{\textcolor{blue}{t}+1,\dots,N-1\}} c_i(X_i(\textcolor{magenta}{\omega}),U_i(\textcolor{magenta}{\omega})), \; \; c_{\textcolor{blue}{t}}(X_{\textcolor{blue}{t}}(\textcolor{magenta}{\omega}),U_{\textcolor{blue}{t}}(\textcolor{magenta}{\omega})), \; \; Z_{\textcolor{blue}{t}}(\textcolor{magenta}{\omega}) \Big\}\right) \\
      & \overset{\eqref{73}}{=} h^s\left(\max\Big\{ c_N(X_N(\textcolor{magenta}{\omega})), \; \; \max_{i \in \{\textcolor{blue}{t}+1,\dots,N-1\}} c_i(X_i(\textcolor{magenta}{\omega}),U_i(\textcolor{magenta}{\omega})), \; \; \max\{c_{\textcolor{blue}{t}}(X_{\textcolor{blue}{t}}(\textcolor{magenta}{\omega}),U_{\textcolor{blue}{t}}(\textcolor{magenta}{\omega})), Z_{\textcolor{blue}{t}}(\textcolor{magenta}{\omega})\} \Big\}\right) \\
      & \overset{\eqref{myZt}}{=} h^s\left(\max\Big\{ c_N(X_N(\textcolor{magenta}{\omega})), \; \; \max_{i \in \{\textcolor{blue}{t}+1,\dots,N-1\}} c_i(X_i(\textcolor{magenta}{\omega}),U_i(\textcolor{magenta}{\omega})), \; \; Z_{\textcolor{blue}{t}+1}(\textcolor{magenta}{\omega}) \Big\}\right)\\
      & \overset{\eqref{42}}{=} Y_{\textcolor{blue}{t}+1}^s(\textcolor{magenta}{\omega}).
\end{align}
Now, let $t = N-1$. For every $\omega \in \Omega$, we have
\begin{align}
    Y_{\textcolor{blue}{N-1}}^s(\textcolor{magenta}{\omega}) & \overset{\eqref{42}}{=} h^s\left(\max\Big\{ c_N(X_N(\textcolor{magenta}{\omega})), \;\; c_{\textcolor{blue}{N-1}}(X_{\textcolor{blue}{N-1}}(\textcolor{magenta}{\omega}),U_{\textcolor{blue}{N-1}}(\textcolor{magenta}{\omega})), \; \; Z_{\textcolor{blue}{N-1}}(\textcolor{magenta}{\omega}) \Big\}\right)\\
    & \overset{\eqref{my74}}{=} h^s\left(\max\Big\{ c_N(X_N(\textcolor{magenta}{\omega})), \;\; \max \{ c_{\textcolor{blue}{N-1}}(X_{\textcolor{blue}{N-1}}(\textcolor{magenta}{\omega}),U_{\textcolor{blue}{N-1}}(\textcolor{magenta}{\omega})), \; \; Z_{\textcolor{blue}{N-1}}(\textcolor{magenta}{\omega}) \} \Big\}\right)\\
    & \overset{\eqref{myZt}}{=} h^s\left(\max\Big\{ c_N(X_N(\textcolor{magenta}{\omega})), \;\; Z_{N}(\textcolor{magenta}{\omega}) \Big\} \right)\\
    & \overset{\eqref{YN}}{=} Y_N^s(\textcolor{magenta}{\omega}).
\end{align}
Therefore, we conclude that $Y_t^s = Y_{t+1}^s$ for every $t \in \mathbb{T} = \{0,1,\dots,N-1\}$.
\end{proof}

In summary, for every $\mathbf{x} \in S$, $\pi \in \Pi$, $t \in \mathbb{T}_N$, and $s \in \mathbb{R}$, we can view $Y_t^s$ \eqref{42}--\eqref{YN} as an extended random variable on $(\Omega,\mathcal{B}_{\Omega},P_{\mathbf{x}}^\pi)$ because $Y_t^s : \Omega \rightarrow \mathbb{R}^*$ is measurable relative to $\mathcal{B}_{\Omega}$ and $\mathcal{B}_{\mathbb{R}^*}$, where $(\Omega,\mathcal{B}_{\Omega},P_{\mathbf{x}}^\pi)$ is a probability space. The property $Y_t^s = Y_{t+1}^s$ for every $t \in \mathbb{T}$ will facilitate the derivation of a dynamic programming algorithm.

\subsubsection{Change-of-Variable Image Measure Theorem}\label{imagemeasuresec}
We paraphrase a change-of-variable image measure theorem from \cite[Th. 1.6.12, p. 50]{ash1972}: Let $(\bar{\Omega}, \mathcal{F})$ and $(\bar{\Omega}_0, \mathcal{F}_0)$ be measurable spaces, and let $T : \bar{\Omega} \rightarrow \bar{\Omega}_0$ be measurable relative to $\mathcal{F}$ and $\mathcal{F}_0$. Suppose that $\mu$ is a measure on $\mathcal{F}$. Define a measure $\mu_0$ on $\mathcal{F}_0$ by
\begin{equation}\label{imagemeasure}
 \mu_0(A) \coloneqq \mu(T^{-1}(A)), \;\;\; A \in \mathcal{F}_0.
 \end{equation}
If $\bar{f} : \bar{\Omega}_0 \rightarrow \mathbb{R}^*$ is measurable relative to $\mathcal{F}_0$ and $\mathcal{B}_{\mathbb{R}^*}$ and $A \in \mathcal{F}_0$, then 
\begin{equation}\label{mymy73}
    \int_{T^{-1}(A)} \bar{f}(T(\omega)) \; \mathrm{d}\mu(\omega) = \int_A \bar{f}(\omega_0) \; \mathrm{d}\mu_0(\omega_0)
\end{equation}
in the sense that if one of the integrals exists, then the other integral exists also, and the two integrals are equal. Some textbooks, e.g., \cite[p. 92]{dudley2018real}, call $\mu_0 = \mu \circ T^{-1}$ \eqref{imagemeasure} the \emph{image measure} of $\mu$ by $T$.

\subsubsection{Relating Integrals with respect to $P_{\mathbf{x}}^\pi$ and $P_{\mathbf{x},\mathcal{X}_t}^\pi$} Let $\mathbf{x} \in S$ and $\pi \in \Pi$ be given. Recall the notation $\mathbb{S} \coloneqq S \times \mathcal{Z}$, where $\chi_t = (x_t,z_t)$ is an arbitrary element of $\mathbb{S}$. 
For every $t \in \mathbb{T}_N$, the probability measure on $(\mathbb{S},\mathcal{B}_{\mathbb{S}})$ induced by $\mathcal{X}_t \coloneqq (X_t,Z_t)$ encodes the process starting from time zero and ending where $\mathcal{X}_t$ may be realized. The symbol $P_{\mathbf{x},\mathcal{X}_t}^\pi$ denotes this induced measure. While $P_{\mathbf{x},\mathcal{X}_t}^\pi$ depends on $\pi$, $t$, $\mathbf{x}$, and $a$, we omit the symbol $a$ from the notation for brevity. Recall that $\mathcal{X}_t$ \eqref{augstatedef} is defined by
\begin{equation*}
    \mathcal{X}_t(\textcolor{magenta}{\omega}) \coloneqq (X_t,Z_t)(\textcolor{magenta}{\omega}) \coloneqq (X_t(\textcolor{magenta}{\omega}),Z_t(\textcolor{magenta}{\omega})), \quad \quad  t \in \mathbb{T}_N, \quad \textcolor{magenta}{\omega} \in \Omega.
\end{equation*}
For any $\underline{\mathbb{S}} \in \mathcal{B}_{\mathbb{S}}$, it holds that
\begin{equation}
  \mathcal{X}_t^{-1}(\underline{\mathbb{S}}) \coloneqq \{ \mathcal{X}_t \in \underline{\mathbb{S}} \} \coloneqq \{ \omega \in \Omega : \mathcal{X}_t(\omega) \in \underline{\mathbb{S}} \} \in \mathcal{B}_{\Omega}
\end{equation}
because $\mathcal{X}_t : \Omega \rightarrow \mathbb{S}$ is measurable relative to $\mathcal{B}_{\Omega}$ and $\mathcal{B}_{\mathbb{S}}$ (Sec. \ref{randomobj}). Thus, we can use the probability measure $P_{\mathbf{x}}^\pi \in \mathcal{P}(\Omega)$ \eqref{keyP} to evaluate the event $\mathcal{X}_t^{-1}(\underline{\mathbb{S}})$. This evaluation defines the induced probability measure $P_{\mathbf{x},\mathcal{X}_t}^\pi$:\footnote{Since we only know the form of $P_\mathbf{x}^\pi$ on measurable rectangles in $\mathcal{B}_{\Omega}$, we only know the form of $P_{\mathbf{x},\mathcal{X}_t}^\pi$ on measurable rectangles in $\mathcal{B}_{S \times \mathcal{Z}}$. We note that
\begin{equation}
    \underline{S} \in \mathcal{B}_S, \; \underline{Z} \in \mathcal{B}_{\mathcal{Z}} \implies \underline{S} \times \underline{Z} \in \mathcal{B}_{S \times \mathcal{Z}}.
\end{equation}
A set of form $\underline{S} \times \underline{Z}$ is called a \emph{measurable rectangle} in $\mathcal{B}_{S \times \mathcal{Z}}$. $\mathcal{B}_{S \times \mathcal{Z}}$ contains measurable rectangles and other forms of subsets of $S \times \mathcal{Z}$.}
\begin{equation}\label{mymy77}
    P_{\mathbf{x},\mathcal{X}_t}^\pi(\underline{\mathbb{S}}) \coloneqq P_\mathbf{x}^\pi(\mathcal{X}_t^{-1}(\underline{\mathbb{S}})), \quad \quad  \underline{\mathbb{S}} \in \mathcal{B}_{\mathbb{S}}.
\end{equation}
We recall from \eqref{P00} that $P_{\mathbf{x},\mathcal{X}_0}^\pi = \delta_{\mathbf{x},a}$. Note that
\begin{align}\label{mymy80}
  \mathcal{X}_t^{-1}(\mathbb{S}) = \{\mathcal{X}_t\in \mathbb{S}\} = \{ \textcolor{magenta}{\omega} \in \Omega : (X_t,Z_t)(\textcolor{magenta}{\omega}) \in \mathbb{S} \} = \Omega, \quad \quad t \in \mathbb{T}_N.
\end{align}

Let $t \in \mathbb{T}_N$ be given, and suppose that $\bar{f} : \mathbb{S} \rightarrow \mathbb{R}^*$ is measurable relative to  
$\mathcal{B}_{\mathbb{S}}$ and $\mathcal{B}_{\mathbb{R}^*}$. We note the following properties:
\begin{itemize}
    \item $(\Omega,\mathcal{B}_\Omega)$ and $(\mathbb{S},\mathcal{B}_{\mathbb{S}})$ are measurable spaces;
    \item $\mathcal{X}_t : \Omega \rightarrow \mathbb{S}$ is measurable relative to $\mathcal{B}_\Omega$ and $\mathcal{B}_{\mathbb{S}}$ (Sec. \ref{randomobj});
    \item $P_{\mathbf{x}}^\pi$ \eqref{keyP} is a probability measure on $(\Omega,\mathcal{B}_{\Omega})$;
    \item $P_{\mathbf{x},\mathcal{X}_t}^\pi = P_\mathbf{x}^\pi \circ \mathcal{X}_t^{-1}$ \eqref{mymy77} is a probability measure on $(\mathbb{S},\mathcal{B}_{\mathbb{S}})$.
\end{itemize}
Therefore, by the change-of-variable image measure theorem (Sec. \ref{imagemeasuresec}), we have
\begin{equation}\label{85my}
   \int_{\Omega} \bar{f}\circ \mathcal{X}_t \; \mathrm{d}P_{\mathbf{x}}^\pi \coloneqq \int_{\Omega} \bar{f}(\mathcal{X}_t(\textcolor{magenta}{\omega})) \; \mathrm{d}P_{\mathbf{x}}^\pi(\textcolor{magenta}{\omega}) \overset{\eqref{mymy80}}{=} \int_{\mathcal{X}_t^{-1}(\mathbb{S})} \bar{f}(\mathcal{X}_t(\textcolor{magenta}{\omega})) \; \mathrm{d}P_{\mathbf{x}}^\pi(\textcolor{magenta}{\omega}) \overset{\eqref{mymy73}}{=} \int_{\mathbb{S}} \bar{f} \; \mathrm{d}P_{\mathbf{x},\mathcal{X}_t}^\pi,
\end{equation}
in the sense that if one of the integrals exists, then the other integral exists as well, and the two integrals are equal. We would like to provide an explicit form for \eqref{85my}.

We consider the case for $t =0$ and the case for $t \in \{1,2,\dots,N\}$ separately. First, if $t =0$, then
\begin{subequations}\label{mymy91}
\begin{equation}\label{91b}
  \bar{f}(\mathbf{x},a) \overset{\eqref{diraceq}}{=} \int_{\mathbb{S}} \bar{f} \;\mathrm{d}\delta_{\mathbf{x},a}  \overset{\eqref{P00}}{=} \int_{\mathbb{S}} \bar{f} \;\mathrm{d}P_{\mathbf{x},\mathcal{X}_0}^\pi \overset{\eqref{85my}}{=} \int_{\Omega} \bar{f} \circ \mathcal{X}_0 \;\mathrm{d}P_{\mathbf{x}}^\pi.
\end{equation}
We are permitted to apply \eqref{diraceq} because $\bar{f} : \mathbb{S} \rightarrow \mathbb{R}^*$ is Borel-measurable and $\mathbb{S}$ (with the product topology) is a Borel space. In the second step, we use $\delta_{\mathbf{x},a} = P_{\mathbf{x},\mathcal{X}_0}^\pi$ \eqref{P00}. 

Now, let $t \in \{1,2,\dots,N\}$ be given, and suppose that $\int_{\mathbb{S}} \bar{f} \;\mathrm{d}P_{\mathbf{x},\mathcal{X}_t}^\pi$ or $\int_{\Omega} \bar{f} \circ \mathcal{X}_t \; \mathrm{d}P_{\mathbf{x}}^\pi$ exists.\footnote{If $\int_{\mathbb{S}} \bar{f} \;\mathrm{d}P_{\mathbf{x},\mathcal{X}_t}^\pi$ exists, then $\int_{\Omega} \bar{f} \circ \mathcal{X}_t \; \mathrm{d}P_{\mathbf{x}}^\pi$ exists and $\int_{\mathbb{S}} \bar{f} \;\mathrm{d}P_{\mathbf{x},\mathcal{X}_t}^\pi = \int_{\Omega} \bar{f} \circ \mathcal{X}_t \; \mathrm{d}P_{\mathbf{x}}^\pi$ by the statement below \eqref{85my}. Similarly, if $\int_{\Omega} \bar{f} \circ \mathcal{X}_t \; \mathrm{d}P_{\mathbf{x}}^\pi$ exists, then $\int_{\mathbb{S}} \bar{f} \;\mathrm{d}P_{\mathbf{x},\mathcal{X}_t}^\pi$ exists and $\int_{\mathbb{S}} \bar{f} \;\mathrm{d}P_{\mathbf{x},\mathcal{X}_t}^\pi = \int_{\Omega} \bar{f} \circ \mathcal{X}_t \; \mathrm{d}P_{\mathbf{x}}^\pi$ by the statement below \eqref{85my}.\label{footnote1212}} Then, 
we have
\begin{align}\label{154bimportant}
 \int_{\Omega} \bar{f} \circ \mathcal{X}_t \; \mathrm{d}P_{\mathbf{x}}^\pi 
& \overset{\eqref{85my}}{=} \int_{\mathbb{S}} \bar{f} \;\mathrm{d}P_{\mathbf{x},\mathcal{X}_t}^\pi \nonumber \\
& \overset{\hphantom{\eqref{85my}}}{=} \int_{(\mathbb{S} \times C)^{\textcolor{blue}{t}} \times \mathbb{S}} \bar{f}(\chi_{\textcolor{blue}{t}}) \; \tilde{q}_{\textcolor{blue}{t}-1}(\mathrm{d}\chi_{\textcolor{blue}{t}}|\chi_{\textcolor{blue}{t}-1},u_{\textcolor{blue}{t}-1})  \cdots \pi_0(\mathrm{d}u_0|\chi_0) \;  \delta_{\mathbf{x},a}(\mathrm{d}\chi_0). 
\end{align}
\end{subequations}
Next, we explain why the last line of \eqref{154bimportant} holds. The key idea is to use the change-of-variable image measure theorem (Sec. \ref{imagemeasuresec}) again but for a different reference measure (to be denoted by $r_{\mathbf{x},t}^\pi$). For convenience, we define
\begin{equation}\label{myomegat}
    \Omega_t \coloneqq (\mathbb{S} \times C)^{t} \times \mathbb{S}, \quad \quad t \in \{1,2,\dots,N\}.
\end{equation}
$\mathbb{S}, C, \dots, \mathbb{S}, C, \mathbb{S}$ is a finite sequence of Borel spaces, $\Omega_t$ \eqref{myomegat} is a Cartesian product of these spaces, $\delta_{\mathbf{x},a} \in \mathcal{P}(\mathbb{S})$ is given, $\tilde{q}_{j}$ is a continuous stochastic kernel under Assumption 1 (Lemma \ref{analysistildeq}), and $\pi_j$ is a Borel-measurable stochastic kernel. Hence, we apply \cite[Prop. 7.28, pp. 140--141]{bertsekas2004stochastic} to guarantee the existence of a unique probability measure $r_{\mathbf{x},t}^\pi \in \mathcal{P}(\Omega_t)$ (which depends on $a$ as well) such that
\begin{equation}\label{myrteq}
     r_{\mathbf{x},t}^\pi(\underline{\mathbb{S}}_0 \times \underline{C}_0 \times \cdots \times \underline{\mathbb{S}}_t) \\
     = \int_{\underline{\mathbb{S}}_0}\int_{\underline{C}_0} \cdots \int_{\underline{\mathbb{S}}_t} \tilde{q}_{t-1}(\mathrm{d}\chi_t|\chi_{t-1},u_{t-1})  \cdots \pi_0(\mathrm{d}u_0|\chi_0) \; \delta_{\mathbf{x},a}(\mathrm{d}\chi_0)
\end{equation}
for every $\underline{\mathbb{S}}_0 \in \mathcal{B}_{\mathbb{S}}, \underline{C}_0 \in \mathcal{B}_{C}, \dots, \underline{\mathbb{S}}_t \in \mathcal{B}_{\mathbb{S}}$, and if $\bar{g} : \Omega_t \rightarrow \mathbb{R}^*$ is Borel-measurable and $\int_{\Omega_t} \bar{g} \; \mathrm{d}r_{\mathbf{x},t}^\pi$ exists, then
\begin{equation}\label{gbarexists}\begin{aligned}
     \int_{\Omega_t} \bar{g}\; \mathrm{d}r_{\mathbf{x},t}^\pi 
      =  \int_{\mathbb{S}}\int_{C} \cdots  \int_{\mathbb{S}} \bar{g}(\chi_0, u_0,\dots, \chi_t) \; \tilde{q}_{t-1}(\mathrm{d}\chi_t|\chi_{t-1},u_{t-1}) \cdots \pi_0(\mathrm{d}u_0|\chi_0) \; \delta_{\mathbf{x},a}(\mathrm{d}\chi_0).
\end{aligned}\end{equation}
We define $H_t : \Omega_t \rightarrow \mathbb{S}$ by
\begin{equation}\label{myHt}
    H_t(\chi_0,u_0,\dots,\chi_t) \coloneqq \chi_t,
\end{equation}
and therefore, for every $\underline{\mathbb{S}} \in \mathcal{B}_{\mathbb{S}}$, it holds that
\begin{equation}\label{my159}\begin{aligned}
    H_t^{-1}(\underline{\mathbb{S}}) \coloneqq \{ \omega_t \in \Omega_t : H_t(\omega_t) \in \underline{\mathbb{S}} \}  & = \{ (\chi_0,u_0,\dots,\chi_t) \in \Omega_t : H_t(\chi_0,u_0,\dots,\chi_t) \in \underline{\mathbb{S}} \} \\ & = \{ (\chi_0,u_0,\dots,\chi_t) \in \Omega_t  : \chi_t \in \underline{\mathbb{S}} \} \\
    & = \{ (\chi_0,u_0,\dots,\chi_t) \in (\mathbb{S} \times C)^{t} \times \mathbb{S} : \chi_t \in \underline{\mathbb{S}} \}\\
    & = (\mathbb{S} \times C)^{t} \times \underline{\mathbb{S}} \in \mathcal{B}_{\Omega_t}.
\end{aligned}\end{equation}
We know that the set $(\mathbb{S} \times C)^{t} \times \underline{\mathbb{S}} \in \mathcal{B}_{\Omega_t}$ because this set is a measurable rectangle. We claim that
\begin{equation}
    P_{\mathbf{x},\mathcal{X}_t}^\pi(\underline{\mathbb{S}}) = r_{\mathbf{x},t}^\pi(H_t^{-1}(\underline{\mathbb{S}})), \quad \quad \underline{\mathbb{S}} \in \mathcal{B}_{\mathbb{S}}.
\end{equation}
Indeed, for every $\underline{\mathbb{S}}  \in \mathcal{B}_{\mathbb{S}}$, we have\footnote{We use the product measure $\tilde{q}_{j}(\cdot|x_j,z_j,u_{j})$ instead of the two measures $q_{j}(\cdot|x_j,u_j)$ and $\overline{q}_{j}(\cdot|x_j,z_j,u_j)$ separately. If we did not use the product measure, then the expression $\int_{\underline{\mathbb{S}}}\overline{q}_{t-1}(\mathrm{d}z_t|x_{t-1},z_{t-1},u_{t-1}) \; q_{t-1}(\mathrm{d}x_t|x_{t-1},u_{t-1})$ would arise in \eqref{my158}. This expression does not quite make sense because $\underline{\mathbb{S}} \in \mathcal{B}_{S \times \mathcal{Z}}$ need \emph{not} take the form $\underline{S} \times \underline{\mathcal{Z}}$ with $\underline{S} \in \mathcal{B}_S$ and $\underline{\mathcal{Z}} \in \mathcal{B}_{\mathcal{Z}}$.}
\begin{equation}\label{my158}\begin{aligned}
  P_{\mathbf{x},\mathcal{X}_t}^\pi(\underline{\mathbb{S}}) & =  P_\mathbf{x}^\pi(\{ \omega \in \Omega : \mathcal{X}_t(\omega) \in \underline{\mathbb{S}} \})\\
    & = P_\mathbf{x}^\pi(\underbrace{\mathbb{S} \times C}_{\text{stage 0}} \times \cdots \times \underbrace{\mathbb{S} \times C}_{\text{stage $t-1$}} \times \underbrace{\underline{\mathbb{S}} \times C}_{\text{stage $t$}} \times \cdots \times \underbrace{\mathbb{S}}_{\text{stage $N$}})\\
    & = \textstyle \int_{\mathbb{S}} \int_{C} \cdots \int_{\mathbb{S}} \int_{C} \int_{\underline{\mathbb{S}}}  \tilde{q}_{t-1}(\mathrm{d}\chi_t|\chi_{t-1},u_{t-1}) \; \pi_{t-1}(\mathrm{d}u_{t-1}|\chi_{t-1}) \; \tilde{q}_{t-2}(\mathrm{d}\chi_{t-1}|\chi_{t-2},u_{t-2}) \cdots \pi_0(\mathrm{d}u_0|\chi_0) \; \delta_{\mathbf{x},a}(\mathrm{d}\chi_0)\\
    & \overset{\eqref{myrteq}}{=} r_{\mathbf{x},t}^\pi((\mathbb{S} \times C)^t \times \underline{\mathbb{S}})\\
    & \overset{\eqref{my159}}{=} r_{\mathbf{x},t}^\pi(H_t^{-1}(\underline{\mathbb{S}})).
\end{aligned}\end{equation}
The third step of \eqref{my158} holds because we use \eqref{keyP} and the innermost integrals evaluate to one. Now, we apply the change-of-variable image measure theorem \cite[Th. 1.6.12, p. 50]{ash1972}. $H_t : \Omega_t \rightarrow \mathbb{S}$ is defined by \eqref{myHt}. It is Borel-measurable, and hence, we write $H_t : (\Omega_t, \mathcal{B}_{\Omega_t}) \rightarrow (\mathbb{S},\mathcal{B}_{\mathbb{S}})$. By our previous discussion, we have $r_{\mathbf{x},t}^\pi \in \mathcal{P}(\Omega_t)$ and $P_{\mathbf{x},\mathcal{X}_t}^\pi(\underline{\mathbb{S}}) = r_{\mathbf{x},t}^\pi(H_t^{-1}(\underline{\mathbb{S}}))$ for every $\underline{\mathbb{S}} \in \mathcal{B}_{\mathbb{S}}$. By  \cite[Th. 1.6.12, p. 50]{ash1972}, if $\varphi : (\mathbb{S},\mathcal{B}_{\mathbb{S}}) \rightarrow (\mathbb{R}^*,\mathcal{B}_{\mathbb{R}^*})$ and $\underline{\mathbb{S}} \in \mathcal{B}_{\mathbb{S}}$, then
\begin{equation}\label{my1622}
    \int_{H_t^{-1}(\underline{\mathbb{S}})} \varphi(H_t(\omega_t)) \; \mathrm{d}r_{\mathbf{x},t}^\pi(\omega_t) = \int_{\underline{\mathbb{S}}} \varphi(\chi_t) \; \mathrm{d}P_{\mathbf{x},\mathcal{X}_t}^\pi(\chi_t)
\end{equation}
in the sense that if one of the integrals exist, then the other does as well, and the two integrals are equal. Now, consider $\underline{\mathbb{S}} = \mathbb{S}$ and $\varphi = \bar{f}$, and recall our assumption that $\int_{\mathbb{S}} \bar{f} \;\mathrm{d}P_{\mathbf{x},\mathcal{X}_t}^\pi$ or $\int_{\Omega} \bar{f} \circ \mathcal{X}_t \; \mathrm{d}P_{\mathbf{x}}^\pi$ exists. %
\begin{itemize}
\item $\int_{\Omega} \bar{f} \circ \mathcal{X}_t \; \mathrm{d}P_{\mathbf{x}}^\pi$ exists $\implies$ $\int_{\mathbb{S}} \bar{f} \;\mathrm{d}P_{\mathbf{x},\mathcal{X}_t}^\pi$ exists and the two integrals are equal (Footnote \ref{footnote1212}).
    \item $\underline{\mathbb{S}} = \mathbb{S}$ and $\varphi = \bar{f}$ $\implies$ $\int_{\underline{\mathbb{S}}} \varphi \; \mathrm{d}P_{\mathbf{x},\mathcal{X}_t}^\pi = \int_{\mathbb{S}} \bar{f} \;\mathrm{d}P_{\mathbf{x},\mathcal{X}_t}^\pi$.
    \item $\int_{\underline{\mathbb{S}}} \varphi \; \mathrm{d}P_{\mathbf{x},\mathcal{X}_t}^\pi = \int_{\mathbb{S}} \bar{f} \;\mathrm{d}P_{\mathbf{x},\mathcal{X}_t}^\pi$ and $\int_{\mathbb{S}} \bar{f} \;\mathrm{d}P_{\mathbf{x},\mathcal{X}_t}^\pi$ exists $\implies$ $\int_{\underline{\mathbb{S}}} \varphi \; \mathrm{d}P_{\mathbf{x},\mathcal{X}_t}^\pi$ exists.
\end{itemize}
Therefore, we have
\begin{equation}\label{my163}
    \int_{\mathbb{S}} \bar{f} \;\mathrm{d}P_{\mathbf{x},\mathcal{X}_t}^\pi \overset{\eqref{my1622}}{=} \int_{\textcolor{blue}{H_t^{-1}(\mathbb{S})}} \bar{f}(H_t(\omega_t)) \; \mathrm{d}r_{\mathbf{x},t}^\pi(\omega_t) = \int_{\textcolor{blue}{\Omega_t}} \bar{f}(H_t(\omega_t)) \; \mathrm{d}r_{\mathbf{x},t}^\pi(\omega_t) = \int_{\Omega_t} \bar{f} \circ H_t \; \mathrm{d}r_{\mathbf{x},t}^\pi,
\end{equation}
and the integrals exist. Since $\bar{f} \circ H_t : \Omega_t \rightarrow \mathbb{R}^*$ is Borel-measurable and $\int_{\Omega_t} \bar{f} \circ H_t \; \mathrm{d}r_{\mathbf{x},t}^\pi$ \eqref{my163} exists, we have
\begin{equation}\label{my164}\begin{aligned}
    \int_{\Omega_t} \bar{f} \circ H_t \; \mathrm{d}r_{\mathbf{x},t}^\pi & \overset{\eqref{gbarexists}}{=}\int_{\mathbb{S}}\int_{C} \cdots  \int_{\mathbb{S}} \bar{f}(\textcolor{blue}{H_t(\chi_0, u_0,\dots, \chi_t)}) \; \tilde{q}_{t-1}(\mathrm{d}\chi_t|\chi_{t-1},u_{t-1}) \cdots \pi_0(\mathrm{d}u_0|\chi_0) \; \delta_{\mathbf{x},a}(\mathrm{d}\chi_0)\\
    & \overset{\eqref{myHt}}{=}\int_{\mathbb{S}}\int_{C} \cdots  \int_{\mathbb{S}} \bar{f}(\textcolor{blue}{\chi_t}) \; \tilde{q}_{t-1}(\mathrm{d}\chi_t|\chi_{t-1},u_{t-1}) \cdots \pi_0(\mathrm{d}u_0|\chi_0) \; \delta_{\mathbf{x},a}(\mathrm{d}\chi_0).
\end{aligned}\end{equation}
Finally, we have
\begin{equation}\begin{aligned}
    \int_{\mathbb{S}} \bar{f} \;\mathrm{d}P_{\mathbf{x},\mathcal{X}_t}^\pi & \overset{\eqref{my163}}{=} \int_{\Omega_t} \bar{f} \circ H_t \; \mathrm{d}r_{\mathbf{x},t}^\pi \\
    & \overset{\eqref{my164}}{=}\int_{\mathbb{S}}\int_{C} \cdots  \int_{\mathbb{S}} \bar{f}(\chi_t) \; \tilde{q}_{t-1}(\mathrm{d}\chi_t|\chi_{t-1},u_{t-1}) \cdots \pi_0(\mathrm{d}u_0|\chi_0) \; \delta_{\mathbf{x},a}(\mathrm{d}\chi_0),
\end{aligned}\end{equation}
completing the proof of the last line of \eqref{154bimportant} under the assumption that $\int_{\mathbb{S}} \bar{f} \;\mathrm{d}P_{\mathbf{x},\mathcal{X}_t}^\pi$ or $\int_{\Omega} \bar{f} \circ \mathcal{X}_t \; \mathrm{d}P_{\mathbf{x}}^\pi$ exists.

\subsection{An Extended Proof for Theorem 1}
For every $t \in \mathbb{T}_N$, we denote a conditional expectation of $Y_t^s$ given $\mathcal{X}_t$ by $\phi_t^{\pi,s} : \mathbb{S} \rightarrow \mathbb{R}^*$ such that
\begin{equation}
    \phi_t^{\pi,s}(x,z) = E^\pi(Y_t^s|\mathcal{X}_t = (x,z)),
\end{equation}
which is unique almost everywhere with respect to $P_{\mathbf{x},\mathcal{X}_t}^\pi$. Next, we study $\phi_t^{\pi,s}$ in Theorem 1.

\emph{Theorem 1 (Properties of $\phi_t^{\pi,s}$):} Let $\mathbf{x} \in S$, $\pi \in \Pi$, and $s \in \mathbb{R}$ be given, and let Assumption 1 hold. Define the function $J_N^s : \mathbb{S} \rightarrow \mathbb{R}^*$ by\footnote{$h^s : \mathbb{R} \rightarrow \mathbb{R}$ and $J_N^s$ is defined by \eqref{defJNs} $\implies$ for every $(x,z) \in \mathbb{S}$, $J_N^s(x, z) \in \mathbb{R}$ $\implies$ for every $(x,z) \in \mathbb{S}$, $J_N^s(x, z) \in \mathbb{R}^*$. Hence, we can view $J_N^s$ as a function from $\mathbb{S}$ to $\mathbb{R}^*$.} 
\begin{equation}\label{defJNs}
    J_N^s(x,z) \coloneqq h^s( \max\{ c_N(x), z \}).
\end{equation}
Then, the following relations hold:
\begin{align}
    E_{\mathbf{x}}^\pi( \max\{ Y - s, 0 \} ) & = \int_{\Omega}  \phi_0^{\pi,s} \circ \mathcal{X}_0\; \mathrm{d}P_{\mathbf{x}}^\pi = \phi_0^{\pi,s}(\mathbf{x},a),
     \label{46}\\
    \int_{\Omega} \phi_N^{\pi,s} \circ \mathcal{X}_N \; \mathrm{d}P_{\mathbf{x}}^\pi & = \int_{\Omega} J_N^s \circ \mathcal{X}_N \; \mathrm{d}P_{\mathbf{x}}^\pi,  \label{47} \\
   \int_{\Omega} \phi_t^{\pi,s} \circ \mathcal{X}_t\; \mathrm{d}P_{\mathbf{x}}^\pi & = \int_{\Omega}  \phi_{t+1}^{\pi,s} \circ \mathcal{X}_{t+1}\; \mathrm{d}P_{\mathbf{x}}^\pi, \quad \quad t \in \mathbb{T}. \label{66}
\end{align}
%

\vspace{3mm}\hspace{-8mm}\begin{proof}
%
We note the following facts. For every $t \in \mathbb{T}_N$,
\begin{itemize}
    \item $Y_t^s$ is an extended random variable on $(\Omega,\mathcal{B}_{\Omega},P_{\mathbf{x}}^\pi)$ (Sec. \ref{analysisYbarts});
    \item $\mathcal{X}_t : \Omega \rightarrow \mathbb{S}$ is a random object, as it is measurable relative to $\mathcal{B}_{\Omega}$ and $\mathcal{B}_{\mathbb{S}}$ (Sec. \ref{randomobj});\footnote{The following statements are equivalent: $\mathcal{X}_t : \Omega \rightarrow \mathbb{S}$ is measurable relative to $\mathcal{B}_{\Omega}$ and $\mathcal{B}_{\mathbb{S}}$; $\mathcal{X}_t : \Omega \rightarrow \mathbb{S}$ is Borel-measurable; and $\mathcal{X}_t : (\Omega,\mathcal{B}_{\Omega}) \rightarrow (\mathbb{S},\mathcal{B}_{\mathbb{S}})$.}
    \item $E_{\mathbf{x}}^\pi(Y_t^s) = \int_\Omega Y_t^s \; \mathrm{d}P_{\mathbf{x}}^\pi$ exists (it does not take the form $+\infty - \infty$) because $Y_t^s(\omega) \geq 0$ for every $\omega \in \Omega$.
\end{itemize}  
Therefore, by \cite[Th. 6.3.3, p. 245]{ash1972}, there is a function 
$\phi_t^{\pi,s} : \mathbb{S} \rightarrow \mathbb{R}^*$, measurable relative to $\mathcal{B}_{\mathbb{S}}$ and $\mathcal{B}_{\mathbb{R}^*}$, such that for every $\underline{\mathbb{S}} \in \mathcal{B}_{\mathbb{S}}$,
\begin{subequations}\label{114}
\begin{equation}\label{114a}
    \int_{\{\mathcal{X}_t\in \underline{\mathbb{S}}\}} Y_t^s \; \mathrm{d}P_{\mathbf{x}}^\pi = \int_{\underline{\mathbb{S}}}  \phi_t^{\pi,s}\; \mathrm{d}P_{\mathbf{x},\mathcal{X}_t}^\pi.
\end{equation}
We define
\begin{equation}\label{114b}
   E^\pi(Y_t^s|\mathcal{X}_t = (x,z)) \coloneqq \phi_t^{\pi,s}(x,z),
\end{equation}
\end{subequations}
which is unique almost everywhere with respect to $P_{\mathbf{x},\mathcal{X}_t}^\pi$ \cite[Th. 6.3.3]{ash1972}. This result holds as a consequence of the Radon-Nikodym Theorem.

First, we write \eqref{114} in a particularly useful form. Let $t \in \mathbb{T}_N$ be given. Consider $\underline{\mathbb{S}} = \mathbb{S}$ in \eqref{114} to find that
\begin{align}\label{my97}
   \textcolor{blue}{\int_{\mathbb{S}}  \phi_t^{\pi,s}\; \mathrm{d}P_{\mathbf{x},\mathcal{X}_t}^\pi} \overset{\eqref{114}}{=} \int_{\{\mathcal{X}_t\in \mathbb{S}\}} Y_t^s \; \mathrm{d}P_{\mathbf{x}}^\pi \overset{\eqref{mymy80}}{=}  \textcolor{blue}{\int_{\Omega} Y_t^s \; \mathrm{d}P_{\mathbf{x}}^\pi}.
\end{align}
Since $\int_{\Omega} Y_t^s \; \mathrm{d}P_{\mathbf{x}}^\pi$ exists, it follows that $\int_{\mathbb{S}}  \phi_t^{\pi,s}\; \mathrm{d}P_{\mathbf{x},\mathcal{X}_t}^\pi$ \eqref{my97} exists. Since $\phi_t^{\pi,s} : \mathbb{S} \rightarrow \mathbb{R}^*$ is measurable relative to $\mathcal{B}_{\mathbb{S}}$ and $\mathcal{B}_{\mathbb{R}^*}$ and $\int_{\mathbb{S}}  \phi_t^{\pi,s}\; \mathrm{d}P_{\mathbf{x},\mathcal{X}_t}^\pi$ exists, we apply the change-of-variable image measure theorem to find that
\begin{equation}\label{my104}
   \int_{\Omega} \phi_t^{\pi,s}\circ \mathcal{X}_t \; \mathrm{d}P_{\mathbf{x}}^\pi  \overset{\eqref{85my}}{=} \textcolor{blue}{\int_{\mathbb{S}} \phi_t^{\pi,s} \; \mathrm{d}P_{\mathbf{x},\mathcal{X}_t}^\pi},
\end{equation}
where the integrals exist. By combining the previous two expressions \eqref{my97}--\eqref{my104}, we have
\begin{equation}\label{mymy105}
    \int_{\Omega} \phi_t^{\pi,s}\circ \mathcal{X}_t \; \mathrm{d}P_{\mathbf{x}}^\pi = \textcolor{blue}{\int_{\Omega} Y_t^s \; \mathrm{d}P_{\mathbf{x}}^\pi}.
\end{equation}
We note that \eqref{mymy105} holds for every $t \in \mathbb{T}_N$ because it has been derived for an arbitrary time index $t \in \mathbb{T}_N$.

Next, we show \eqref{66}, i.e., 
\begin{align*}
 \int_{\Omega} \phi_t^{\pi,s} \circ \mathcal{X}_t\; \mathrm{d}P_{\mathbf{x}}^\pi & = \int_{\Omega}  \phi_{t+1}^{\pi,s} \circ \mathcal{X}_{t+1}\; \mathrm{d}P_{\mathbf{x}}^\pi, \quad \quad  t \in \mathbb{T}.
\end{align*}
Let $t \in \mathbb{T}$ be given. From Lemma \ref{lemma3pt3}, we have $Y_t^s = Y_{t+1}^s$, and therefore,
\begin{equation}\label{fromlemma3pt3}
    \int_{\Omega} Y_t^s \; \mathrm{d}P_{\mathbf{x}}^\pi = \int_{\Omega} Y_{t+1}^s \; \mathrm{d}P_{\mathbf{x}}^\pi,
\end{equation}
where the integrals exist because $Y_j^s$ is nonnegative and Borel-measurable for every $j$. Since $t \in \mathbb{T} = \{0,1,\dots,N-1\}$, it holds that $t+1 \in \mathbb{T}_N = \{0,1,\dots,N\}$. Since \eqref{mymy105} applies to any time index in $\mathbb{T}_N$, we have
\begin{align}\label{my98}
 \int_{\Omega} \phi_{t+1}^{\pi,s}\circ \mathcal{X}_{t+1} \; \mathrm{d}P_{\mathbf{x}}^\pi   \overset{\eqref{mymy105}}{=}  \int_{\Omega} Y_{t+1}^s \; \mathrm{d}P_{\mathbf{x}}^\pi.
\end{align}
We show \eqref{66} by combining prior steps:
\begin{equation}
  \int_{\Omega} \phi_t^{\pi,s}\circ \mathcal{X}_t \; \mathrm{d}P_{\mathbf{x}}^\pi \overset{\eqref{mymy105}}{=}  \int_{\Omega} Y_t^s \; \mathrm{d}P_{\mathbf{x}}^\pi \overset{\eqref{fromlemma3pt3}}{=} \int_{\Omega} Y_{t+1}^s \; \mathrm{d}P_{\mathbf{x}}^\pi \overset{\eqref{my98}}{=} \int_{\Omega} \phi_{t+1}^{\pi,s}\circ \mathcal{X}_{t+1} \; \mathrm{d}P_{\mathbf{x}}^\pi,
\end{equation}
noting that $t \in \mathbb{T}$ is arbitrary.
%

To show \eqref{47}, note that the function $J_N^s \circ \mathcal{X}_N : \Omega \rightarrow \mathbb{R}^*$ is given by
\begin{equation}\label{my109}
    J_N^s(\mathcal{X}_N(\textcolor{magenta}{\omega} )) = J_N^s(X_N(\textcolor{magenta}{\omega}),Z_N(\textcolor{magenta}{\omega})) = h^s(\max\{ c_N(X_N(\textcolor{magenta}{\omega})), Z_N(\textcolor{magenta}{\omega}) \}) = Y_N^s(\textcolor{magenta}{\omega}),
\end{equation}
by applying the definitions for $\mathcal{X}_N$ \eqref{augstatedef}, $J_N^s$ \eqref{defJNs}, and $Y_N^s$ \eqref{YN}. Also, by considering $t =N$ in \eqref{mymy105}, we have
\begin{align}
     \int_{\Omega} \phi_N^{\pi,s}\circ \mathcal{X}_N \; \mathrm{d}P_{\mathbf{x}}^\pi \overset{\eqref{mymy105}}{=} \int_{\Omega} Y_N^s \; \mathrm{d}P_{\mathbf{x}}^\pi \overset{\eqref{my109}}{=} \int_{\Omega} J_N^s \circ \mathcal{X}_N \; \mathrm{d}P_{\mathbf{x}}^\pi,
\end{align}
which shows \eqref{47}.

To show that $E_{\mathbf{x}}^\pi( \max\{ Y - s, 0 \} ) = \int_{\Omega}  \phi_0^{\pi,s} \circ \mathcal{X}_0\; \mathrm{d}P_{\mathbf{x}}^\pi = \phi_0^{\pi,s}(\mathbf{x},a)$ \eqref{46} holds, first we recall from Lemma \ref{lemma3pt2} that
\begin{equation}\label{92}
     E_{\mathbf{x}}^\pi(\max\{Y-s,0\}) =  E_{\mathbf{x}}^\pi(Y_0^s).
\end{equation}
By considering $t = 0$ in \eqref{mymy105}, we have
\begin{equation}\label{my113}
    \int_{\Omega} \phi_0^{\pi,s}\circ \mathcal{X}_0 \; \mathrm{d}P_{\mathbf{x}}^\pi \overset{\eqref{mymy105}}{=} \int_{\Omega} Y_0^s \; \mathrm{d}P_{\mathbf{x}}^\pi = E_{\mathbf{x}}^\pi(Y_0^s).
\end{equation}
Note that $\phi_0^{\pi,s}: \mathbb{S} \rightarrow \mathbb{R}^*$ is measurable relative to $\mathcal{B}_{\mathbb{S}}$ and $\mathcal{B}_{\mathbb{R}^*}$, and the following equalities hold:
\begin{equation}\label{my114}
    E_{\mathbf{x}}^\pi(Y_0^s) \overset{\eqref{my113}}{=} \int_{\Omega} \phi_0^{\pi,s}\circ \mathcal{X}_0 \; \mathrm{d}P_{\mathbf{x}}^\pi \overset{\eqref{my104}}{=} \textcolor{blue}{\int_{\mathbb{S}} \phi_0^{\pi,s} \; \mathrm{d}P_{\mathbf{x},\mathcal{X}_0}^\pi}.
\end{equation}
We apply \eqref{91b} to find that
\begin{equation}\label{mymy115}
   \textcolor{blue}{ \int_{\mathbb{S}} \phi_0^{\pi,s} \; \mathrm{d}P_{\mathbf{x},\mathcal{X}_0}^\pi} \overset{\eqref{91b}}{=} \phi_0^{\pi,s}(\mathbf{x},a).
\end{equation}
By \eqref{92} and \eqref{my114}--\eqref{mymy115}, we conclude that
\begin{equation}
  E_{\mathbf{x}}^\pi(\max\{Y-s,0\})  \overset{\eqref{92}}{=}  E_{\mathbf{x}}^\pi(Y_0^s) = \int_{\Omega} \phi_0^{\pi,s}\circ \mathcal{X}_0 \; \mathrm{d}P_{\mathbf{x}}^\pi = \phi_0^{\pi,s}(\mathbf{x},a),
\end{equation}
which shows \eqref{46}.
\end{proof}

The next result is useful for Theorem 2.
\subsection{Analysis of Lower Semi-continuous Bounded Below Functions}
Variations of the lemma in this section can be found in the literature, e.g., see \cite[Lemma 7.14 (a), p. 147]{bertsekas2004stochastic} and \cite[Th. A6.6, pp. 390--391]{ash1972}.\footnote{Another example is \cite[Prop. D.5, pp. 182--183]{hernandez2012discrete}.} Our proof combines techniques from these textbooks. We present the technical details of the proof in one place for convenience. The notation $\mathcal{C}(\mathcal{M})$ denotes the Banach space of bounded, real-valued, and continuous functions on $\mathcal{M}$, where $\mathcal{M}$ is a metrizable space.
\begin{lemma}\label{oldlemma3}
Let $\mathcal{M}$ be a metrizable space. Suppose that $J : \mathcal{M} \rightarrow \mathbb{R}^*$ is lower semi-continuous (l.s.c.) and bounded below by zero. Then, there is a sequence $\{ J_m : m \in \mathbb{N} \}$ in $\mathcal{C}(\mathcal{M})$ such that $0 \leq J_m \uparrow J$, i.e.,
\begin{enumerate}
    \item $0 \leq J_m(x) \leq J_{m+1}(x)  \leq J(x)$ for every $x \in \mathcal{M}$ and $m \in \mathbb{N}$, and
    \item $\underset{m \rightarrow +\infty}{\lim} J_m(x) = J(x)$ for every $x \in \mathcal{M}$.
\end{enumerate}
\end{lemma}

\begin{remark}[Generalization of Lemma \ref{oldlemma3}]
Before proving the lemma, we note a generalization. Let $v :\mathcal{M} \rightarrow \mathbb{R}^*$ be l.s.c. and bounded below by $\underline{b} \in \mathbb{R}$. We would like to show that there is a sequence $\{ v_m : m \in \mathbb{N} \}$ in $\mathcal{C}(\mathcal{M})$ such that $\underline{b} \leq v_m \uparrow v$. Define $J \coloneqq v - \underline{b}$, which is l.s.c. and bounded below by 0. By Lemma \ref{oldlemma3}, there is a sequence $\{ J_m : m \in \mathbb{N} \}$ in $\mathcal{C}(\mathcal{M})$ such that $0 \leq J_m \uparrow J$. Now, define $v_m \coloneqq J_m + \underline{b}$. Then, $\{ v_m : m \in \mathbb{N} \}$ is a sequence in $\mathcal{C}(\mathcal{M})$ such that $\underline{b} \leq (J_m + \underline{b}) \uparrow (J + \underline{b})$, which is equivalent to $\underline{b} \leq v_m \uparrow v$.
\end{remark}


A proof for Lemma \ref{oldlemma3} follows.

\hspace{-8mm}\begin{proof}
Let $\rho$ be a metric on $\mathcal{M}$. We recall that $J : \mathcal{M} \rightarrow \mathbb{R}^*$ is l.s.c. $\iff$ for any sequence $\{ x_n : n \in \mathbb{N} \}$ in $\mathcal{M}$ converging to $x \in \mathcal{M}$,\footnote{$\rho(x_n,x) \rightarrow 0$ as $n \rightarrow +\infty$.} it holds that $\underset{n \rightarrow +\infty}{\liminf} J(x_n) \geq J(x)$.

There are two cases to consider. The first case is that $J(x) = +\infty$ for every $x \in \mathcal{M}$.\footnote{We know that $J(x) > -\infty$ for every $x \in \mathcal{M}$ because $J$ is bounded below.} In this case, we define $J_m : \mathcal{M} \rightarrow \mathbb{R}$ by
\begin{equation}
    J_m(x) = m, \quad \quad  x \in \mathcal{M}, \quad   m \in \mathbb{N},
\end{equation}
which implies that
\begin{equation}
    0 \leq J_m(x) \leq J_{m+1}(x) \leq J(x), \quad \quad   x \in \mathcal{M}, \quad   m \in \mathbb{N},
\end{equation}
because $0\leq m \leq m+1 < +\infty$ for every $m \in \mathbb{N}$.
For every $m \in \mathbb{N}$, $J_m$ is constant and finite, and therefore $J_m$ is continuous and bounded, i.e., $J_m \in \mathcal{C}(\mathcal{M})$. Finally, 
\begin{equation}
    \lim_{m \rightarrow +\infty} J_m(x) = \lim_{m \rightarrow +\infty} m = +\infty = J(x), \quad \quad  x \in \mathcal{M},
\end{equation}
which completes the proof in the first case.

Now, in the second case, there exists an $x_0 \in \mathcal{M}$ such that $J(x_0) < +\infty$. Define
\begin{equation}
    g_m(x) \coloneqq \inf_{y \in \mathcal{M}} \left( J(y) + m \rho(x,y) \right), \quad \quad  x \in \mathcal{M}, \quad  m \in \mathbb{N}.
\end{equation}
Since $m\rho(x,y) \geq 0$ for every $(x,y) \in \mathcal{M} \times \mathcal{M}$ and $m \in \mathbb{N}$, and since $J(y) \geq 0$ for every $y \in \mathcal{M}$, we have
\begin{equation}
 0 \leq  J(y)  \leq J(y) + m \rho(x,y), \quad \quad   y \in \mathcal{M}, \quad x \in \mathcal{M}, \quad  m \in \mathbb{N}.
\end{equation}
Thus, zero is a lower bound for the set $\{ J(y) + m \rho(x,y) : y \in \mathcal{M}\}$ for every $x \in \mathcal{M}$ and $m \in \mathbb{N}$, which implies
\begin{equation}\label{my1888}
    0\leq  \underbrace{\inf \{ J(y) + m \rho(x,y) : y \in \mathcal{M}\}}_{g_m(x)}, \quad \quad   x \in \mathcal{M},  \quad m \in \mathbb{N}.
\end{equation}
Since $x_0 \in \mathcal{M}$, $J(x_0) < +\infty$, and metrics are real-valued, we have
\begin{equation}\label{my1899}
   0 \overset{\eqref{my1888}}{\leq} \underbrace{\inf \{ J(y) + m \rho(x,y) : y \in \mathcal{M}\}}_{g_m(x)} \leq J(x_0) + m \rho(x,x_0) < +\infty, \quad \quad  x \in \mathcal{M}, \quad  m \in \mathbb{N}.
\end{equation}
Thus, $g_m(x) \in \mathbb{R}$ for every $x \in \mathcal{M}$ and $m \in \mathbb{N}$. (In \eqref{my1899}, for example, we have written ``$x \in \mathcal{M}, m \in \mathbb{N}$,'' which means for every $x \in \mathcal{M}$ and for every $m \in \mathbb{N}$. In the rest of the proof, we use the symbol $\forall$.)

To show that $g_m \leq g_{m+1}$ for every $m \in \mathbb{N}$, note that since $\rho(x,y) \geq 0$ for every $(x,y) \in \mathcal{M} \times \mathcal{M}$ and $0 \leq m \leq m + 1$ for every $m \in \mathbb{N}$, we have 
\begin{equation}
 J(y) + m \rho(x,y) \leq J(y) + (m+1) \rho(x,y) \; \; \; \;\;\; \; \;\;\;\forall y \in \mathcal{M}\; \; \forall x \in \mathcal{M}\; \; \forall m \in \mathbb{N}.
\end{equation}
By taking infima over $y \in \mathcal{M}$, we obtain
\begin{equation}\label{my186}
    \underbrace{\inf_{y \in \mathcal{M}} (J(y) + m \rho(x,y))}_{g_m(x)} \leq \underbrace{\inf_{y \in \mathcal{M}} (J(y) + (m+1) \rho(x,y))}_{g_{m+1}(x)}\; \; \; \;\;\; \; \;\;\;\forall x \in \mathcal{M} \; \; \forall m \in \mathbb{N}.
\end{equation}
To show that $g_m \leq J$, note that
\begin{equation}\label{my187}
    \underbrace{\inf_{y \in \mathcal{M}} \; (J(y) + m \rho(x,y))}_{g_m(x)} \leq J(x) + m \rho(x,x) = J(x) \; \;\;\; \; \; \; \;\;\;\forall x \in \mathcal{M}\; \; \forall m \in \mathbb{N},
\end{equation}
which we obtained by setting $y = x$ in the objective of $g_m(x)$.

In summary, by \eqref{my1888}, \eqref{my186}, and \eqref{my187}, it holds that
\begin{equation}\label{my188}
    0 \leq g_m(x) \leq g_{m+1}(x) \leq J(x) \; \; \; \;\; \; \; \;\;\;\forall x \in \mathcal{M}\; \; \forall m \in \mathbb{N},
\end{equation}
where $g_m$ is finite for every $m \in \mathbb{N}$ by \eqref{my1899}.

For any $x \in \mathcal{M}$, $\{g_m(x)\}_{m = 1}^\infty \subseteq \mathbb{R}$ is an increasing sequence that is bounded above by $J(x) \in \mathbb{R}^*$. Thus, the limit of $\{g_m(x)\}_{m = 1}^\infty$ exists in $\mathbb{R}^*$ (it may be $+\infty$), and the limit is less than or equal to $J(x)$. Therefore,
\begin{equation}\label{my194}
    \lim_{m \rightarrow +\infty} g_m(x) \leq J(x) \; \; \;\;\; \; \; \;\;\;\forall x \in \mathcal{M}.
\end{equation}

For any $m \in \mathbb{N}$, to show that $g_m$ is (uniformly) continuous (with respect to $\rho$), we will show that
$$\forall \epsilon > 0 \; \exists \delta > 0 \text{ s.t. } \forall (x,z) \in \mathcal{M} \times \mathcal{M}, \; \rho(x,z) \leq \delta \implies |g_m(x) - g_m(z)| \leq \epsilon.$$
Using the procedure on p. 126 of \cite{bertsekas2004stochastic} (a symmetry argument using the definition of $g_m$), we have that
\begin{equation}\label{my189}
   |g_m(x) - g_m(z)| \leq m \rho(x,z) \; \; \;\; \; \;\;\; \;\;\forall (x,z)\in \mathcal{M} \times \mathcal{M}.
\end{equation}
Let $\epsilon > 0$ be given, and set $\delta \coloneqq \frac{\epsilon}{m}$. Suppose that $(x,z)\in \mathcal{M} \times \mathcal{M}$ satisfies $\rho(x,z) \leq \delta$. Then, we have
\begin{equation}
    |g_m(x) - g_m(z)| \overset{\eqref{my189}}{\leq} m \rho(x,z) \leq m \delta = m \frac{\epsilon}{m} = \epsilon.
\end{equation}
Thus, for every $m \in \mathbb{N}$, $g_m$ is (uniformly) continuous (with respect to $\rho$).

We will show that \eqref{my194} holds with equality by considering two cases. In the first case, we assume that $J$ is finite-valued. Let $x \in \mathcal{M}$ be given. For every $m \in \mathbb{N}$, $g_m(x) \in \mathbb{R}$, which implies (by using the definition of the infimum) that
\begin{equation}\label{my149}
    \forall \epsilon > 0 \; \exists y_m \in  \mathcal{M} \text{ s.t. }J(y_m) + m\rho(x,y_m) \leq g_m(x) + \epsilon.
\end{equation}
Note that $y_m$ depends on $\epsilon$ and $x$, which we do not write explicitly for brevity. We will construct a sequence using \eqref{my149}.
Let $\epsilon > 0$ be given. Since $g_1(x) \in \mathbb{R}$, we have
\begin{equation}
    \exists y_1 \in  \mathcal{M} \text{ s.t. }J(y_1) + 1\rho(x,y_1) \leq g_1(x) + \epsilon.
\end{equation}
Since $g_2(x) \in \mathbb{R}$, we have
\begin{equation}
    \exists y_2 \in  \mathcal{M} \text{ s.t. }J(y_2) + 2\rho(x,y_2) \leq g_2(x) + \epsilon.
\end{equation}
By repeating this process, we obtain a sequence $\{ y_m : m \in \mathbb{N} \}$ in $\mathcal{M}$ such that
\begin{equation}\label{my196}
    J(y_m) + m\rho(x,y_m) \leq g_m(x) + \epsilon \; \; \;\; \; \;\;\; \;\; \forall m \in \mathbb{N}.
\end{equation}
Moreover, since $0 \leq J(y_m)$ and $g_m(x) \leq J(x)$ for every $m \in \mathbb{N}$, we have
\begin{equation}
 m\rho(x,y_m)   \leq  J(y_m) + m\rho(x,y_m) \leq g_m(x) + \epsilon \leq J(x) + \epsilon \; \; \;\; \; \;\;\; \;\; \forall m \in \mathbb{N}.
\end{equation}
Therefore,
\begin{equation}
0 \leq   m\rho(x,y_m) \leq  J(x) + \epsilon \; \; \;\; \; \;\;\; \;\; \forall m \in \mathbb{N},
\end{equation}
where we also use the fact that $m\rho(x,y_m) \geq 0$ for every $m \in \mathbb{N}$. Since $m \in \mathbb{N}$ is positive and finite,
\begin{equation}\label{my156}
        0   \leq    \rho(x,y_m) \leq  \frac{J(x) + \epsilon}{m} \; \; \;\; \; \;\;\; \;\; \forall m \in \mathbb{N}.
\end{equation}
Since $J(x)$ is finite, it follows that
\begin{equation}\label{my157}
    \lim_{m \rightarrow +\infty} \frac{J(x) + \epsilon}{m} = 0.
\end{equation}
The statements \eqref{my156} and \eqref{my157} imply that the limit of $\{ \rho(x,y_m) \}_{m=1}^\infty$ exists and equals zero. This is because
\begin{equation}
         0   \leq  \liminf_{m \rightarrow +\infty}  \rho(x,y_m) \leq \liminf_{m \rightarrow +\infty} \frac{J(x) + \epsilon}{m} = 0
\end{equation}
and
\begin{equation}
         0   \leq  \limsup_{m \rightarrow +\infty}  \rho(x,y_m) \leq \limsup_{m \rightarrow +\infty} \frac{J(x) + \epsilon}{m} = 0,
\end{equation}
and therefore,
\begin{equation}
    \liminf_{m \rightarrow +\infty}  \rho(x,y_m) = \limsup_{m \rightarrow +\infty}  \rho(x,y_m) = 0,
\end{equation}
which allows us to conclude that
\begin{equation}\label{77}
    \lim_{m \rightarrow +\infty}  \rho(x,y_m) = 0.
\end{equation}

Moreover, since $J$ is lower semi-continuous and by \eqref{77}, we have
\begin{equation}\label{my162}
     J(x) \leq \liminf_{m \rightarrow +\infty} J(y_m) .
\end{equation}
By using \eqref{my196}, $0 \leq m\rho(x,y_m)$, and $g_m(x) \leq J(x)$, we have
\begin{equation}
J(y_m)   \leq  J(y_m) + m\rho(x,y_m) \leq g_m(x) + \epsilon \leq J(x) + \epsilon \; \; \;\;\;\;\;\;\;\; \forall m \in \mathbb{N},
\end{equation}
which implies that
\begin{equation}\label{my198}
     J(y_m) \leq g_m(x) + \epsilon \leq J(x) + \epsilon \; \; \;\;\;\;\;\;\;\; \forall m \in \mathbb{N}.
\end{equation}
By \eqref{my162}, \eqref{my198}, and the existence of the limit of $\{g_m(x)\}_{m=1}^\infty$, we have
\begin{equation}
 J(x)   \leq \liminf_{m \rightarrow +\infty}  J(y_m) \leq \lim_{m \rightarrow +\infty} g_m(x) + \epsilon \leq J(x) + \epsilon.
\end{equation}
Since $J(x) \in \mathbb{R}$, it follows that
\begin{equation}
    | -J(x)  + \lim_{m \rightarrow +\infty} g_m(x) | \leq \epsilon.
\end{equation}
Since the above analysis holds for any $\epsilon > 0$, we conclude that
\begin{equation}\label{limitingresult}
    \lim_{m \rightarrow +\infty} g_m(x) = J(x). 
\end{equation}
Since the above analysis holds for any $x \in \mathcal{M}$, we have that $\underset{m \rightarrow +\infty}{\lim} g_m(x) = J(x)$ for every $x \in \mathcal{M}$, under the assumption that $J$ is finite-valued.

Now, suppose that $J$ is not necessarily finite-valued. The function $h \colon [0, +\infty] \to [0, \frac{\pi}{2}]$ defined by $h(x) = \arctan(x)$ is increasing and continuous (Fig. \ref{tangraphs}). The inverse of $h$ exists and is increasing and continuous; the inverse is $h^{-1}: [0, \frac{\pi}{2}] \rightarrow [0, +\infty]$ such that $h^{-1}(y) = \tan(y)$ (Fig. \ref{tangraphs}). Since the range of $h$ is $[0, \frac{\pi}{2}]$, the composition $h \; \circ \; J : \mathcal{M} \rightarrow [0, \frac{\pi}{2}]$ is finite-valued and bounded below by 0.
\begin{figure}[h]
\centerline{\includegraphics[width=0.6\textwidth]{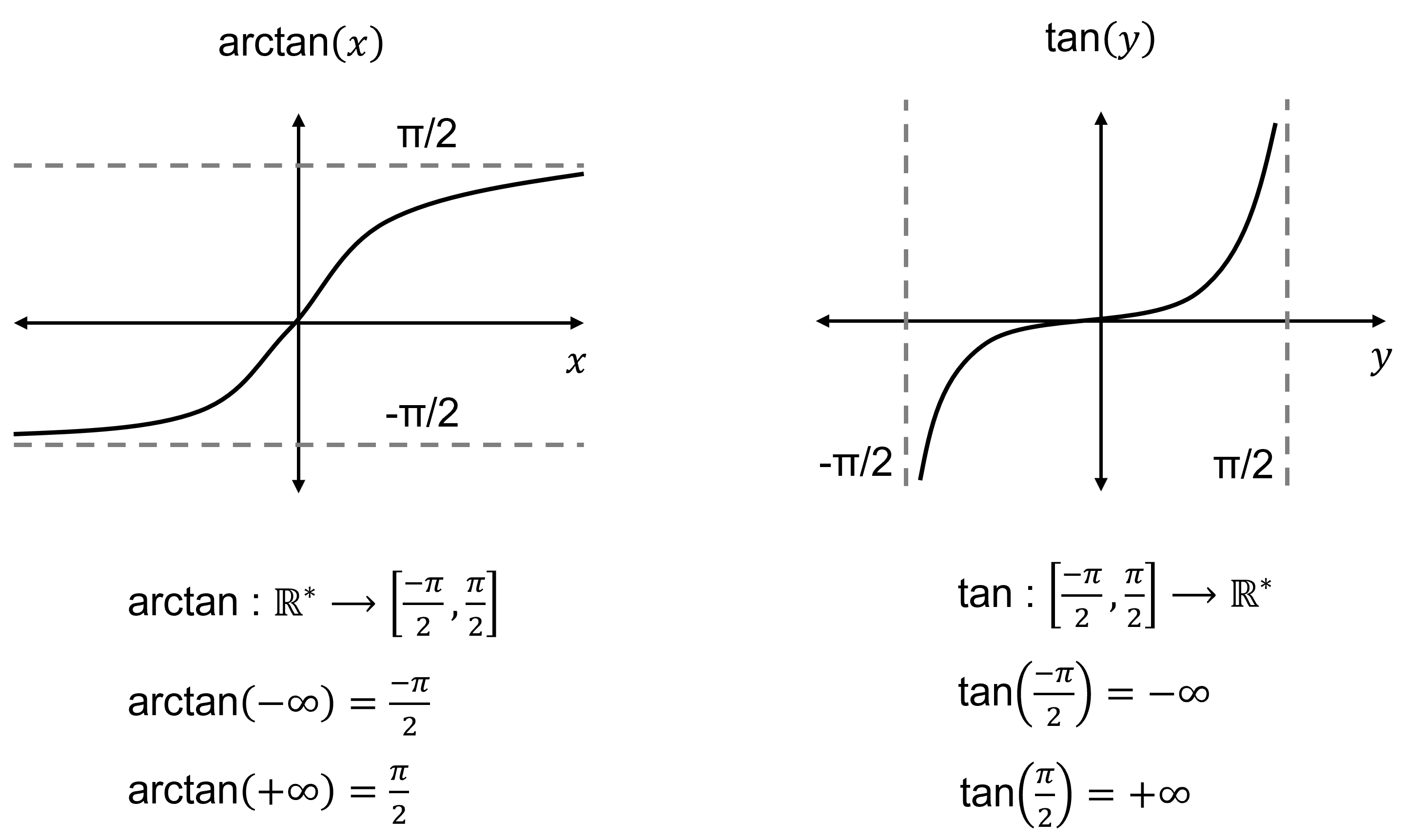}}
\caption{Illustrations of the tangent and arctangent functions, where the domain of the tangent function is restricted to $[-\frac{\pi}{2}, \frac{\pi}{2}]$. In our proof, we use the part of these functions in the nonnegative quadrant, i.e., $h \colon [0, +\infty] \to [0, \frac{\pi}{2}]$ such that $h(x) = \arctan(x)$, and $h^{-1}: [0, \frac{\pi}{2}] \rightarrow [0, +\infty]$ such that $h^{-1}(y) = \tan(y)$.}
\label{tangraphs}
\end{figure}
As a consequence of $h$ being increasing, continuous, and finite-valued, and $J : \mathcal{M} \rightarrow \mathbb{R}^*$ being l.s.c., the composition $h \circ J$ is also l.s.c. To show this explicitly, let $\{x_n\}_{n=1}^\infty \subseteq \mathcal{M}$ converge to $x \in \mathcal{M}$, i.e., $\rho(x_n,x) \rightarrow 0$, and we will show that
\begin{equation}\label{toshowlemmatan}
    \liminf_{n \rightarrow +\infty} h(J(x_n)) \geq h(J(x)).
\end{equation}
Since $\{x_n\}_{n=1}^\infty$ converges to $x$ and $J$ is l.s.c., it holds that
\begin{equation}
  +\infty  \geq \liminf_{n \rightarrow +\infty} J(x_n) \geq J(x) \geq 0.
\end{equation}
Since $h$ is increasing and its domain is $[0, +\infty]$, we have
\begin{equation}\label{1722}
    h\Big( \liminf_{n \rightarrow +\infty} J(x_n) \Big) \geq h(J(x)).
\end{equation}
Now,
\begin{equation}\label{1733}
    \liminf_{n \rightarrow +\infty} J(x_n) \coloneqq \sup_{n \in \mathbb{N}} \inf_{k \geq n} J(x_k) = \lim_{n \rightarrow +\infty} \inf_{k \geq n} J(x_k).
\end{equation}
The second inequality holds because
\begin{equation}
   \inf_{k \geq 1} J(x_k) \leq \inf_{k \geq 2} J(x_k) \leq \inf_{k \geq 3} J(x_k) \leq \cdots
\end{equation}
By \eqref{1733} and since $h$ is continuous,
\begin{equation}\label{176}
     h\Big( \liminf_{n \rightarrow +\infty} J(x_n) \Big) = h\Big( \lim_{n \rightarrow +\infty} \inf_{k \geq n} J(x_k) \Big) = \lim_{n \rightarrow +\infty} h\Big( \inf_{k \geq n} J(x_k) \Big).
\end{equation}
Let $n \in \mathbb{N}$ be given. Note that
\begin{equation}
    +\infty \geq J(x_k) \geq \inf_{k \geq n} J(x_k) \geq 0  \; \; \;\;\;\;\;\; \;\;\forall k \geq n,
\end{equation}
and since $h : [0,+\infty] \rightarrow [0, \frac{\pi}{2}]$ is increasing, it holds that
\begin{equation}
   h( J(x_k)) \geq h\Big(\inf_{k \geq n} J(x_k)\Big) \; \; \;\;\;\;\;\; \;\;\forall k \geq n.
\end{equation}
Now, $h(\inf_{k \geq n} J(x_k)) \in \mathbb{R}$ is a lower bound for $\{ h( J(x_k)) : k \geq n\}$, and so it is less than the greatest lower bound,
\begin{equation}\label{179}
    \underbrace{\inf_{k \geq n} h( J(x_k))}_{\text{greatest lower bound}} \geq \;\;\;\underbrace{h\Big(\inf_{k \geq n} J(x_k)\Big)}_{\text{a lower bound}}.
\end{equation}
Since we have derived \eqref{179} for any $n \in \mathbb{N}$, it holds for every $n \in \mathbb{N}$,
\begin{equation}\label{112}
     \inf_{k \geq n} h( J(x_k)) \geq h\Big(\inf_{k \geq n} J(x_k)\Big) \; \; \;\;\;\;\;\; \;\;\forall n \in \mathbb{N}.
\end{equation}
The limit of the left side is the limit inferior, and the limit of the right side exists by \eqref{176}, and thus,
\begin{equation}\label{1811}
  \liminf_{n \rightarrow +\infty} h( J(x_n)) = \lim_{n \rightarrow +\infty} \inf_{k \geq n} h( J(x_k)) \overset{\eqref{112}}{\geq}  \lim_{n \rightarrow +\infty} h\Big(\inf_{k \geq n} J(x_k)\Big) \overset{\eqref{176}}{=}   h\Big( \liminf_{n \rightarrow +\infty} J(x_n) \Big).
\end{equation}
Finally, we derive
\begin{equation}
     \liminf_{n \rightarrow +\infty} h( J(x_n)) \overset{\eqref{1811}}{\geq}  h\Big( \liminf_{n \rightarrow +\infty} J(x_n) \Big) \overset{\eqref{1722}}{\geq} h(J(x)),
\end{equation}
which shows that $h \circ J$ is lower semi-continuous.

Since $h \circ J$ is finite-valued, l.s.c., and bounded below by 0, there is a sequence of continuous functions $f_m : \mathcal{M} \rightarrow \mathbb{R}$ such that 
\begin{enumerate}
    \item $0 \leq f_m(x) \leq f_{m+1}(x) \leq h(J(x)) \leq \frac{\pi}{2}$ for every $m \in \mathbb{N}$ and $ x \in \mathcal{M}$,\footnote{Recall that $h : [0,+\infty] \rightarrow [0, \frac{\pi}{2}]$ is bounded above by $\frac{\pi}{2}$.} and
    \item $\textcolor{blue}{\underset{m \rightarrow +\infty}{\lim} f_m(x)} = \textcolor{blue}{h(J(x))}$ for every $ x \in \mathcal{M}$.
\end{enumerate}
Recall that $h^{-1} : [0, \frac{\pi}{2}] \rightarrow [0, +\infty]$ such that $h^{-1}(y) = \tan(y)$ is continuous and increasing (Fig. \ref{tangraphs}). It follows that
\begin{enumerate}
    \item $0=h^{-1}(0) \leq h^{-1}(f_m(x)) \leq h^{-1}(f_{m+1}(x)) \leq J(x)$ for every $m \in \mathbb{N}$ and $x \in \mathcal{M}$, and
    \item $\underset{m \rightarrow +\infty}{\lim} h^{-1}(f_m(x)) = h^{-1}\left(\textcolor{blue}{\underset{m \rightarrow +\infty}{\lim} f_m(x)}\right) = h^{-1}\big(\textcolor{blue}{h(J(x))}\big)= J(x)$ for every $x \in \mathcal{M}$.
\end{enumerate}
In summary, $h^{-1} \circ f_m : \mathcal{M} \rightarrow \mathbb{R}^*$ is continuous and $0 \leq (h^{-1} \circ f_m) \uparrow J$.

For any $m \in \mathbb{N}$, we define $J_m : \mathcal{M} \rightarrow \mathbb{R}$ by
\begin{equation}\label{my115}
    J_m(x) \coloneqq \min\{m, h^{-1}(f_m(x))\},
\end{equation}
which is a composition of continuous functions, and therefore is continuous. Each $J_m$ is bounded because
\begin{equation}\label{my116}
   0 \leq  J_m(x) \leq m \;\;\; \;\;\;\;\;\;\;\forall x \in \mathcal{M}\; \; \forall m \in \mathbb{N}.
\end{equation}
It holds that
\begin{equation}
    0 \leq J_m(x) \leq J_{m+1}(x) \leq J(x) \;\;\;\;\;\;\;\;\;\; \forall x \in \mathcal{M}\; \; \forall m \in \mathbb{N}.
\end{equation}
$J_m \leq J_{m+1}$ holds because
\begin{align}
    J_m(x) & \overset{\eqref{my115}}{\leq} m \; \leq \; \textcolor{magenta}{m+1} \\
    J_m(x) & \overset{\eqref{my115}}{\leq} h^{-1}(f_m(x)) \; \leq \; \textcolor{magenta}{h^{-1}(f_{m+1}(x))}
\end{align}
and therefore,
\begin{equation}
    J_m(x) \leq \min\{\textcolor{magenta}{m+1}, \textcolor{magenta}{h^{-1}(f_{m+1}(x))}\} = J_{m+1}(x).
\end{equation}
$J_{m+1} \leq J$ holds because
\begin{equation}
    J_{m+1}(x) \leq h^{-1}(f_{m+1}(x)) \leq J(x).
\end{equation}
Finally, since $\min$ is continuous, we have for every $x \in \mathcal{M}$,
\begin{equation}
 \lim_{m \rightarrow +\infty} J_m(x) =   \lim_{m \rightarrow +\infty} \min\{m, h^{-1}(f_m(x)) \} = \min\Big\{\lim_{m \rightarrow +\infty} m, \lim_{m \rightarrow +\infty} h^{-1}(f_m(x)) \Big\} = \min\{ +\infty, J(x) \} = J(x).
\end{equation}
The last equality holds because
\begin{equation}
    \min\{ +\infty, J(x) \} = \begin{cases} J(x) & \text{if } J(x) < +\infty \\ +\infty & \text{if }J(x) = +\infty \end{cases}.
\end{equation}
In summary, each $J_m : \mathcal{M} \rightarrow \mathbb{R}$ is continuous and bounded and $0 \leq J_m \uparrow J$, where $J$ need not be finite-valued. This concludes the proof of Lemma \ref{oldlemma3}.\end{proof}

We will use Lemma \ref{oldlemma3} to show that key properties are preserved under integration, which is needed for Theorem 2.
\subsection{Analysis of Properties under Integration}
Recall the notation $\mathbb{S} = S \times \mathcal{Z}$ and $\mathcal{Z} = [a,b] \subset \mathbb{R}$. We consider the following conditions: 
\begin{enumerate}
    \item For every $t$, $f_t$ and $c_t$ are continuous functions, and $p_t(\cdot|\cdot,\cdot)$ is a continuous stochastic kernel on $D$ given $S \times C$;
    \item For every $t$, $a \leq c_t \leq b$, where $a \in \mathbb{R}$ and $b \in \mathbb{R}$.
\end{enumerate}

\begin{lemma}\label{oldlemma4}
Let Conditions (i)--(ii) hold. If $v : \mathbb{S} \rightarrow \mathbb{R}^*$ is lower semi-continuous (l.s.c.) and bounded below by zero, then the function $g_{v,t} : \mathbb{S} \times C \rightarrow \mathbb{R}^*$ defined by
\begin{equation}\label{124}
   g_{v,t}(x, z, u) \coloneqq \int_{D} v\big(f_t(x,u,w), \max\{ z, c_t(x, u) \}\big) \; p_t(\mathrm{d} w|x,u)
\end{equation}
is l.s.c. and bounded below by zero.
\end{lemma}

Lemma \ref{oldlemma4} is different from \cite[Prop. 7.31, p. 148]{bertsekas2004stochastic} because the functions $f_t$, $\max$, and $c_t$ appear in the integral in \eqref{124}. Therefore, we cannot say that Lemma \ref{oldlemma4} holds by this proposition immediately. To prove Lemma \ref{oldlemma4}, we use Lemma \ref{oldlemma3} from the previous subsection and two other results, which are stated below.
\begin{lemma}\label{contintegrallemma}
Let Conditions (i)--(ii) hold. If $v \in \mathcal{C}(\mathbb{S})$, then the function $g_{v,t} : \mathbb{S} \times C \rightarrow \mathbb{R}$ defined by \eqref{124}
is continuous.
\end{lemma}
\begin{lemma}\label{convlemma}
Let Conditions (i)--(ii) hold. Let $v_m : \mathbb{S} \rightarrow \mathbb{R}^*$ be Borel-measurable for every $m \in \mathbb{N}$, $v : \mathbb{S} \rightarrow \mathbb{R}^*$ be Borel-measurable, and $\underline{b} \in \mathbb{R}$. Suppose that $\underline{b} \leq v_m \uparrow v$ holds, i.e.,
$\underline{b} \leq v_m \leq v_{m+1} \leq v$ for every $m \in \mathbb{N}$ and $\underset{m \rightarrow +\infty}{\lim} v_m({\tilde x},{\tilde z}) = v({\tilde x},{\tilde z})$ for every $({\tilde x},{\tilde z}) \in \mathbb{S}$. Then, for every $(x,z,u) \in \mathbb{S} \times C$, we have
\begin{equation}
    \lim_{m \rightarrow +\infty}  \int_{D} v_m\big(f_t(x,u,w),\max\{z,c_t(x,u)\}\big) \; p_t(\mathrm{d}w|x,u) = \int_{D} v\big(f_t(x,u,w),\max\{z,c_t(x,u)\}\big) \; p_t(\mathrm{d}w|x,u).
\end{equation}
\end{lemma}
In short, Lemma \ref{convlemma} holds by an application of the Monotone Convergence Theorem \cite[Th. 1.6.7, p. 47]{ash1972}.

First, we prove Lemma \ref{oldlemma4}, and then we prove the supporting results.

\hspace{-8mm}\begin{proof}
Since $v(x',s') \geq 0$ for every $(x',s') \in \mathbb{S}$, we have
\begin{equation}
    v\bigl(f_t(x,u,w), \max\{ z, c_t(x, u) \}\bigr) \geq 0 \; \; \;\;\;\;\;\;\; \forall (x,z,u,w) \in \mathbb{S} \times C \times D.
\end{equation}
For every $(x,z,u) \in \mathbb{S} \times C$, it holds that
\begin{enumerate}
    \item $v(f_t(x,u,\cdot), \max\{ z, c_t(x, u) \}): D \rightarrow \mathbb{R}^*$ is Borel-measurable,
    \item $v(f_t(x,u,\cdot), \max\{ z, c_t(x, u) \}): D \rightarrow \mathbb{R}^*$ is nonnegative, and
    \item $(D, \mathcal{B}_{D}, p_t(\cdot|x,u))$ is a probability space.
\end{enumerate}
By the above three items, the integral
\begin{equation}
g_{v,t}(x,z,u) \coloneqq  \int_{D} v\bigl(f_t(x,u,w), \max\{ z, c_t(x, u) \}\bigr) \; p_t(\mathrm{d} w|x,u)
\end{equation}
exists and is nonnegative for every $(x,z,u) \in \mathbb{S} \times C$. Therefore, $g_{v,t}$ is bounded below by zero.

To prove that $g_{v,t}$ is l.s.c., it suffices to show that if $\{(x_n,z_n,u_n) : n \in \mathbb{N} \}$ is a sequence in $\mathbb{S} \times C$ converging to $(x,z,u) \in \mathbb{S} \times C$, then
\begin{equation}
    \liminf_{n \rightarrow +\infty} g_{v,t}(x_n,z_n,u_n) \geq g_{v,t}(x,z,u).
\end{equation}
$\mathbb{S}$ is a metrizable space,\footnote{$\mathbb{S} = S \times \mathcal{Z}$ is a metrizable space because $S$ and $\mathcal{Z}$ are Borel spaces, a Cartesian product of Borel spaces with the product topology is a Borel space, and a Borel space is metrizable \cite[pp. 118--119, Prop. 7.13]{bertsekas2004stochastic}.} and $v : \mathbb{S} \rightarrow \mathbb{R}^*$ is l.s.c. and bounded below by zero by assumption. Therefore, by Lemma \ref{oldlemma3}, there is a sequence $\{ v_m : m \in \mathbb{N} \}$ in $\mathcal{C}(\mathbb{S})$ such that $0 \leq v_m \uparrow v$, i.e., 
\begin{enumerate}
    \item $0 \leq v_m\leq v_{m+1}  \leq v$ for every $m \in \mathbb{N}$, and
    \item $\underset{m \rightarrow +\infty}{\lim} v_m({\tilde x}, {\tilde z}) = v({\tilde x}, {\tilde z})$ for every $({\tilde x}, {\tilde z}) \in \mathbb{S}$. 
\end{enumerate}

Let $m \in \mathbb{N}$ and $n \in \mathbb{N}$ be given, and consider a probability space $(D, \mathcal{B}_{D}, p_t(\cdot|x_n,u_n))$. Since $v \geq v_m \geq 0$, we have
\begin{equation}
    v\bigl(f_t(x_n,u_n,w), \max\{ z_n, c_t(x_n, u_n) \}\bigr) \geq v_m\bigl(f_t(x_n,u_n,w), \max\{ z_n, c_t(x_n, u_n) \}\bigr) \geq 0 \; \; \;\;\;\;\;\;\;\; \forall w \in D.
\end{equation}
Since $v$, $f_t$, $\max$, and $c_t$ are Borel-measurable functions, the functions
\begin{equation}\begin{aligned}
     v\bigl(f_t(x_n,u_n,\cdot), \max\{ z_n, c_t(x_n, u_n) \}\bigr) &: D \rightarrow \mathbb{R}^*\\
     v_m\bigl(f_t(x_n,u_n,\cdot), \max\{ z_n, c_t(x_n, u_n) \}\bigr)&: D \rightarrow \mathbb{R}^*
\end{aligned}\end{equation}
are also Borel-measurable. It follows that
\begin{equation}\label{177}\begin{aligned}
    g_{v,t}(x_n,z_n,u_n) & \coloneqq  \int_{D} v\bigl(f_t(x_n,u_n,w), \max\{ z_n, c_t(x_n, u_n) \}\bigr) \; p_t(\mathrm{d} w|x_n,u_n) \\
    & \hphantom{:}\geq \underbrace{ \int_{D} v_m\bigl(f_t(x_n,u_n,w), \max\{ z_n, c_t(x_n, u_n) \}\bigr) \; p_t(\mathrm{d} w|x_n,u_n)}_{g_{v_m,t}(x_n,z_n,u_n)},
\end{aligned}\end{equation}
where all the integrals exist. Since the inequality \eqref{177} was derived for arbitrary $n \in \mathbb{N}$ and $m \in \mathbb{N}$, we have
\begin{equation}\label{132}
     g_{v,t}(x_n,z_n,u_n) \geq g_{v_m,t}(x_n,z_n,u_n) \; \; \;\;\;\;\;\;\;\; \forall m \in \mathbb{N}\; \; \forall n \in \mathbb{N}.
\end{equation}

For any $m \in \mathbb{N}$, we have $v_m \in \mathcal{C}(\mathbb{S})$, which implies that $g_{v_m,t} : \mathbb{S} \times C \rightarrow \mathbb{R}$ is continuous (Lemma \ref{contintegrallemma}). Therefore, we have
\begin{equation}\label{180}\begin{aligned}
    \forall m \in \mathbb{N}, \;\;\;\;\; \liminf_{n \rightarrow +\infty} g_{v,t}(x_n,z_n,u_n) \overset{\eqref{132}}{\geq} \liminf_{n \rightarrow +\infty} g_{v_m,t}(x_n,z_n,u_n)
      = \lim_{n \rightarrow +\infty} g_{v_m,t}(x_n,z_n,u_n)
      = g_{v_m,t}(x,z,u),
\end{aligned}\end{equation}
where we use $(x_n,z_n,u_n) \rightarrow (x,z,u)$.

Since $v_m : \mathbb{S} \rightarrow \mathbb{R}^*$ is Borel-measurable for every $m \in \mathbb{N}$, $v : \mathbb{S} \rightarrow \mathbb{R}^*$ is Borel-measurable, and $0 \leq v_m \uparrow v$, we use Lemma \ref{convlemma} to conclude that
\begin{equation}\label{181}
    \lim_{m \rightarrow +\infty} \underbrace{ \int_{D} v_m\big(f_t(x,u,w),\max\{z,c_t(x,u)\}\big) \; p_t(dw|x,u)}_{g_{v_m,t}(x,z,u)} = \underbrace{\int_{D} v\big(f_t(x,u,w),\max\{z,c_t(x,u)\}\big) \; p_t(dw|x,u)}_{g_{v,t}(x,z,u)}.
\end{equation}
Finally, by \eqref{180} and \eqref{181}, it holds that
\begin{equation}
    \liminf_{n \rightarrow +\infty} g_{v,t}(x_n,z_n,u_n) \geq \lim_{m \rightarrow +\infty} g_{v_m,t}(x,z,u) =  g_{v,t}(x,z,u),
\end{equation}
which shows that $g_{v,t}$ is lower semi-continuous.
\end{proof}
\subsubsection{Proof of Lemma \ref{contintegrallemma}}
Recall that $f_t$ and $c_t$ are continuous functions, and $p_t(\cdot|\cdot,\cdot)$ is a continuous stochastic kernel. We will show that if $v \in \mathcal{C}(\mathbb{S})$, then the function $g_{v,t} : \mathbb{S} \times C \rightarrow \mathbb{R}$ defined by \eqref{124} and provided below:
\begin{equation*}
      g_{v,t}(x, z, u) = \int_{D} v\bigl(f_t(x,u,w), \max\{ z, c_t(x, u) \}\bigr) \; p_t(\mathrm{d} w|x,u)
\end{equation*}
is continuous by applying \cite[Prop. 7.30, p. 145]{bertsekas2004stochastic} to our problem setting.
%

\hspace{-8mm}\begin{proof}
The spaces $\mathbb{S} \times C$, $S \times C$, and $D$ are separable and metrizable.\footnote{$\mathbb{S} \times C$, $S \times C$, and $D$ are Borel spaces, and therefore, they are separable and metrizable \cite[p. 118]{bertsekas2004stochastic}. $\mathcal{Z}$ is a closed set in $\mathbb{R}$ $\implies$ $\mathcal{Z} \in \mathcal{B}_\mathbb{R}$ $\implies$ $\mathcal{Z}$ is a Borel space. $S$, $C$, and $D$ are Borel spaces by the assumed system model. A countable Cartesian product of Borel spaces with the product topology is a Borel space \cite[Prop. 7.13, p. 119]{bertsekas2004stochastic}.} $p_t(\cdot|\cdot,\cdot)$ is a continuous stochastic kernel on $D$ given $S \times C$ by assumption. Recall that $\mathcal{P}(D)$ is the space of probability measures on $(D, \mathcal{B}_D)$ with the weak topology. The following statements are equivalent:
\begin{enumerate}
\item $p_t(\cdot|\cdot,\cdot)$ is a continuous stochastic kernel on $D$ given $S \times C$.
    \item The function $\gamma_t : S \times C \rightarrow \mathcal{P}(D)$ defined by $\gamma_t(x,u) \coloneqq p_t(\cdot|x,u)$ is continuous \cite[Def. 7.12, p. 134]{bertsekas2004stochastic}.
    \item For any $\{(x_n,u_n)\}_{n \in \mathbb{N}} \subseteq S \times C$ converging to $(x,u) \in S \times C$, it holds that $\{\gamma_t(x_n,u_n)\}_{n \in \mathbb{N}} \subseteq \mathcal{P}(D)$ converges to $\gamma_t(x,u) \in \mathcal{P}(D)$ in the weak topology.
    \item  For any $(x_n,u_n) \rightarrow (x,u)$, it holds that $\int_D \phi(w) \; p_t(\mathrm{d}w|x_n,u_n) \rightarrow \int_D \phi(w) \; p_t(\mathrm{d}w|x,u)$ for any continuous bounded function $\phi : D \rightarrow \mathbb{R}$, i.e., $\phi \in \mathcal{C}(D)$ \cite[Prop. 7.21, p. 128]{bertsekas2004stochastic}.
\end{enumerate}
The stochastic kernel $\bar{p}_t(\cdot|\cdot,\cdot,\cdot)$ on $D$ given $\mathbb{S} \times C$ defined by
\begin{equation}\label{137}
    \bar{p}_t(\mathrm{d} w|x,z,u) \coloneqq p_t(\mathrm{d} w|x,u) \; \; \;\;\;\;\;\;\;\; \forall (x,z,u) \in \mathbb{S} \times C
\end{equation}
is continuous. To see this, let $(x_n,z_n,u_n) \rightarrow (x,z,u)$ and a continuous bounded function $\phi : D \rightarrow \mathbb{R}$ be given. Then,
\begin{equation}
    \int_D \phi(w) \; \bar{p}_t(\mathrm{d}w|x_n,z_n,u_n) = \int_D \phi(w) \; p_t(\mathrm{d}w|x_n,u_n) \rightarrow \int_D \phi(w) \; p_t(\mathrm{d}w|x,u) = \int_D \phi(w) \; \bar{p}_t(\mathrm{d}w|x,z,u),
\end{equation}
where the equalities hold by \eqref{137}; the limit holds because $p_t(\cdot|\cdot,\cdot)$ is a continuous stochastic kernel on $D$ given $S \times C$, $(x_n,u_n) \rightarrow (x,u)$, and $\phi : D \rightarrow \mathbb{R}$ is continuous and bounded.

If $h_t \in \mathcal{C}(\mathbb{S} \times C \times D)$, then the function $\nu_t : \mathbb{S} \times C \rightarrow \mathbb{R}$ defined by
\begin{equation}
   \nu_t(x,z,u) \coloneqq \int_{D} h_t(x,z,u,w) \; \bar{p}_t(\mathrm{d} w|x,z,u) = \int_{D} h_t(x,z,u,w) \; p_t(\mathrm{d}w|x,u)
\end{equation}
is continuous by \cite[Prop. 7.30]{bertsekas2004stochastic}. Hence, it suffices to show that
\begin{equation}
    h_t(x,z,u,w) \coloneqq v\bigl( f_t(x,u,w), \max\{ z, c_t(x, u) \} \bigr)
\end{equation}
satisfies $h_t \in \mathcal{C}(\mathbb{S} \times C \times D)$. Consider the function
\begin{align}
    h_{1,t} \colon \mathbb{S} \times C \times D &\to \mathbb{S} \\
    (x, z, u, w) &\mapsto \bigl( f_t(x,u,w), \max\{ z, c_t(x, u) \} \bigr).
\end{align}
Since $f_t$, $c_t$, and $\max$ are continuous, it holds that 
\begin{align}
    \lim_{n \to \infty} h_{1,t}(x_n, z_n, u_n, w_n) &= \lim_{n \to \infty} \bigl( f_t(x_n,u_n,w_n), \max\{ z_n, c_t(x_n, u_n) \} \bigr) \\
    &= \bigl( \lim_{n \to \infty} f_t(x_n, u_n, w_n), \lim_{n \to \infty} \max\{ z_n, c_t(x_n, u_n) \} \bigr) \\
    &= \bigl( f_t(x, u, w),  \max\{ z, c_t(x, u) \} \bigr)
\end{align}
for any sequence $\{(x_n, z_n, u_n, w_n)\}_{n \in \mathbb{N}}$ in $\mathbb{S} \times C \times D$ converging to a point $(x, z, u, w) \in \mathbb{S} \times C \times D$. Therefore, $h_{1,t}$ is continuous. The function $h_t$ can be written as $v \circ h_{1,t} \colon \mathbb{S} \times C \times D \to \mathbb{S} \to \mathbb{R}$, where $\circ$ denotes composition. Since the composition of continuous functions is again a continuous function \cite[Prop.~3.1.8]{sutherland1975introduction}, $h_t$ is continuous. Moreover, it holds that 
\begin{equation}
    \sup_{\mathbb{S} \times C \times D} h_t = \sup_{\mathbb{S} \times C \times D} v \circ h_{1,t} \leq \sup_{\mathbb{S}} v < +\infty,
\end{equation}
where the last inequality holds because $v$ is bounded. Hence, $h_t$ is bounded. Since $h_t : \mathbb{S} \times C \times D \rightarrow \mathbb{R}$ is continuous and bounded, we have that $h_t \in \mathcal{C}(\mathbb{S} \times C \times D)$, which concludes the proof of Lemma \ref{contintegrallemma}.
\end{proof}
\subsubsection{Proof of Lemma \ref{convlemma}}
Recall that $f_t$ and $c_t$ are continuous functions, and $p_t(\cdot|\cdot,\cdot)$ is a continuous stochastic kernel. 
Let $v_m : \mathbb{S} \rightarrow \mathbb{R}^*$ be Borel-measurable for every $m \in \mathbb{N}$, $v : \mathbb{S} \rightarrow \mathbb{R}^*$ be Borel-measurable, and $\underline{b} \in \mathbb{R}$. Suppose that $\underline{b} \leq v_m \uparrow v$ holds, i.e.,
$\underline{b} \leq v_m \leq v_{m+1} \leq v$ $\forall m \in \mathbb{N}$ and $\underset{m \rightarrow +\infty}{\lim} v_m({\tilde x},{\tilde z}) = v({\tilde x},{\tilde z})$ for every $({\tilde x},{\tilde z}) \in \mathbb{S}$. Under these conditions, we will show that for every $(x,z,u) \in \mathbb{S} \times C$
\begin{equation}
    \lim_{m \rightarrow +\infty}  \int_{D} v_m\big(f_t(x,u,w),\max\{z,c_t(x,u)\}\big) \; p_t(dw|x,u) = \int_{D} v\big(f_t(x,u,w),\max\{z,c_t(x,u)\}\big) \; p_t(dw|x,u).
\end{equation}
\hspace{-4mm}\begin{proof} We use the Extended Monotone Convergence Theorem \cite[Th. 1.6.7, p. 47]{ash1972}: Let $(\bar{\Omega},\mathcal{F},\mu)$ be a measure space. Let $g_1, g_2, \dots, g, h$ be functions from $\bar{\Omega}$ to $\mathbb{R}^*$, which are measurable relative to $\mathcal{F}$ and $\mathcal{B}_{\mathbb{R}^*}$. If $g_n(\omega) \geq h(\omega)$ for every $\omega \in \bar{\Omega}$ and $n \in \mathbb{N}$, $\int_{\bar{\Omega}} h(\omega) \mu(\mathrm{d}\omega) > -\infty$, and $g_n \uparrow g$,\footnote{$g_n \uparrow g$ means $g_{n}(\omega) \leq g_{n+1}(\omega) \leq g(\omega)$ for every $\omega \in \bar{\Omega}$ and $n \in \mathbb{N}$, and $\underset{n \rightarrow +\infty}{\lim} g_n(\omega) = g(\omega)$ for every $\omega \in \bar{\Omega}$.} then $\int_{\bar{\Omega}} g_n(\omega) \mu(\mathrm{d}\omega) \uparrow \int_{\bar{\Omega}} g(\omega) \mu(\mathrm{d}\omega)$.

Let $(x,z,u) \in \mathbb{S} \times C$ be given. We use the probability space $(D, \mathcal{B}_{D}, p_t(\cdot|x,u))$. Define the functions $g_{m,t}^{x,u,z} : D \rightarrow \mathbb{R}^*$ for every $m \in \mathbb{N}$, $g_t^{x,u,z} : D \rightarrow \mathbb{R}^*$, and $h : D \rightarrow \mathbb{R}^*$ as follows:
\begin{align}
    g_{m,t}^{x,u,z}(w) & \coloneqq v_m\big(f_t(x,u,w),\max\{z,c_t(x,u)\}\big) \label{148}\\
    g_t^{x,u,z}(w) & \coloneqq v\big(f_t(x,u,w),\max\{z,c_t(x,u)\}\big) \label{149}\\
    h(w) & \coloneqq \bar{b} \label{1500}.
\end{align}
The functions \eqref{148}--\eqref{1500} are measurable relative to $\mathcal{B}_{D}$ and $\mathcal{B}_{\mathbb{R}^*}$. $f_t$, $\max$, $c_t$, and $h$ are continuous, which implies that they are Borel-measurable. $v_m$ and $v$ are Borel-measurable, and the composition of Borel-measurable functions is Borel-measurable.

Recall that $\underline{b} \leq v_m({\tilde x}, {\tilde z}) \leq v_{m+1}({\tilde x}, {\tilde z}) \leq v({\tilde x}, {\tilde z})$ for every $({\tilde x}, {\tilde z}) \in \mathbb{S}$ and $m \in \mathbb{N}$. Therefore, for every $w \in D$ and $m \in \mathbb{N}$, we have
\begin{equation}\begin{aligned}
    \underbrace{\bar{b}}_{h(w)} \leq \underbrace{v_m\big(f_t(x,u,w),\max\{z,c_t(x,u)\}\big)}_{g_{m,t}^{x,u,z}(w)}  \leq \underbrace{v_{m+1}\big(f_t(x,u,w),\max\{z,c_t(x,u)\}\big)}_{g_{m+1,t}^{x,u,z}(w)}  \leq \underbrace{v\big(f_t(x,u,w),\max\{z,c_t(x,u)\}\big)}_{g_t^{x,u,z}(w)}. 
\end{aligned}\end{equation}

Recall that for every $({\tilde x},{\tilde z}) \in \mathbb{S}$, we have $\underset{m \rightarrow +\infty}{\lim} v_m({\tilde x},{\tilde z}) = v({\tilde x},{\tilde z})$. Let $w \in D$ be given. Then, $f_t(x,u,w) \in S$ and $\max\{z,c_t(x,u)\} \in \mathcal{Z}$. Therefore,
\begin{equation}\label{152}
  \lim_{m \rightarrow +\infty} v_m\big(f_t(x,u,w),\max\{z,c_t(x,u)\}\big) = v\big(f_t(x,u,w),\max\{z,c_t(x,u)\}\big).
\end{equation}
Since $w \in D$ in \eqref{152} is arbitrary and by using the definitions \eqref{148}--\eqref{149}, we conclude that
\begin{equation}
     \lim_{m \rightarrow +\infty} g_{m,t}^{x,u,z}(w) = g_t^{x,u,z}(w) \;\;\;\;\;\;\;\;\;\; \forall w \in D.
\end{equation}
By using the definition \eqref{1500}, we have $\int_{D} h(w)\; p_t(\mathrm{d}w|x,u) = \int_{D} \bar{b} \; p_t(\mathrm{d}w|x,u) = \bar{b} > -\infty$.

To summarize, we are working on the probability space $(D, \mathcal{B}_{D}, p_t(\cdot|x,u))$, and the following properties hold:
\begin{itemize}
    \item $g_{1,t}^{x,u,z}, g_{2,t}^{x,u,z}, \dots, g_t^{x,u,z}
    $, and $h$ are functions from $D$ to $\mathbb{R}^*$, which are measurable relative to $\mathcal{B}_D$ and $\mathcal{B}_{\mathbb{R}^*}$.
    \item $g_{m,t}^{x,u,z} \geq h$ for every $m \in \mathbb{N}$, $\int_{D} h(w)\; p_t(\mathrm{d}w|x,u) > - \infty$, and $g_{m,t}^{x,u,z} \uparrow g_t^{x,u,z}$.
\end{itemize}
Thus, by the Extended Monotone Convergence Theorem, it holds that 
\begin{equation}\label{17555}
    \lim_{m \rightarrow +\infty}  \int_{D} g_{m,t}^{x,u,z}(w) \; p_t(\mathrm{d}w|x,u) = \int_{D} g_t^{x,u,z}(w) \; p_t(\mathrm{d}w|x,u).
\end{equation}
Since we have derived the equality \eqref{17555} for an arbitrary $(x,z,u) \in \mathbb{S} \times C$, it holds for every $(x,z,u) \in \mathbb{S} \times C$, which completes the proof of Lemma \ref{convlemma}.
\end{proof}

\subsection{Background about Measurable Selection}\label{lscmeas}
We use a measurable selection result \cite[Prop. 7.33, p. 153]{bertsekas2004stochastic} to prove Theorem 2 in the main paper. Variations of \cite[Prop. 7.33]{bertsekas2004stochastic} can be found in other texts, e.g., \cite{hernandez2012discrete} and \cite{dynkin1979controlled}. To understand how \cite[Prop. 7.33]{bertsekas2004stochastic} applies to our setting, we state a special case below. 

\begin{remark}[Special case of Prop. 7.33, Bertsekas and Shreve, 1996]
Let $X$ and $Y$ be metrizable spaces, where $Y$ is compact. Assume that $g : X \times Y \rightarrow \mathbb{R}^*$ is lower semi-continuous (l.s.c.). Let $g^* : X \rightarrow \mathbb{R}^*$ be defined by
\begin{equation}\label{1555}
    g^*(x) \coloneqq \inf_{y \in Y} g(x,y).
\end{equation}
Then, $g^*$ is l.s.c., and there is a Borel-measurable function $\phi : X \rightarrow Y$ such that
\begin{equation}
    g(x,\phi(x)) = g^*(x) \; \; \;\;\;\;\;\;\;\; \forall x \in X.
\end{equation}
\end{remark}
\vspace{3mm}\begin{remark}
If $g$ is bounded below by $\underline{b} \in \mathbb{R}$, then $g^*$ is also bounded below by $\underline{b}$.
\end{remark}
\subsection{An Extended Proof for Theorem 2}
We use the previous results to prove Theorem 2. 

\emph{Theorem 2 (DP on $\mathbb{S}$):}
Let Assumption 1 hold, and let $s \in \mathbb{R}$ be given. Recall that $J_N^s : \mathbb{S} \rightarrow \mathbb{R}^*$ is given by $J_N^s(x, z) = h^s(\max\{c_N(x),z\})$ \eqref{defJNs}. For $t = N-1,\dots,1,0$, we define $J_t^s : \mathbb{S} \rightarrow \mathbb{R}^*$ recursively by
\begin{subequations}\label{27}
\begin{equation}\label{27b}
    J_t^s(x, z) \coloneqq \inf_{u \in C} v_t^s(x,z,u),
\end{equation}
where $v_t^s : \mathbb{S} \times C \rightarrow \mathbb{R}^*$ depends on $J_{t+1}^s$ as follows:
\begin{align}
    v_t^s(x,z,u) 
    & \coloneqq \int_{D} J_{t+1}^s\bigl(f_t(x,u,w),  \max\{c_t(x, u),z \}\bigr) \; p_t(\mathrm{d} w|x,u). \label{27c} 
\end{align}
\end{subequations}
Then, for every $t \in \mathbb{T}_N$, $J_t^s$ is lower semi-continuous (l.s.c.) and bounded below by zero. For every $t \in \mathbb{T}$, there is a Borel-measurable function $\kappa_t^s : \mathbb{S} \rightarrow C$ such that
\begin{equation}\label{mykappaeq}
J_t^s(x, z) = v_t^s(x,z,\kappa_t^s(x,z)) \;\;\;\;\; \;\;\;\;\;\forall (x,z) \in \mathbb{S}.
\end{equation}
For every $(x,z) \in \mathbb{S}$, let $\delta_{\kappa_t^s(x,z)}$ denote the Dirac measure on $(C,\mathcal{B}_C)$ concentrated at the point $\kappa_t^s(x,z) \in C$.\footnote{Recall that $\mathcal{P}(C)$ is the space of probability measures on $(C, \mathcal{B}_C)$ with the weak topology. $\delta_{\kappa_t^s}$ is a Borel-measurable stochastic kernel on $C$ given $\mathbb{S}$ because the function $\gamma_t : \mathbb{S} \rightarrow \mathcal{P}(C)$ defined by $\gamma_t(x,z) \coloneqq \delta_{\kappa_t^s(x,z)}$ is Borel-measurable. $\gamma_t$ is a composition of Borel-measurable functions. The function $\nu : C \rightarrow \mathcal{P}(C)$, where $\nu(u) \coloneqq \delta_u$ is the Dirac measure on $(C, \mathcal{B}_C)$ concentrated at the point $u \in C$, is continuous by \cite[Corollary 7.21.1, p. 130]{bertsekas2004stochastic}. The function $\kappa_t^s : \mathbb{S} \rightarrow C$ is Borel-measurable by Theorem 2.\label{footnote7}} We define $\pi^s \coloneqq (\delta_{\kappa_0^s},\delta_{\kappa_1^s},\dots,\delta_{\kappa_{N-1}^s})$, which is an element of $\Pi$. Then, for every $\mathbf{x} \in S$, we have
    \begin{equation}\label{my28}
         J_0^s(\mathbf{x}, a) = V^{s}(\mathbf{x}) = E_{\mathbf{x}}^{\pi^s}( \max\{Y - s, 0 \}).
    \end{equation}

\hspace{-7mm}\begin{proof}
The proof has two parts.
\subsubsection{Properties of the dynamic programming iterates}\label{part1}
We proceed by induction. $J_N^s$ is continuous because $c_N$ is continuous, $\max$ is continuous, and a composition of continuous functions is continuous. Since $J_N^s$ is continuous, it is also l.s.c. $J_N^s$ is bounded below by zero because $\max\{y,0\} \geq 0$ for every $y \in \mathbb{R}$. Now, assume (the induction hypothesis) that for some $t \in\{ N-1,\dots,1,0\}$, $J_{t+1}^s : \mathbb{S} \rightarrow \mathbb{R}^*$ is l.s.c. and bounded below by zero. Then, by Lemma \ref{oldlemma4} and Assumption 1, the function
$v_t^s : \mathbb{S} \times C \rightarrow \mathbb{R}^*$ defined by
\begin{equation*}
    v_t^s(x, z, u) \overset{\eqref{27c}}{=} \int_{D} J_{t+1}^s\bigl(f_t(x,u,w), \max\{c_t(x, u), z \}\bigr) \; p_t(\mathrm{d} w|x,u)
\end{equation*}
is l.s.c. and bounded below by zero. Moreover, the function $J_t^s : \mathbb{S} \rightarrow \mathbb{R}^*$ defined by 
\begin{equation*}\begin{aligned}
    J_t^s(x, z) 
    \overset{\eqref{27b}}{=} \inf_{u \in C}\;  v_t^s(x, z, u)
\end{aligned}\end{equation*}
is l.s.c. and bounded below by zero, where we use the compactness of $C$ in particular and apply \cite[Prop. 7.33]{bertsekas2004stochastic}; the reader may refer to Sec. \ref{lscmeas} for details. Since we have shown the induction step, we conclude that $J_t^s$ is l.s.c. and bounded below by zero for every $t \in \{ N, \dots,1,0\}$.

Let $t \in \{0,1,\dots,N-1\}$ be given. Since $v_t^s : \mathbb{S} \times C \rightarrow \mathbb{R}^*$ is l.s.c., $\mathbb{S}$ and $C$ are metrizable spaces, and $C$ is compact, there is a Borel-measurable function $\kappa_t^s : \mathbb{S} \rightarrow C$ such that \eqref{mykappaeq} holds, which we repeat below:
\begin{equation*}
    J_t^s(x, z) \overset{\eqref{27b}}{=} \inf_{u \in C}\;  v_t^s(x, z, u) = v_t^s(x, z, \kappa_t^s(x,z)) \; \; \; \;\;\; \; \; \;\; \forall (x,z) \in \mathbb{S}
\end{equation*}
by an application of \cite[Prop. 7.33]{bertsekas2004stochastic} (Sec. \ref{lscmeas}). 

We define $\pi^s \coloneqq (\delta_{\kappa_0^s}, \delta_{\kappa_1^s}, \dots, \delta_{\kappa_{N-1}^s})$. For every $(x,z) \in \mathbb{S}$, $\delta_{\kappa_t^s(x,z)} \in \mathcal{P}(C)$ is the Dirac measure on $(C, \mathcal{B}_C)$ that is concentrated at the point $\kappa_t^s(x,z) \in C$. $\delta_{\kappa_t^s}$ is a Borel-measurable stochastic kernel on $C$ given $\mathbb{S}$ (Footnote \ref{footnote7}). Since $\pi^s$ is a tuple of $N$ Borel-measurable stochastic kernels on $C$ given $\mathbb{S}$, $\pi^s$ is an element of $\Pi$.
\subsubsection{Optimality}
Our goal is to prove that \eqref{my28} holds:
\begin{equation*}
         \forall \mathbf{x} \in S, \;\;\;\;\;\;\;\;\; J_0^s(\mathbf{x}, a) = V^{s}(\mathbf{x}) = E_{\mathbf{x}}^{\pi^s}( \max\{Y - s, 0 \} ).
\end{equation*}
We recall the results from Theorem 1. For every $\mathbf{x} \in S$ and $\pi \in \Pi$, the following statements hold:
\begin{align*}
    E_{\mathbf{x}}^\pi( \max\{ Y - s, 0 \} ) & \overset{\eqref{46}}{=} \int_{\Omega}  \phi_0^{\pi,s} \circ \mathcal{X}_0\; \mathrm{d}P_{\mathbf{x}}^\pi \overset{\eqref{46}}{=} \phi_0^{\pi,s}(\mathbf{x},a),
     \\
    \int_{\Omega} \phi_N^{\pi,s} \circ \mathcal{X}_N \; \mathrm{d}P_{\mathbf{x}}^\pi & \overset{\eqref{47}}{=} \int_{\Omega} J_N^s \circ \mathcal{X}_N \; \mathrm{d}P_{\mathbf{x}}^\pi ,  \\
\forall t \in \mathbb{T}, \;\;\;\;\;   \int_{\Omega} \phi_t^{\pi,s} \circ \mathcal{X}_t\; \mathrm{d}P_{\mathbf{x}}^\pi & \overset{\eqref{66}}{=} \int_{\Omega}  \phi_{t+1}^{\pi,s} \circ \mathcal{X}_{t+1}\; \mathrm{d}P_{\mathbf{x}}^\pi,
\end{align*}
where for every $t \in \mathbb{T}_N$, $\phi_t^{\pi,s} : \mathbb{S} \rightarrow \mathbb{R}^*$ is a Borel-measurable function that characterizes the conditional expectation of $Y_t^s$ given $\mathcal{X}_t =(X_t,Z_t)$. We will explain why it suffices to show that
\begin{subequations}\label{toshow}
\begin{align}
   \forall t \in \mathbb{T}_N \; \; \;\forall \mathbf{x} \in S \; \; \;  \forall \pi \in \Pi,    \; \; \; \; \; \int_{\Omega} \phi_t^{\pi,s} \circ \mathcal{X}_t\; \mathrm{d}P_{\mathbf{x}}^\pi & \geq \int_{\Omega}  J_t^s\circ \mathcal{X}_t\; \mathrm{d}P_{\mathbf{x}}^\pi ,\label{toshow1} \\
   \forall t \in \mathbb{T}_N  \; \; \; \forall \mathbf{x} \in S, \; \; \; \int_{\Omega}  \phi_t^{\pi^s,s} \circ \mathcal{X}_t\; \mathrm{d}P_{\mathbf{x}}^{\pi^s} & = \int_{\Omega}  J_t^s \circ \mathcal{X}_t \; \mathrm{d}P_{\mathbf{x}}^{\pi^s}. \label{toshow2}
\end{align}
\end{subequations}
(Since $J_{i}^s\circ \mathcal{X}_{i} : (\Omega,\mathcal{B}_{\Omega}) \rightarrow (\mathbb{R}^*, \mathcal{B}_{\mathbb{R}^*})$ is nonnegative for every $i \in \mathbb{T}_N$ and $P_{\mathbf{x}}^\pi$ is a probability measure on $(\Omega,\mathcal{B}_{\Omega})$ for every $\mathbf{x} \in S$ and $\pi \in \Pi$, the integral $\int_{\Omega}  J_{i}^s\circ \mathcal{X}_{i}\; \mathrm{d}P_{\mathbf{x}}^\pi$ exists and is nonnegative for every $i \in \mathbb{T}_N$, $\mathbf{x} \in S$, and $\pi \in \Pi$.) If \eqref{toshow} holds, then by considering $t = 0$, we find that
\begin{subequations}\label{my116}
\begin{align}
 \forall \mathbf{x} \in S\;\;\;  \forall \pi \in \Pi,    \; \; \; \; \;  E_{\mathbf{x}}^\pi( \max\{ Y - s, 0 \} ) & \overset{\eqref{46}}{=} \int_{\Omega} \phi_0^{\pi,s} \circ \mathcal{X}_0\; \mathrm{d}P_{\mathbf{x}}^\pi  \overset{\eqref{toshow1}}{\geq} \int_{\Omega}  J_0^s \circ \mathcal{X}_0\; \mathrm{d}P_{\mathbf{x}}^\pi \overset{\eqref{91b}}{=} J_0^s(\mathbf{x},a)\label{153}\\
  \forall \mathbf{x} \in S,\;\;\;\; E_{\mathbf{x}}^{\pi^s}( \max\{ Y - s, 0 \} ) & \overset{\eqref{46}}{=}  \int_{\Omega}  \phi_0^{\pi^s,s} \circ \mathcal{X}_0\; \mathrm{d}P_{\mathbf{x}}^{\pi^s} \overset{\eqref{toshow2}}{=} \int_{\Omega}  J_0^s  \circ \mathcal{X}_0\; \mathrm{d}P_{\mathbf{x}}^{\pi^s} \overset{\eqref{91b}}{=} J_0^s(\mathbf{x},a). \label{154}
\end{align}
\end{subequations}
($J_0^s : \mathbb{S} \rightarrow \mathbb{R}^*$ is measurable relative to $\mathcal{B}_{\mathbb{S}}$ and $\mathcal{B}_{\mathbb{R}^*}$ because it is l.s.c.)
We can write \eqref{my116} more concisely as follows:
\begin{equation}\label{almosthere}
 \forall \mathbf{x} \in S\;\;\; \forall \pi \in \Pi, \; \; \; \; \;   E_{\mathbf{x}}^\pi( \max\{ Y - s, 0 \} ) \geq \textcolor{magenta}{J_0^s(\mathbf{x},a) = E_{\mathbf{x}}^{\pi^s}( \max\{ Y - s, 0 \} )},
\end{equation}
where the last quantity is bounded below by 0. We take the infimum over $\pi \in \Pi$ in \eqref{almosthere} to find that
\begin{equation}\label{almostalmost}
  \forall \mathbf{x} \in S,\;\;\;   \textcolor{blue}{V^s({\mathbf x}) = \inf_{\pi \in \Pi} E_{\mathbf{x}}^\pi( \max\{ Y - s, 0 \} )} \geq \textcolor{magenta}{J_0^s(\mathbf{x},a) = E_{\mathbf{x}}^{\pi^s}( \max\{ Y - s, 0 \} )} \geq \textcolor{blue}{\inf_{\pi \in \Pi} E_{\mathbf{x}}^\pi( \max\{ Y - s, 0 \} ) =  V^s({\mathbf x})},
\end{equation}
which shows the desired statement: for every $\mathbf{x} \in S$, $\textcolor{magenta}{J_0^s(\mathbf{x}, a)} = \textcolor{blue}{V^{s}(\mathbf{x})} = \textcolor{magenta}{E_{\mathbf{x}}^{\pi^s}( \max\{Y - s, 0 \} )}$ \eqref{my28}. In summary, if \eqref{toshow} holds, then the desired statement \eqref{my28} holds, and the proof is complete.

To show \eqref{toshow}, we proceed by induction. For the base case ($t=N$), we recall from Theorem 1 that
\begin{align*}
  \forall \mathbf{x} \in S\;\;\;  \forall \pi \in \Pi,\;\;\;\;  \int_{\Omega} \phi_N^{\pi,s} \circ \mathcal{X}_N \; \mathrm{d}P_{\mathbf{x}}^\pi  \overset{\eqref{47}}{=} \int_{\Omega} J_N^s \circ \mathcal{X}_N \; \mathrm{d}P_{\mathbf{x}}^\pi ,
\end{align*}
which implies 
\begin{align}
   \forall \mathbf{x} \in S\;\;\; \forall \pi \in \Pi,\;\;\;\;  \int_{\Omega} \phi_N^{\pi,s} \circ \mathcal{X}_N \; \mathrm{d}P_{\mathbf{x}}^\pi & \geq \int_{\Omega} J_N^s \circ \mathcal{X}_N \; \mathrm{d}P_{\mathbf{x}}^\pi ,\\
  \forall \mathbf{x} \in S,\;\;\;  \int_{\Omega} \phi_N^{\pi^s,s} \circ \mathcal{X}_N \; \mathrm{d}P_{\mathbf{x}}^{\pi^s} & = \int_{\Omega} J_N^s \circ \mathcal{X}_N \; \mathrm{d}P_{\mathbf{x}}^{\pi^s}. \label{showingbasecase2}
\end{align}

Now, assume (the induction hypothesis for \eqref{toshow1}) that for some $t \in \{N-1,\dots,1,0\}$, it holds that 
\begin{equation}\label{inductionhyp}
 \forall \mathbf{x} \in S\;\;\;   \forall \pi \in \Pi,    \; \; \; \;\;\; \int_{\Omega} \phi_{t+1}^{\pi,s} \circ \mathcal{X}_{t+1}\; \mathrm{d}P_{\mathbf{x}}^\pi  \geq \int_{\Omega}  J_{t+1}^s\circ \mathcal{X}_{t+1}\; \mathrm{d}P_{\mathbf{x}}^\pi.
\end{equation}
We will show that
\begin{equation}\label{183}
 \forall \mathbf{x} \in S\;\;\;   \forall \pi \in \Pi,    \; \; \;  \;\;\; \int_{\Omega} \phi_{t}^{\pi,s} \circ \mathcal{X}_{t}\; \mathrm{d}P_{\mathbf{x}}^\pi  \geq \int_{\Omega}  J_{t}^s\circ \mathcal{X}_{t}\; \mathrm{d}P_{\mathbf{x}}^\pi
\end{equation}
to prove \eqref{toshow1} by induction. Let $\mathbf{x} \in S$ and $\pi = (\pi_0,\pi_1,\dots,\pi_{N-1}) \in \Pi$ be given. By Theorem 1 \eqref{66} and the induction hypothesis \eqref{inductionhyp}, we have
\begin{align}\label{184}
  \textcolor{blue}{ \int_{\Omega} \phi_t^{\pi,s} \circ \mathcal{X}_t\; \mathrm{d}P_{\mathbf{x}}^\pi} & \overset{\eqref{66}}{=} \int_{\Omega}  \phi_{t+1}^{\pi,s} \circ \mathcal{X}_{t+1}\; \mathrm{d}P_{\mathbf{x}}^\pi
   \overset{\eqref{inductionhyp}}{\geq} \int_{\Omega}  J_{t+1}^s\circ \mathcal{X}_{t+1}\; \mathrm{d}P_{\mathbf{x}}^\pi.
\end{align}
To show that \eqref{183} holds, it suffices to show that
\begin{equation}\label{230}
    \int_{\Omega}  J_{t+1}^s\circ \mathcal{X}_{t+1}\; \mathrm{d}P_{\mathbf{x}}^\pi \geq \textcolor{blue}{\int_{\Omega} J_{t}^s \circ \mathcal{X}_{t} \; \mathrm{d}P_{\mathbf{x}}^\pi}.
\end{equation}
Since $t+1 \in \{1,2,\dots,N\}$, $J_{t+1}^s : (\mathbb{S},\mathcal{B}_{\mathbb{S}}) \rightarrow (\mathbb{R}^*,\mathcal{B}_{\mathbb{R}^*})$, and the integral $\int_{\Omega} J_{t+1}^s\circ \mathcal{X}_{t+1}\; \mathrm{d}P_{\mathbf{x}}^\pi$ exists, we have
\begin{align}
 \int_{\Omega} J_{t+1}^s\circ \mathcal{X}_{t+1} \; \mathrm{d}P_{\mathbf{x}}^\pi 
       & \overset{\eqref{154bimportant}}{=}  \int_{(\mathbb{S} \times C)^{\textcolor{blue}{t}+1} \times \mathbb{S}}  J_{t+1}^s(\chi') \; \tilde{q}_{\textcolor{blue}{t}}(\mathrm{d}\chi'|\chi_{\textcolor{blue}{t}},u_{\textcolor{blue}{t}}) \; \pi_{t}(\mathrm{d}u_{t}|\chi_t) \; \tilde{q}_{t-1}(\mathrm{d}\chi_{t}|\chi_{t-1},u_{t-1}) \cdots  \delta_{\mathbf{x},a}(\mathrm{d}\chi_0) \nonumber \\
       & \overset{\hphantom{\eqref{154bimportant}}}{=}  \int_{(\mathbb{S} \times C)^{\textcolor{blue}{t}} \times \mathbb{S}} \int_{C} \int_{\mathbb{S}}  J_{t+1}^s(\chi') \; \tilde{q}_{\textcolor{blue}{t}}(\mathrm{d}\chi'|\chi_{\textcolor{blue}{t}},u_{\textcolor{blue}{t}}) \; \pi_{t}(\mathrm{d}u_{t}|\chi_t) \; \tilde{q}_{t-1}(\mathrm{d}\chi_{t}|\chi_{t-1},u_{t-1}) \cdots  \delta_{\mathbf{x},a}(\mathrm{d}\chi_0). \label{225}
\end{align}
In the last line, we use
\begin{equation}
    (\mathbb{S} \times C)^{\textcolor{blue}{t}+1} \times \mathbb{S} = (\mathbb{S} \times C)^{\textcolor{blue}{t}} \times \mathbb{S} \times C \times \mathbb{S}.
\end{equation}
For convenience, we define $v_t^{s,\pi} : \mathbb{S} \rightarrow \mathbb{R}^*$ by
\begin{equation}\label{vtpi}
   \textcolor{blue}{ v_t^{s,\pi}(\chi)} \coloneqq \textcolor{magenta}{\int_C \textcolor{black}{v_t^s(\chi, u)} \; \pi_t(\mathrm{d}u|\chi)} = \textcolor{magenta}{\int_C \textcolor{black}{\int_{\mathbb{S}} J_{t+1}^s(\chi') \; \tilde{q}_t(\mathrm{d} \chi'|\chi,u) } \; \pi_t(\mathrm{d}u|\chi)}.
\end{equation}
$v_t^{s,\pi}$ is Borel-measurable by an application of \cite[Prop. 7.29, p. 144]{bertsekas2004stochastic} and nonnegative.\footnote{$\mathbb{S}$ and $C$ are Borel spaces. $\pi_t$ is a Borel-measurable stochastic kernel on $C$ given $\mathbb{S}$ because $\pi \in \Pi$. $v_t^s : \mathbb{S} \times C \rightarrow \mathbb{R}^*$ is Borel-measurable because it is l.s.c. (Sec. \ref{part1}). For every $(\chi,u) \in \mathbb{S} \times C$, it holds that $v_t^s(\chi, u) \geq J_t^s(\chi) \geq 0$.} Next, we show the second equality in \eqref{vtpi}. Let $(x,z,u) = (\chi,u) \in \mathbb{S} \times C$ be given. Since $J_{t+1}^s : (\mathbb{S},\mathcal{B}_{\mathbb{S}}) \rightarrow (\mathbb{R}^*,\mathcal{B}_{\mathbb{R}^*})$ is nonnegative, the integral $\int_{\mathbb{S}} J_{t+1}^s \; \mathrm{d}\tilde{q}_t(\cdot|\chi,u)$ exists. We apply the Classical Fubini Theorem \cite[Th. 2.6.6, p. 103]{ash1972} using the product measure $\tilde{q}_t(\cdot|x,z,u) \in \mathcal{P}(\mathbb{S})$ of $q_t(\cdot|x,u) \in \mathcal{P}(S)$ and $\overline{q}_t(\cdot|x,z,u) \in \mathcal{P}(\mathcal{Z})$ to find
\begin{equation}\begin{aligned}
    \int_{\mathbb{S}} J_{t+1}^s \; \mathrm{d}\tilde{q}_t(\cdot|x,z,u) & = \int_{S} \int_{\mathcal{Z}} J_{t+1}^s(x',z') \; \overline{q}_t(\mathrm{d}z'|x,z,u) \; q_t(\mathrm{d}x'|x,u).
\end{aligned}\end{equation}
Since $\overline{q}_t(\cdot|x,z,u) = \delta_{\max\{c_t(x, u), z \}}$, we have
\begin{equation}\label{292}
    \int_{\mathbb{S}} J_{t+1}^s \; \mathrm{d}\tilde{q}_t(\cdot|x,z,u) = \int_{S} \int_{\mathcal{Z}} J_{t+1}^s(x',z') \; \delta_{\max\{c_t(x, u), z \}}(\mathrm{d}z') \; q_t(\mathrm{d}x'|x,u).
\end{equation}
For every $x' \in S$, $J_{t+1}^s(x',\cdot) : \mathcal{Z} \rightarrow \mathbb{R}^*$ is Borel-measurable, and thus,
\begin{equation}\label{293}
    \int_{\mathcal{Z}} J_{t+1}^s(x',z') \; \delta_{\max\{c_t(x, u), z \}}(\mathrm{d}z') \overset{\eqref{diraceq}}{=} J_{t+1}^s(x',\max\{c_t(x, u), z \}).
\end{equation}
By \eqref{292} and \eqref{293}, we have
\begin{equation}
    \int_{\mathbb{S}} J_{t+1}^s \; \mathrm{d}\tilde{q}_t(\cdot|x,z,u) = \int_{S} J_{t+1}^s(x',\max\{c_t(x, u), z \}) \; q_t(\mathrm{d}x'|x,u).
\end{equation}
Finally, we use the definition
\begin{equation}
    q_t(\underline{S}|x,u) = p_t(\{w \in D : f_t(x,u,w) \in \underline{S}\} |x,u), \quad \quad \underline{S} \in \mathcal{B}_{S},
\end{equation}
and the change-of-variable image measure theorem \cite[Th. 1.6.12, p. 50]{ash1972} to find that
\begin{equation}\label{my296}\begin{aligned}
    \int_{\mathbb{S}} J_{t+1}^s \; \mathrm{d}\tilde{q}_t(\cdot|x,z,u) & \overset{\hphantom{\eqref{27c}}}{=} \int_{D} J_{t+1}^s(f_t(x,u,w),\max\{c_t(x, u), z \}) \; p_t(\mathrm{d}w|x,u) \\
    & \overset{\eqref{27c}}{=} v_t^s(x, z, u).
\end{aligned}\end{equation}
Recalling the notation $\chi = (x,z)$, we have
\begin{equation}
    v_t^s(\chi, u) \overset{\eqref{my296}}{=} \int_{\mathbb{S}} J_{t+1}^s \; \mathrm{d}\tilde{q}_t(\cdot|\chi,u),
\end{equation}
which implies the second equality in \eqref{vtpi}.

By substituting $v_t^{s,\pi}$ \eqref{vtpi} into \eqref{225}, we have
\begin{align}
 & \int_{\Omega} J_{t+1}^s\circ \mathcal{X}_{t+1} \; \mathrm{d}P_{\mathbf{x}}^\pi \label{my299}\\ 
   &    = \begin{cases} \int_{(\mathbb{S} \times C)^{t} \times \mathbb{S}} \textcolor{blue}{v_t^{s,\pi}(\chi_t)} \; \tilde{q}_{t-1}(\mathrm{d}\chi_{t}|\chi_{t-1},u_{t-1}) \cdots \pi_0(\mathrm{d}u_0|\chi_0)\;  \delta_{\mathbf{x},a}(\mathrm{d}\chi_0), & \text{if }t \in \{1,2,\dots,N-1\}, \\ \int_{\mathbb{S}} \textcolor{blue}{v_0^{s,\pi}(\chi_0)} \; \delta_{\mathbf{x},a}(\mathrm{d}\chi_0),  & \text{if }t = 0.\end{cases} \nonumber
\end{align}
Now, for $t = 0$, we have
\begin{equation}\label{my300}
    \int_{\Omega} J_{1}^s\circ \mathcal{X}_{1} \; \mathrm{d}P_{\mathbf{x}}^\pi \overset{\eqref{my299}}{=} \int_{\mathbb{S}} v_0^{s,\pi} \; \mathrm{d}\delta_{\mathbf{x},a} \overset{\eqref{91b}
}{=} \int_{\Omega} v_0^{s,\pi} \circ \mathcal{X}_0 \; \mathrm{d}P_{\mathbf{x}}^\pi.
\end{equation}
We are permitted to apply \eqref{91b} in particular because $v_0^{s,\pi} : \mathbb{S} \rightarrow \mathbb{R}^*$ is Borel-measurable. 

Next, we consider $t \in \{1,2,\dots,N-1\}$. Since $v_t^{s,\pi} : (\mathbb{S}, \mathcal{B}_{\mathbb{S}}) \rightarrow (\mathbb{R}^*,\mathcal{B}_{\mathbb{R}^*})$ is nonnegative and $\mathcal{X}_t : (\Omega,\mathcal{B}_{\Omega}) \rightarrow (\mathbb{S},\mathcal{B}_{\mathbb{S}})$, the map $v_t^{s,\pi} \circ \mathcal{X}_t : (\Omega,\mathcal{B}_{\Omega}) \rightarrow (\mathbb{R}^*,\mathcal{B}_{\mathbb{R}^*})$ is nonnegative. Hence, the integral $\int_{\Omega} v_t^{s,\pi} \circ \mathcal{X}_t \; \mathrm{d}P_{\mathbf{x}}^\pi$ exists, and we have
\begin{align}
 \int_{\Omega} v_t^{s,\pi} \circ \mathcal{X}_t \; \mathrm{d}P_{\mathbf{x}}^\pi 
& \overset{\eqref{154bimportant}}{=} \int_{(\mathbb{S} \times C)^{\textcolor{blue}{t}} \times \mathbb{S}} v_t^{s,\pi}(\chi_{\textcolor{blue}{t}}) \; \tilde{q}_{\textcolor{blue}{t}-1}(\mathrm{d}\chi_{\textcolor{blue}{t}}|\chi_{\textcolor{blue}{t}-1},u_{\textcolor{blue}{t}-1})  \cdots \pi_0(\mathrm{d}u_0|\chi_0) \;  \delta_{\mathbf{x},a}(\mathrm{d}\chi_0)\\
& \overset{\eqref{my299}}{=} \int_{\Omega} J_{t+1}^s\circ \mathcal{X}_{t+1} \; \mathrm{d}P_{\mathbf{x}}^\pi. \label{my302}
\end{align}
By \eqref{my300} and \eqref{my302}, we conclude that
\begin{equation}\label{my228}
    \int_{\Omega} J_{t+1}^s\circ \mathcal{X}_{t+1} \; \mathrm{d}P_{\mathbf{x}}^\pi 
       = \int_{\Omega} v_t^{s,\pi} \circ \mathcal{X}_{t} \; \mathrm{d}P_{\mathbf{x}}^\pi, \quad \quad t \in \{0, 1,2,\dots,N-1\}.
\end{equation}
Now, for every $(\chi,u) \in \mathbb{S} \times C$, it holds that $v_t^s(\chi, u) \geq J_t^s(\chi) \geq 0$, $v_t^s(\chi, \cdot) : C \rightarrow \mathbb{R}^*$ is Borel-measurable, and $\pi_t(\cdot|\chi)$ is a probability measure on $(C,\mathcal{B}_C)$. 
Therefore, we have\footnote{We paraphrase Proposition 1.24(c) from \cite[pp. 19--20]{rudin1987realcomplex}: Let $(X,\mathcal{M},\mu)$ be a measure space, $f : X \rightarrow [0,+\infty]$ be measurable, and $E \in \mathcal{M}$. If $0 \leq c < +\infty$, then $\int_E cf \; \mathrm{d}\mu = c \int_E f \; \mathrm{d}\mu$. Next, we refer to \cite[Exercise 13, p. 32]{rudin1987realcomplex}: Show that Proposition 1.24(c) is also true when $c = +\infty$. These two statements are useful for us because we would like to evaluate $\int_C J_t^s(\chi) \; \pi_t(\mathrm{d}u|\chi)$, where $0 \leq J_t^s(\chi) \leq +\infty$. From the previous discussion, we have $\int_C J_t^s(\chi) \; \pi_t(\mathrm{d}u|\chi) = J_t^s(\chi) \int_C \pi_t(\mathrm{d}u|\chi) = J_t^s(\chi)$. \label{rudinfootnote}}
\begin{equation}\label{187}
  \forall \chi \in \mathbb{S}, \;\;\;\;\;  v_t^{s,\pi}(\chi) \overset{\eqref{vtpi}}{=} \int_C v_t^s(\chi, u) \; \pi_t(\mathrm{d}u|\chi) \geq \int_C J_t^s(\chi) \; \pi_t(\mathrm{d}u|\chi) = J_t^s(\chi) \geq 0.
\end{equation}
Since $v_t^{s,\pi} \geq J_t^s \geq 0$, we also have
\begin{equation}\label{237}
  \forall \omega \in \Omega, \;\;\;\;\;\;\;\;\;\;  v_t^{s,\pi}(\mathcal{X}_t(\omega)) \geq J_t^s(\mathcal{X}_t(\omega)) \geq 0.
\end{equation}
Note that $v_t^{s,\pi} \circ \mathcal{X}_t : \Omega \rightarrow \mathbb{R}^*$ and $J_t^s \circ \mathcal{X}_t : \Omega \rightarrow \mathbb{R}^*$ are Borel-measurable because $v_t^{s,\pi}$ and $J_t^s$ are l.s.c. and $\mathcal{X}_t$ is Borel-measurable. All together, we have
\begin{align}
    \int_{\Omega} J_{t+1}^s \circ \mathcal{X}_{t+1} \; \mathrm{d}P_{\mathbf{x}}^\pi \overset{\eqref{my228}}{=} \int_{\Omega}  v_t^{s,\pi} \circ \mathcal{X}_t \; \mathrm{d}P_{\mathbf{x}}^\pi \overset{\eqref{237}}{\geq} \int_{\Omega} J_t^s \circ \mathcal{X}_t\; \mathrm{d}P_{\mathbf{x}}^\pi \geq 0,
\end{align}
which shows \eqref{230}, and therefore, the first induction step \eqref{183} is complete.


We provide a similar induction argument for \eqref{toshow2}. The base case $t=N$ holds by \eqref{showingbasecase2}. Assume (the induction hypothesis for \eqref{toshow2}) that for some $t \in \{N-1,\dots,1,0\}$, it holds that
\begin{equation}\label{inductionhyp22}
  \forall \mathbf{x} \in S, \;\;\;\;\;  \int_{\Omega}  \phi_{t+1}^{\pi^s,s} \circ \mathcal{X}_{t+1}\; \mathrm{d}P_{\mathbf{x}}^{\pi^s}  = \int_{\Omega}  J_{t+1}^s \circ \mathcal{X}_{t+1} \; \mathrm{d}P_{\mathbf{x}}^{\pi^s},
\end{equation}
and we will show that
\begin{equation}\label{183b}
   \forall \mathbf{x} \in S, \;\;\;\;\; \int_{\Omega}  \phi_t^{\pi^s,s} \circ \mathcal{X}_t\; \mathrm{d}P_{\mathbf{x}}^{\pi^s}  = \int_{\Omega}  J_t^s \circ \mathcal{X}_t \; \mathrm{d}P_{\mathbf{x}}^{\pi^s}
\end{equation}
to prove \eqref{toshow2} by induction. Let $\mathbf{x} \in S$ be given. Since $\pi^s \in \Pi$ and $t \in \mathbb{T}$, we use the relation \eqref{66} from Theorem 1 and the induction hypothesis \eqref{inductionhyp22} to find that
\begin{equation}
    \int_{\Omega} \phi_t^{\pi^s,s} \circ \mathcal{X}_t\; \mathrm{d}P_{\mathbf{x}}^{\pi^s}  \overset{\eqref{66}}{=} \int_{\Omega}  \phi_{t+1}^{\pi^s,s} \circ \mathcal{X}_{t+1}\; \mathrm{d}P_{\mathbf{x}}^{\pi^s} \overset{\eqref{inductionhyp22}}{=} \textcolor{blue}{\int_{\Omega}  J_{t+1}^s \circ \mathcal{X}_{t+1} \; \mathrm{d}P_{\mathbf{x}}^{\pi^s}}.
\end{equation}
Therefore, to prove the desired statement \eqref{183b}, it suffices to show that
\begin{equation}\label{243}
    \textcolor{blue}{\int_{\Omega} J_{t+1}^s \circ \mathcal{X}_{t+1} \; \mathrm{d}P_{\mathbf{x}}^{\pi^s}} = \int_{\Omega} J_{t}^s \circ \mathcal{X}_{t} \; \mathrm{d}P_{\mathbf{x}}^{\pi^s}.
\end{equation}
Now, the function $v_t^{s,\pi^s}$ is equivalent to $J_t^s$ because for every $\chi \in \mathbb{S}$,
\begin{equation}\label{my236}
    \textcolor{magenta}{v_t^{s,\pi^s}}(\chi) \overset{\eqref{vtpi}}{=} \int_C v_t^s(\chi, u) \; \delta_{\kappa_t^s(\chi)}(\mathrm{d}u) \overset{\eqref{diraceq}}{=} v_t^s(\chi, \kappa_t^s(\chi)) \overset{\eqref{mykappaeq}}{=} \textcolor{magenta}{J_t^s}(\chi),
\end{equation}
noting that $\delta_{\kappa_t^s(\chi)}$ is a Dirac measure on $(C,\mathcal{B}_C)$ and using the expression for $J_t^s$ from \eqref{mykappaeq}. By applying the expression for $\int_{\Omega} J_{t+1}^s \circ \mathcal{X}_{t+1} \; \mathrm{d}P_{\mathbf{x}}^{\pi}$ from \eqref{my228} to the particular policy $\pi = \pi^s$, we find that
\begin{equation}
    \int_{\Omega} J_{t+1}^s\circ \mathcal{X}_{t+1} \; \mathrm{d}P_{\mathbf{x}}^{\pi^s}
     \overset{\eqref{my228}}{=} \int_{\Omega} \textcolor{magenta}{v_t^{s,\pi^s}} \circ \mathcal{X}_{t} \; \mathrm{d}P_{\mathbf{x}}^{\pi^s} \overset{\eqref{my236}}{=} \int_{\Omega} \textcolor{magenta}{J_t^s} \circ \mathcal{X}_{t} \; \mathrm{d}P_{\mathbf{x}}^{\pi^s},
\end{equation}
which shows \eqref{243} and therefore shows \eqref{183b}, completing the proof of Theorem 2. \end{proof}

\bibliography{references}

%% file: main.bbl
\begin{thebibliography}{00}
\bibitem{bertsekas1971minimax} D. P. Bertsekas and I. B. Rhodes, ``On the minimax reachability of target sets and target tubes,'' \emph{Automatica}, vol. 7, no. 2, pp. 233--247, 1971.


\bibitem{lygeros2011} K. Margellos and J. Lygeros, ``Hamilton-Jacobi formulation for reach-avoid differential games,'' \emph{IEEE Trans. Autom. Control}, vol. 56, no. 8, pp. 1849--1861,  2011.


\bibitem{chen2018hamilton} M. Chen and C. J. Tomlin, ``Hamilton-Jacobi reachability: Some recent theoretical advances and applications in unmanned airspace management,'' \emph{Annu. Rev. Control Rob. Auton. Syst.}, vol. 1, no. 1, pp. 333--358, 2018.

\bibitem{sylviathesis} S. L. Herbert, Safe Real-World Autonomy in Uncertain and Unstructured Environments (Doctoral dissertation). Technical Report No. UCB/EECS--2020--147. University of California Berkeley, 2020.

\bibitem{ding2013} J. Ding, M. Kamgarpour, S. Summers, A. Abate, J. Lygeros, and C. Tomlin, ``A stochastic games framework for verification and control of discrete time stochastic hybrid systems,'' \emph{Automatica}, vol. 49, pp. 2665--2674, 2013.  

\bibitem{yang2018dynamic} I. Yang, ``A dynamic game approach to distributionally robust safety specifications for stochastic systems,'' \emph{Automatica}, vol. 94, pp. 94--101, 2018.

\bibitem{abate2008probabilistic} A. Abate, M. Prandini, J. Lygeros, and S. Sastry, ``Probabilistic reachability and safety for controlled discrete time stochastic hybrid systems,'' \emph{Automatica}, vol. 44, no. 11, pp. 2724--2734, 2008.

\bibitem{summers2010verification} S. Summers and J. Lygeros, ``Verification of discrete time stochastic hybrid systems: A stochastic reach-avoid decision problem,'' \emph{Automatica}, vol. 46, no. 12, pp. 1951--1961, 2010.

\bibitem{samuelson2018safety} S. Samuelson and I. Yang, ``Safety-aware optimal control of stochastic systems using Conditional Value-at-Risk,'' in \emph{Proc. Am. Control Conf.}, pp. 6285--6290, 2018.

\bibitem{chapmanACC} M. P. Chapman, J. Lacotte, A. Tamar, D. Lee, K. M. Smith, V. Cheng, J. F. Fisac, S. Jha, M. Pavone, and C. J. Tomlin, ``A risk-sensitive finite-time reachability approach for safety of stochastic dynamic systems,'' in \emph{Proc. Am. Control Conf.}, pp. 2958--2963, 2019.

\bibitem{Safaoui2020} S. Safaoui, L. Lindemann, D. V. Dimarogonas, I. Shames, and T. H. Summers, ``Control design for risk-based signal temporal logic specifications,'' \emph{IEEE Control Syst. Lett.}, vol. 4, no. 4, pp. 1000--1005, 2020.

\bibitem{chapmantac2021} M. P. Chapman, R. Bonalli, K. Smith, I. Yang, M. Pavone, and Claire J. Tomlin, ``Risk-sensitive safety analysis using Conditional Value-at-Risk,'' \emph{IEEE Trans. Autom. Control}, in press, 2022. 

\bibitem{lindemann2021reactive} L. Lindemann, G. J. Pappas, and D. V. Dimarogonas, ``Reactive and risk-aware control for signal temporal logic,'' \emph{IEEE Trans. Autom. Control}, in press, 2022.

\bibitem{howardmat1972}
R. A. Howard and J. E. Matheson, ``Risk-sensitive Markov decision processes,'' \emph{Manage. Sci.}, vol. 18, no. 7, pp. 356--369, 1972.

\bibitem{jacobson1973} D. H. Jacobson, ``Optimal stochastic linear systems with exponential performance criteria and their relation to deterministic differential games,'' \emph{IEEE Trans. Autom. Control}, vol. 18, no. 2, pp. 124--131, 1973.

\bibitem{whittle1981} P. Whittle, ``Risk-sensitive linear/quadratic/Gaussian control,'' \emph{Adv. Appl. Probab.}, vol. 13, no. 4, pp. 764--777, 1981.

\bibitem{bauerlerieder}
N. B{\"a}uerle and U. Rieder, ``More risk-sensitive Markov decision processes,'' \emph{Math. Oper. Res.}, vol. 39, no. 1, pp. 105--120, 2014.

\bibitem{saldi2020}
N. Saldi, T. Ba\c{s}ar, and M. Raginsky, ``Approximate Markov-Nash equilibria for discrete-time risk-sensitive mean-field games,'' \emph{Math. Oper. Res}, vol. 45, no. 4, pp. 1596--1620, 2020.

\bibitem{chapmansmith2021} 
M. P. Chapman and K. M. Smith, ``Classical risk-averse control for a finite-horizon Borel model,'' \emph{IEEE Control Syst. Lett.}, vol. 6, pp. 1525--1530, 2021.

\bibitem{smithchapman2021} K. M. Smith and M. P. Chapman, ``On Exponential Utility and Conditional Value-at-Risk as risk-averse performance criteria,'' under review for \emph{IEEE Trans. Control Syst. Technol.}, arXiv preprint arXiv:2108.01771.

\bibitem{ruszczynski2010risk} A. Ruszczy{\'n}ski, ``Risk-averse dynamic programming for Markov decision processes,'' \emph{Math. Program.}, vol. 125, no. 2, pp. 235--261, 2010.

\bibitem{singh2018} S. Singh, Y. Chow, A. Majumdar, M. Pavone, ``A framework for time-consistent, risk-sensitive model predictive control: Theory and algorithms,'' \emph{IEEE Trans. Autom. Control}, vol. 64, no. 7, pp. 2905--2912, 2018.

\bibitem{bauerle2020markov}
N. B{\"a}uerle and A. Glauner, ``Markov decision processes with recursive risk measures,'' \emph{Eur. J. Oper. Res.}, vol. 296, no. 3, pp. 953--966, 2022.

\bibitem{shapiro2009lectures} A. Shapiro, D. Dentcheva, and A. Ruszczy{\'n}ski, \emph{Lectures on Stochastic Programming: Modeling and Theory}. MOS-SIAM, 2009.

\bibitem{bauerle2020minimizing}
N. B{\"a}uerle and A. Glauner, ``Minimizing spectral risk measures applied to Markov decision processes,'' \emph{Math. Methods Oper. Res.}, vol. 94, no. 1, pp. 35--69, 2021.

\bibitem{Artzner}
P. Artzner, F. Delbaen, J. M. Eber, and D. Heath, ``Coherent measures of risk,'' \emph{Math. Finance}, vol. 9., no. 3, pp. 203--228, 1999.

\bibitem{rockafellar2002conditional} R. T. Rockafellar and S. Uryasev, ``Conditional Value-at-Risk for general loss distributions,'' \emph{J. Bank. Financ.}, vol. 26, no. 7, pp. 1443--1471, 2002.

\bibitem{bauerleott} N. B{\"a}uerle and J. Ott, ``Markov decision processes with Average-Value-at-Risk criteria,''
\emph{Math. Methods Oper. Res.}, vol. 74, no. 3, pp. 361--379, 2011.

\bibitem{borkar}
V. Borkar and R. Jain, ``Risk-constrained Markov decision processes,'' \emph{IEEE Trans. Autom. Control}, vol. 59, no. 9, pp. 2574--2579, 2014.

\bibitem{haskell}
W. B. Haskell and R. Jain, ``A convex analytic approach to risk-aware Markov decision processes,'' \emph{SIAM J. Control Optim.}, vol. 53, no. 3, pp. 1569--1598, 2015.

\bibitem{extremevaluetheory}
L. de Haan and A. Ferreira, \emph{Extreme Value Theory: An Introduction}, New York: Springer, 2006.

\bibitem{bertsekas2004stochastic}
D. P. Bertsekas and S. Shreve, \emph{Stochastic Optimal Control: The Discrete-Time Case}, Belmont: Athena Scientific, 1996.

\bibitem{jaskiew2017}
H. Asienkiewicz and A. Ja\'{s}kiewicz, ``A note on a new class of recursive utilities in Markov decision processes,'' \emph{Applicationes Mathematicae}, vol. 44, no. 2, pp. 149--161, 2017.


\bibitem{wangchapman}
Y. Wang and M. P. Chapman, ``Risk-averse autonomous systems: A brief history and recent developments from the perspective of optimal control,'' \emph{J. Artif. Intell.}, in press, 2022.
%

\bibitem{Tsiamis}
A. Tsiamis, D. S. Kalogerias, L. F. Chamon, A. Ribeiro, and G. J.
Pappas, ``Risk-constrained linear-quadratic regulators,'' in \emph{Proc. IEEE Conf. Decis. Control}, pp. 3040--3047, 2020.









\bibitem{folland2013real}
G. B. Folland, \emph{Real Analysis: Modern Techniques and Their Applications}, 2nd Edition, New York: John Wiley \& Sons, Inc., 1999.

\bibitem{dudley2018real} R. M. Dudley, \emph{Real Analysis and Probability}. Boca Raton: CRC Press, 1989.




\bibitem{dynkin1979controlled}
E. B. Dynkin and A. A. Yushkevich,  \emph{Controlled Markov Processes}, Springer, 1979.










\bibitem{rudin1987realcomplex} W. Rudin, \emph{Real and Complex Analysis}, 3rd Edition, New York: McGraw-Hill Book Company, 1987.





\bibitem{van2015distributionally} B. P. G. Van Parys, D. Kuhn, P. J. Goulart, and M. Morari, ``Distributionally robust control of constrained stochastic systems,'' \emph{IEEE Trans. Automat. Control}, vol. 61, no. 2, pp. 430--442, 2015.






\bibitem{hernandez2012discrete}
O. Hern{\'a}ndez-Lerma and J. B. Lasserre, \emph{Discrete-Time Markov Control Processes: Basic Optimality Criteria}, New York: Springer, 1996.

\bibitem{ash1972}
R. B. Ash, \emph{Real Analysis and Probability}, New York: Academic Press, 1972.















\bibitem{smartwater} A. Mullapudi, B. P. Wong, and B. Kerkez, ``Emerging investigators series: Building a theory for smart stormwater systems,'' \emph{Environmental Science: Water Research \& Technology}, vol. 3, no. 1, pp. 66--77, 2017.

\bibitem{sustech} M. P. Chapman, K. M. Smith, V. Cheng, D. L. Freyberg, and C. J. Tomlin, ``Reachability analysis as a design tool for stormwater systems,'' in \emph{Proc. IEEE Conf. Technol. Sustain.}, pp. 1--8, 2018.


\bibitem{swmm} L. A. Rossman, \emph{Storm Water Management Model User's Manual}, Version 5.0. National Risk Management Research Laboratory, Office of Research and Development, US EPA, Cincinnati, 2010.


\bibitem{petrucci2013}
G. Petrucci, E. Rioust, J.-F. Deroubaix, and B. Tassin, ``Do stormwater source control policies deliver the right hydrologic outcomes?,'' \emph{Journal of Hydrology}, vol. 485, pp. 188--200, 2013.

\bibitem{emerson2005}
C. H. Emerson, C. Welty, and R. G. Traver, ``Watershed-Scale Evaluation of a System of Storm Water Detention Basins,'' \emph{Journal of Hydrologic Engineering}, vol. 10, no. 3, pp. 237--242, 2005.

\end{thebibliography}
